\documentclass[10pt,a4paper]{report}

    \def\href#1#2{#2}

\usepackage{fancyheadings}

\usepackage{ifthen}
\usepackage{enumerate}
\usepackage{amssymb}
\usepackage[mathscr]{eucal}
\usepackage[errorshow]{tracefnt}
\usepackage{amsmath}
\usepackage{amssymb}
\usepackage{array}
\usepackage{amsthm}
\usepackage{eucal}
\usepackage{eufrak}
\usepackage{epsf}
\usepackage{epsfig}
\usepackage{pgf,pgfarrows,pgfautomata,pgfheaps,pgfnodes,pgfshade}
\usepackage[apple mac]{inputenc}

\usepackage{picins}

\usepackage{subfigure}

\newcommand\be{\begin{equation}}
\newcommand\bea{\begin{eqnarray}}
\newcommand\ee{\end{equation}}
\newcommand\eea{\end{eqnarray}}

\newcommand\bref[1]{(\ref{#1})}

\newcommand\h{\frac{1}{2}}

\newcommand\f[2]{\frac{#1}{#2}}

\newcommand{\nn}{\nonumber \\}

\newcommand\bra[1]{\langle #1 |}
\newcommand\ket[1]{| #1 \rangle}

\newcommand\p{\phantom{a}}

\newcommand{\Regge}{\alpha'}

\newcommand{\rom}[1]{\mathrm{#1}}

\newcommand{\beq}{\begin{equation}}
\newcommand{\eeq}{\end{equation}}
\newcommand{\beqa}{\begin{eqnarray}}
\newcommand{\eeqa}{\end{eqnarray}}
\newcommand{\beqar}{\begin{eqnarray*}}
\newcommand{\eeqar}{\end{eqnarray*}}

\def\f{\frac}

\newcommand{\n}{\nabla }

\RequirePackage{hyperref}

\numberwithin{equation}{section}

\begin{document}

 \pagestyle{empty}

\begin{center}

\vspace*{2cm}

\noindent {\LARGE\textsf{\textbf{Modave Lectures on Fuzzballs and Emission from \\ \vskip 0.3cm the D1-D5 System}}}
\vskip 2truecm


\begin{center}
{\large \textsf{\textbf{Borun D. Chowdhury$^\Diamond$ and Amitabh Virmani$^\sharp$}}} \\
\vskip 1truecm
     $\Diamond$ {\it   Instituut voor Theoretische Fysica, \\
     Universiteit van Amsterdam,\\
    Valckenierstraat 65, \\
   1018 XE Amsterdam, The Netherlands\\
        [3mm]e-mail:
        {\tt B.D.Chowdhury@uva.nl}} \\
\vskip 1truecm
    $^\sharp$   {\it   {Physique Th\'eorique et Math\'ematique, \\
        Universit\'e Libre
de Bruxelles \& International Solvay Institutes, \\
        ULB-Campus Plaine C.P. 231, B-1050, Bruxelles, Belgium \\
        [3mm]e-mail:}
        {\tt avirmani@ulb.ac.be}} \\
\end{center}
\vskip .5 cm

\small{Based on lectures given by AV at the Fifth
International Modave Summer School on Mathematical Physics, held in
Modave, Belgium, August 2009.}

\vskip 1 cm
\centerline{\sffamily\bfseries Abstract}
\end{center}

\noindent
These lecture notes present an introduction to the fuzzball proposal and emission from the D1-D5 system which is geared to an audience of graduate students and
others with little background in the subject.  The presentation begins with a discussion of the Penrose process and Hawking radiation. The fuzzball proposal
is then introduced,  and the two- and three-charge systems are reviewed.  In the three-charge case details are not discussed. A detailed
discussion of emission calculations for D1-D5-P black holes and for certain non-extremal
fuzzballs from both the gravity and CFT perspectives is included. We explicitly demonstrate how seemingly different emission processes in gravity,
namely, Hawking radiation and superradiance from
D1-D5-P black holes, and ergoregion emission from certain non-extremal fuzzballs, are only different manifestations of the same phenomenon in the CFT.
\newpage

\pagestyle{plain}
\tableofcontents
\chapter{Introduction}
Black holes provide a deep and satisfying connection between gravitational physics, thermodynamics, and quantum mechanics. They also provide the best theoretical laboratory
for studying problems of quantum gravity.  In this regard, they not only pose well-defined questions, but also spring forth numerous deep puzzles. Perhaps the most
debated and the sharpest of these puzzles is the Hawking information paradox. The information paradox puts quantum mechanics and general relativity in sharp
contrast. It leads us to conclude that pure quantum states can evolve into mixed states  in processes involving formation and complete evaporation of black holes.
Such  evolutions are forbidden by the usual rules of quantum mechanics. A working theory of quantum gravity must resolve issues raised by the information paradox.

String theory is a quantum theory of gravity.  It has provided significant insights in answering many puzzles surrounding properties of black holes. One of the greatest successes of string theory
is the Sen-Strominger-Vafa counting of the microscopic configurations of the D1-D5-P system, and thereby reproducing the Bekenstein-Hawking entropy of the
corresponding black hole \cite{Sen:1995in,Strominger:1996sh}. After Strominger and Vafa several authors have counted microscopic configurations at weak
coupling using brane physics and have reproduced the Bekenstein-Hawking entropy of
 a variety
of extremal and near-extremal black holes (see, e.g., reviews \cite{Peet:1997es, Peet:2000hn, David:2002wn, Mathur:2005ai} and references therein). These calculations---important as they are---do not tell us anything about how
the individual microstates look like in the strong coupling description, and in particular, in the gravity
description. These calculations also do not shed much light on the Hawking information paradox or on the intrinsic nature of the
Bekenstein-Hawking entropy  from a purely gravitational perspective.

To address these and related questions Samir Mathur and his collaborators have put forward a bold proposal---the so-called Fuzzball Proposal.
In this proposal, a black hole
geometry is an effective description of an ensemble of states. In the framework of this
proposal, quantum gravity effects are not confined to the Planck length. Typical states
of the ensemble have structure up to the scale of the horizon. Radiation
from non-extremal would-be black holes does not happen by a pair production process from
the vacuum, but rather happens from the surfaces of the black hole's microstates. Consequently,
the emitted radiation is capable of carrying information, and seemingly the information
paradox does not arise.

The fuzzball proposal has met with most success in the case of the extremal two-charge string theory
black holes. For the two-charge system all microstates have been identified in the gravity description and it is
shown that typical states have a size that scales as a function of charges as it would for
a would-be black hole. For other black holes the situation is much less developed.

In this review we discuss the fuzzball proposal and related ideas.  Our presentation  is geared to an audience of graduate students and others with
little background in the subject. The review is organized as follows; for more detailed overviews see the chapter introductions.
We begin with a discussion of the Penrose process and Hawking radiation in chapter \ref{infoparadox}. In chapter \ref{Fuzzballs}
the fuzzball proposal is introduced, and the two- and three-charge systems are reviewed. Our discussion of the three-charge system should be regarded as a starting
tutorial on the subject rather than a review. A detailed discussion of emission calculations for D1-D5-P black holes and for certain non-extremal
fuzzballs from both the gravity and CFT perspectives is included in chapter \ref{emission}. This chapter is the main emphasis of the review. Here we explicitly
demonstrate how seemingly different emission processes in gravity, namely, Hawking radiation
and superradiance from D1-D5-P black holes, and ergoregion emission from certain
non-extremal fuzzballs, are only different manifestations of the same phenomenon in the dual
CFT.


\chapter{The Information Paradox}
\label{infoparadox}
In this chapter we review the Hawking information paradox.
\section{Penrose Process}
\label{penrose}
Historically the roots of the Hawking's remarkable discovery lie in the Penrose process, which we very briefly review in this section. For more details we refer the reader to standard references such as \cite{Wald:1984rg}. Our presentation follows \cite{Jacobson:2003vx}. This discussion also sets the stage for our later chapters where we discuss superradiant scattering from a rotating black hole.

\begin{figure}[ht] 
\begin{center}
\subfigure[]{
	\includegraphics[width=6.3cm]{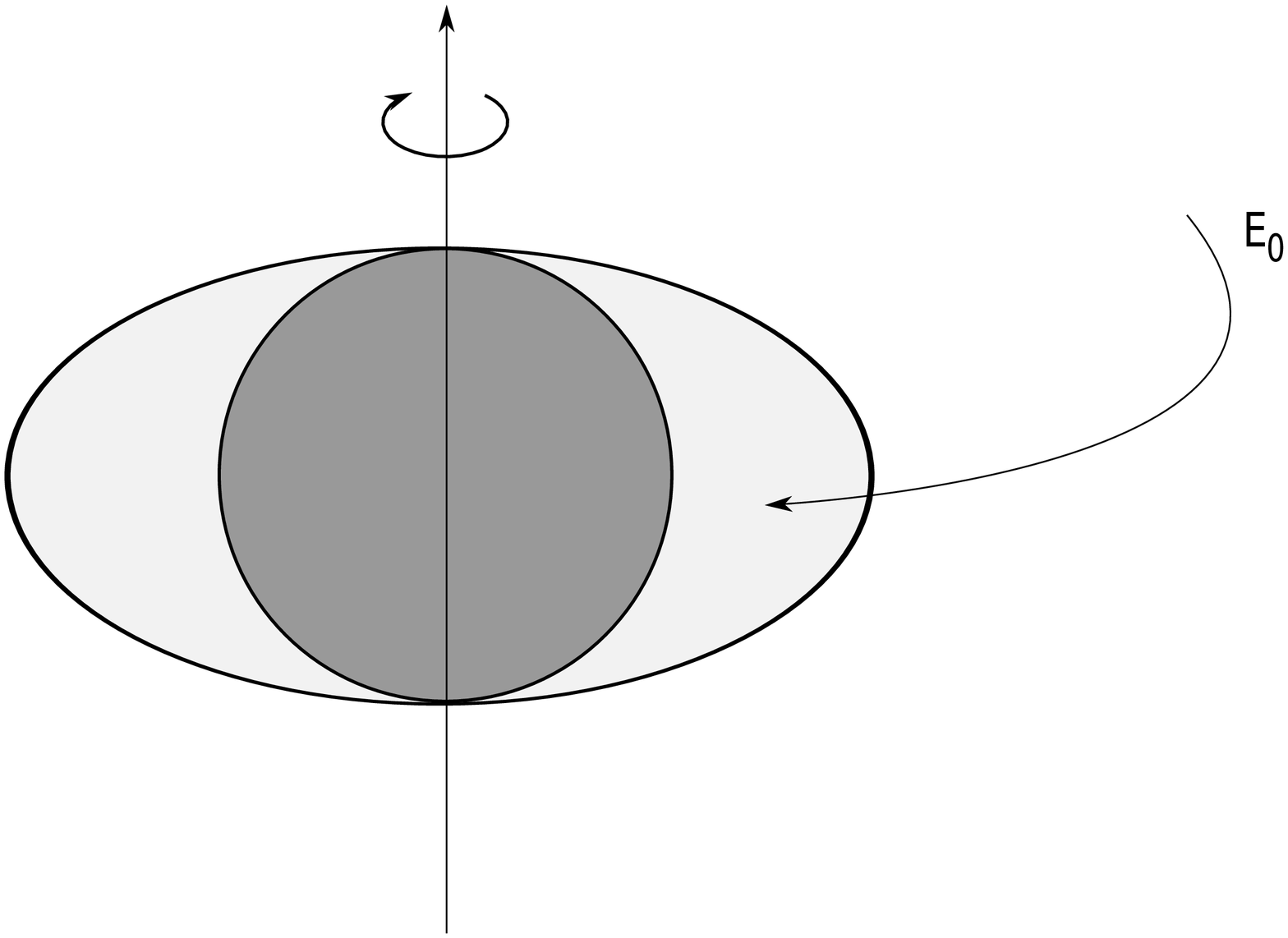}}
\hspace{15pt}
\subfigure[]{
	\includegraphics[width=6.3cm]{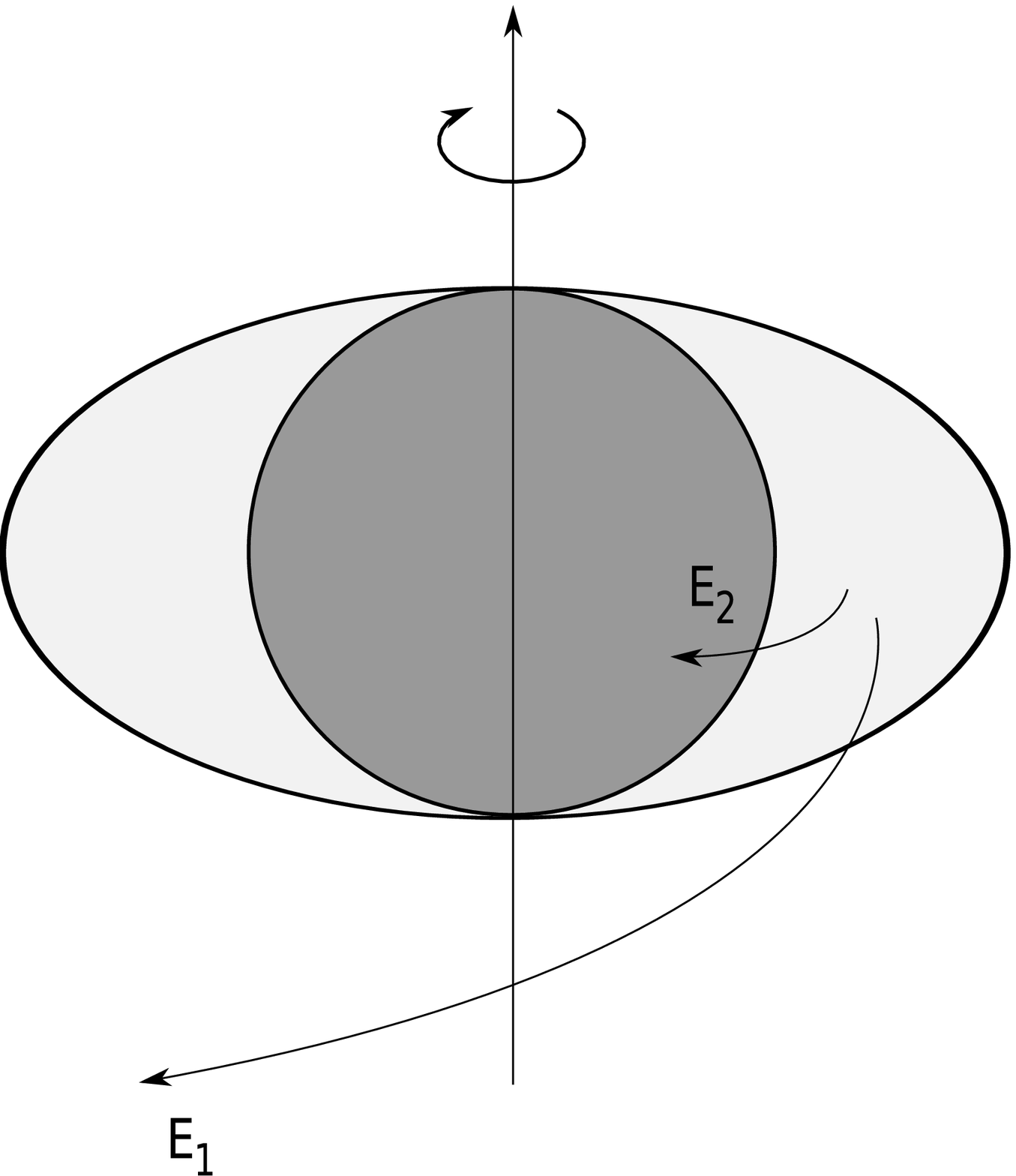}}
   \caption{The Penrose process. In (a) a particle of energy $E_0>0$ is sent into the ergoregion. In (b) the particle splits into two particles; one with energy $E_1 > E_0$ escaping out,  and the other with energy $E_2<0$ falling into the horizon.}
   \label{fig:Penrose}
   \end{center}
\end{figure}

Let us start by recalling that for the stationary Kerr metric the asymptotic time translation
Killing field $t^\mu$ becomes spacelike even in a region outside the event horizon. The region  outside the horizon where it is
spacelike is called the ergoregion, and the surface where it becomes null is called the ergosphere. For a point particle with four momentum $p^{\mu}$ propagating in the Kerr geometry, the conserved Killing energy is $E = t^{\mu} p_{\mu}$. For physical particles the Killing energy is positive provided the Killing vector $t^{\mu}$ is also future directed and timelike. However, where $t^{\mu}$ becomes spacelike, i.e., in the ergoregion, some physical four-momenta can have negative Killing energy. This fact is at the heart of the Penrose process. We depict the Penrose process in figure \ref{fig:Penrose}. In this process, a particle of energy $E_0 > 0$ is sent from outside into the ergoregion of a rotating black hole where it breaks up into two pieces with Killing energies $E_1$ and $E_2$. The total energy is conserved $E_0 = E_1 + E_2$, but the energy $E_2$ is arranged to be negative. The particle with energy $E_1$ comes out whereas the particle with energy $E_2$ falls into the black hole. As a result,  $E_1 > E_0$;  more energy comes out than entered. The black hole absorbs the negative energy $E_2$ and
hence loses mass. It also loses angular momentum. In the Penrose process the area of the horizon either increases or remains unchanged.

A very similar phenomenon occurs for a classical field scattering from a rotating black hole spacetime. A scattering in which a wave-packet of more energy than the one sent-in comes out is called a superradiant scattering. This is similar to the lasing action\footnote{In chapter \ref{emission} we will see that the superradiant scattering is exactly the lasing action when interpreted in the CFT language.}.  The analogy with lasers immediately suggests that the amplification of the energy of the wave-packet is stimulated emission of quanta from the rotating black hole. Thus, it is expected that quantum fields exhibit not only stimulated emission but also spontaneous emission in a rotating black hole spacetime; though the spontaneous emission is not captured in the classical description of the scattering. Hawking in his pioneering work showed that spontaneous quantum emission occurs even in a non-rotating black hole spacetime\footnote{A history of these developments can be found in, e.g., \cite{Page:2004xp}.}. This phenomenon is called the Hawking effect.
It is a pair creation
process in which one particle of the pair has negative energy and it falls inside the horizon, while the other particle  has positive energy and it stays outside and escapes to infinity. The Hawking radiation comes out with a thermal spectrum
at the temperature $T_H = \frac{\hbar c^3 }{8 \pi G M k_B}$. We review the Hawking effect in more detail in the next section. In the following we work with Planck units: $\hbar=c=G=k_B=1$.

\section{Hawking Radiation}

Hawking in his pioneering work \cite{Hawking:1974sw} showed that black holes spontaneously emit thermal radiation. In this section we rederive Hawking's result. Our derivation closely follows the one given in \cite{Wald:1984rg, Kiefer:1999az}. See also \cite{Jacobson:2003vx, Wald:1995yp, Traschen:1999zr, Ross:2005sc}.

\subsection{Quantum Field Theory in Curved Spacetime}
We start with a brief review of some basic concepts related to quantum fields in curved spacetime. Let us consider a free massless scalar field on a curved background satisfying the Klein-Gordon equation
\begin{equation}
\f{1}{\sqrt{-g}} \partial_\mu(\sqrt{-g} \partial^\mu \Phi)=0~.
\end{equation}
For a solution $A$ of this equation we define the conjugate momenta as $\pi_A = n^\mu \partial_\mu A$, where $n^\mu$ is the vector normal to a Cauchy surface $\Sigma$.
If $A$ and $B$ are two solutions to the Klein-Gordon equation, we define the Klein-Gordon inner product as
\bea
\langle A , B \rangle &=& - i \int_{\Sigma} d\Sigma^\mu ( A \, \partial_\mu B^* -  \partial_\mu A \, B^*) \nonumber \\
&=& - i \int_{\Sigma} dV  ( A \, \pi^*_B  -  \pi_A \, B^*)~.
\eea
This inner product is conserved on-shell but is not positive definite.

One can now choose a complete set of solutions  $\{u_k, u_k^* \}$ of the Klein-Gordon equation (here $k$ is a generalized index which would be momentum in flat spacetime) normalized as
\bea
\langle u_k, u_{k'} \rangle = \delta(k - k'), \qquad \langle u^*_k, u^*_{k'} \rangle = - \delta(k -k'), \qquad \langle u_k, u^*_{k'} \rangle = 0~.
\eea
We now expand the field $\Phi$ in this complete basis as
\be
\Phi = \int d \mu (k) ( a_k u_k +  a_k^\dagger u_k^*)~, \label{expansion}
\ee
with $d\mu(k)$ being the measure used in the normalization of the delta functions
$
\int d\mu(k) \delta(k)=1.$ The coefficients in the expansion \eqref{expansion} can be extracted using the inner product
\be
a_k = \langle u_k , \Phi \rangle, \qquad  a_k^\dagger = -\langle u^*_k , \Phi \rangle~.
\ee

To canonically quantize the scalar field we simply write the equal time commutation relation
\be
[\Phi(x,t),\Pi(y,t)] = i \delta(x-y)~.
\ee
This commutation relation leads to the mode algebra
\be
[a_k, a^\dagger_{k'}] = \delta(k-k'), \qquad  [a_k, a_{k'}] =0, \qquad [a^\dagger_k, a^\dagger_{k'}] =0~.
\ee
As in standard quantum mechanics, we define the Fock vacuum $\ket{0}_u$ for the $u$-basis as the state annihilated by all annihilation operators $a_k$
\be
a_k \ket{0}_u =0~.
\ee
Excited states are obtained by acting with the creation operators $a^\dagger_{k}$ on the vacuum $\ket{0}_u$. The number of particles in an excited state is the eigenvalue of the number operator
$N_k = a_k^\dagger a_k\,$.

One can also consider a different complete set of solutions $\{v_k, v^*_{k'}\}$ normalized in the same way and expand $\Phi$ in terms of this set with annhilation and creation operators $b_k, b^\dagger_k$. The vacuum with respect to the $v$-modes is defined as
\be
b_k \ket{0}_v =0~,
\ee
and in general
\be
\p_v \langle 0|0 \rangle_u \ne 1~.
\ee
Expansion of a $v$-mode in terms of the $u$-basis takes the following form
\be
v_k = \int d\mu(k')  \left(\alpha_{kk'} u_{k'} + \beta_{kk'} u^*_{k'} \right)~,
\ee
where $\alpha_{kk'}, \beta_{kk'}$ are the so-called Bogoliubov coefficients. They obey
\bea
\int d \mu(k') (\alpha_{k k'} \alpha^*_{ k'' k'} - \beta_{k k'} \beta^*_{k'' k'}) &=& \delta(k-k'')~, \nn
\int d \mu(k')  (\beta_{kk'} \alpha_{k'' k'} - \alpha_{k k'} \beta_{k'' k'}) &=& 0~. \label{BogolubovRelations}
\eea
From these relations we get
\be
\p_u \bra{0} b_k^\dagger b_k \ket{0}_u =  \int d\mu(k') \beta_{kk'} \beta^*_{kk'}~, \label{BogolubovRelations2}
\ee
and
\be
\p_v \bra{0} a_k^\dagger a_k \ket{0}_v =  \int d\mu(k') \beta_{k'k} \beta^*_{k'k}~.
\ee
Thus, the Fock vacuum $\ket{0}_u$ is in general populated with the $v$-modes. Similarly, the Fock vacuum $\ket{0}_v$ is populated with the $u$-modes.

Different choices for the complete set of solutions for the Klein-Gordon equation correspond to different choices of the observer.  It is easy to verify that in flat spacetime  the vacuum states for all inertial observers are the same. However, the vacuum states for a uniformly accelerated observer and an inertial observer are not the same.
A  uniformly accelerated observer sees  the vacuum state of an inertial observer in thermal equilibrium at a non-zero temperature \cite{Unruh:1976db}.
This is known as the Unruh effect.

Let us emphasize that in a general curved spacetime there is no preferred set of solutions $\{ u_k, u^*_{k'} \}$, and hence there is no natural choice for the vacuum state. As a consequence, there is no natural notion of particles. However, in stationary spacetimes the notion of particles is physically and mathematically well defined. If the spacetime admits a globally well defined timelike Killing vector $\xi$ with its norm bounded away from zero,  then there are distinguished modes of positive (and negative) frequencies whose Lie derivative along the Killing vector obey
\be
\mathcal L_\xi u_k = \mp i \omega_k u_k~.
\ee

The  analysis presented above is at the heart of the Hawking effect as we review now. By looking at the infalling vacuum in terms of the outgoing modes at late times it leads us to conclude that black holes radiate with a thermal spectrum.

\subsection{Hawking Radiation}
For the Schwarzschild black hole
\be
ds^2= - \left(1- \f{2M}{r}\right)dt^2 + \f{dr^2}{\left(1- \f{2 M}{r}\right)} + r^2 d\theta^2 + r^2 \sin^2 \theta d \phi^2~,
\ee
we study the Hawking effect by looking at the s-wave of a scalar field in the black-hole background. Since we are working with the s-wave, from now on we restrict our attention to only the temporal and radial part of the metric. To solve the Klein-Gordon equation in the Schwarzschild background we take the ansatz
\be
\Phi = e^{-i \omega t} \f{f(r)}{r}~,
\ee
which gives the wave equation
\be
\f{\partial^2 f(r_*)}{\partial r_*^2} + (\omega^2 - V(r) ) f(r_*)=0~,
\ee
where
\be
V(r)=\left(1- \f{2 M}{r}\right) \f{2M}{r^3}~. \label{BHPotential}
\ee
The tortoise radial coordinate $r_*$ is defined as \be r_*= r+ 2M \log \left|\f{r}{2M}-1\right|~.\ee
Next, we define the null coordinates
\be
u=t- r_*~, \qquad v= t+r_*~,
\ee
and the conformally related Kruskal coordinates as
\be
U= - e^{- \kappa u}~, \qquad V= e^{\kappa v}~,
\ee
where
$
\kappa = \f{1}{4M}
$ is the surface gravity of the Schwarzschild black hole.  Note that the potential \bref{BHPotential} goes to zero near the horizon $( r_* \to  - \infty)$ and also near infinity $(r_* \to \infty)$. Thus in these two regions we have the positive frequency solutions as
\be
f_\omega^{(out)}(u)=\f{1}{\sqrt{2 \omega}} e^{-i \omega u} , \qquad f_\omega^{(in)}(v) = \f{1}{\sqrt{2 \omega}} e^{-i \omega v}~. \label{modes}
\ee

\begin{figure}[ht] 
\begin{center}
	\includegraphics[width=6.3cm]{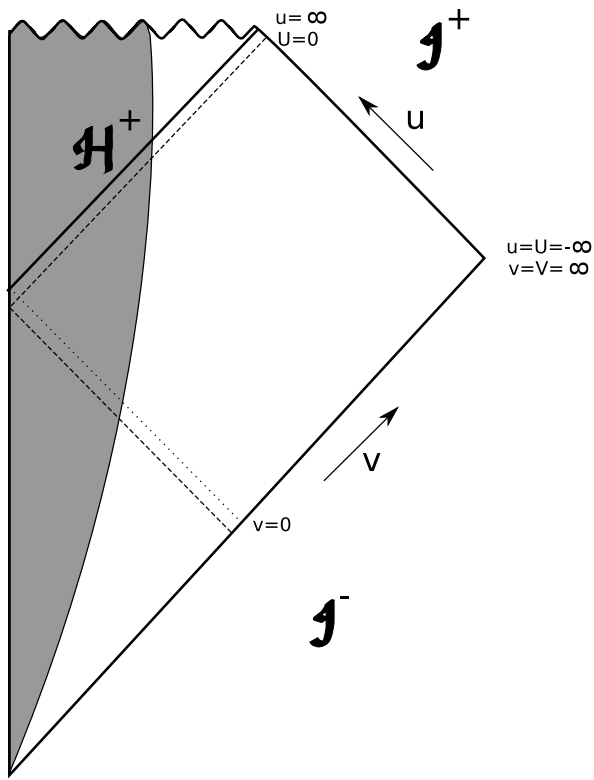}
   \caption{A Penrose diagram of a spherically symmetric spacetime in which gravitational collapse to a Schwarzschild black hole takes place. Outgoing waves starting from $\mathcal{J}^+$
   propagated backwards end up on $\mathcal{J}^-$.}
   \label{fig:collapse}
   \end{center}
\end{figure}

Our strategy to see the Hawking effect is to look at the infalling vacuum and study how it looks in terms of the outgoing modes at late times in
a collapsing black hole spacetime.  This is most easily done by looking at the `time reversed' situation, that is, by propagating outgoing waves
\emph{backward} in time and studying their decomposition in terms of ingoing positive and negative frequency modes at past null infinity.
To this end let us start by looking at the Penrose diagram in figure \ref{fig:collapse} of a black hole formed in a gravitational collapse.
Notice from equation (\ref{modes}) that at late times, $ u\to \infty$, the phase of the outgoing modes oscillate rapidly. Therefore geometrical optics
approximation is valid, and it becomes better and better as $ u\to \infty$.  Outgoing waves propagated backwards starting from $\mathcal{J}^+$ all end up on $\mathcal{J}^-$. This is illustrated in figure \ref{fig:collapse}. The approximate waveform on $\mathcal{J}^-$ can be obtained as follows.

It can be easily shown that at late times near the future null infinity the metric is
\be
ds^2 \simeq - dU dV~.
\ee
So $U=\lambda$ is the affine parameter and we get
\be
f_\omega^{(out)}=\f{1}{\sqrt{2 \omega}} e^{-i \omega u} = \f{1}{\sqrt{2 \omega}} e^{i \f{ \omega}{\kappa} \log (-\lambda)}~.
\ee
From  figure \ref{fig:collapse} we see that the horizon is at $U=\lambda=0$. By definition this is the last ray which makes it out of the black hole.
We trace back this ray to past null infinity. We call this ray $\gamma_\rom{H}$. We define $v=0$ to be the point where $\gamma_\rom{H}$ intersects $\mathcal{J}^-$. The metric at past null infinity is
\be
ds^2 = -du dv~,
\ee
so the affine parameter is simply $v$. Now recall that the behavior of geodesics sufficiently close to $\gamma_\rom{H}$ will be  well described by the geodesic deviation vector. The geodesic deviation vector describing geodesics near $\gamma_{\rom{H}}$ propagates linearly along $\gamma_\rom{H}$.  As a result, we see that the wave-form near $v=0$ will behave as a function of $v$ in the same way as it behaves as a function of the affine parameter along the geodesic tangent to the geodesic deviation vector at any other point on $\gamma_{\rom{H}}$. In particular, we can choose that point to be near $U=0$ at  future null infinity. Thus the projection of the wave-form at future null infinity onto  past null infinity is given by
\be
f^{(out)}_\omega(v)= \left\{
\begin{array}{rl}
0 & v>0 \\
e^{i \f{\omega}{\kappa} \log (-v)} &  v < 0~.
\end{array} \right. \label{Eqn:OutMode}
\ee

Now we want to express these (projected) outgoing modes in terms of ingoing modes on  past null infinity
\be
f^{(out)}_\omega (v) = \int_0^\infty \f{d\omega'}{2\pi} \f{1}{\sqrt{2 \omega'}} ( \alpha_{\omega \omega'} e^{-i \omega' v} +  \beta_{\omega \omega'} e^{i \omega' v} )~.
\ee
Instead of directly evaluating this we look for Fourier modes of \bref{Eqn:OutMode}
\bea
\tilde f_{\omega \omega'} &=& \int_{-\infty}^\infty e^{i \omega' v} f_\omega(v)
= \int_0^\infty dv e^{-i \omega' v} v^{i \f{\omega}{\kappa}}~. \label{integral}
\eea
This integral can be explicitly evaluated. Keeping $\omega' >0$ we can see that (for details see e.g. Appendix A of \cite{Wald:1975kc})~.
\be
\tilde f_{\omega(-\omega')}= - e^{-\pi \f{\omega}{\kappa} } \tilde f_{\omega \omega'} , \qquad \omega'>0
\ee
From this is follows that
\be
\beta_{\omega \omega'} = - e^{-\pi \f{\omega}{\kappa}} \alpha_{\omega \omega'}~. \label{betaAlphaRelation}
\ee
Using \bref{BogolubovRelations}, \bref{BogolubovRelations2} and \bref{betaAlphaRelation} one finds
\be
N_\omega = \int \f{d \omega'}{2 \pi} |\beta_{\omega \omega'}|^2 = t \f{1}{e^{2 \pi \f{\omega}{\kappa}}-1}~.
\ee
In arriving at this expression we have replaced $\delta(0)$ on the right hand side by the large time cutoff $t$ \cite{Giddings:1992ff}.
The total number of particles emitted is therefore infinite. However this infinity is because the black hole radiates for an infinite amount of time. In this calculation the mass decrease of the black hole because  of the radiation is not taken into account. Looking at the number of particles emitted per unit time in the frequency range $\omega$ and $\omega + d \omega$ we see that the black hole radiates at the temperature
\be
T_\rom{BH} = \f{1}{8\pi M}~.
\ee

One can see from the potential \bref{BHPotential} that there is a potential barrier centered around $r=\f{8 M}{3}> 2M$. Thus only a fraction of the flux passes through the barrier. Taking this into account \cite{DeWitt:1975ys} as well as the effect of different angular momenta the total luminosity is given by
\be
L =  - \f{dM}{dt}= \int_0^\infty \f{d \omega}{2\pi} \sum_{l=0}^\infty (2l+1) \f{ \omega \  \Gamma_{\omega l}}{e^{2 \pi \f{\omega}{\kappa}}-1} \label{graybody}
\ee
where $\Gamma_{\omega l }$ are the so-called `greybody  factors.'  They encode the deviation from the blackbody spectrum. They are the fraction of the incoming quanta that fall into the black hole. They are also the fraction of the outgoing quanta emitted at the horizon that make it out to the asymptotic region. We will explicitly calculate these factors for the string theory D1-D5 black holes in section \ref{Section:BHGravityRadiation}.

We can now estimate the lifetime of a black hole by ignoring the greybody factors and using the Stefan-Boltzman law
\be
\f{dM}{dt} = - 4 \pi (2M)^2 \f{1}{(8 \pi M)^4} \propto -\f{1}{M^2}~.
\ee
This gives the lifetime
\be
t \sim M^3~.
\ee
Observe that  the black hole completely evaporates away after a long but finite time.


For rotating black holes almost the same analysis goes through. The most important modification arises from the fact that the Killing vector normal to the horizon is $\partial_t + \Omega_\rom{H} \partial_\phi$ rather than simply $\partial_t$, where $\Omega_\rom{H}$ is the angular velocity of the horizon. This essentially has the effect of the replacement $\omega$ to $\omega - m \,  \Omega_\rom{H}$ in \bref{graybody}. Here $m$ is the azimuthal number of the outgoing wave at $\mathcal{J}^+$. The expression for luminosity now reads
\be
L =  - \f{dM}{dt}= \int_0^\infty \f{d \omega}{2\pi} \sum_{l=0}^\infty \sum_{m=-l}^l \f{\omega \  \Gamma_{\omega l}}{e^{2 \pi \f{\omega-m \Omega_\rom{H}}{\kappa}}-1}~.
\ee
For $\omega - m \Omega_\rom{H}<0$ the integrand is negative; though the equation formally continues to hold. Modes for which $\omega - m \Omega_\rom{H}<0$ are called superradiant modes.
For these modes  the scattering cross-section is negative. As a result, a wave analogue of the Penrose process occurs. The phenomenon is called superradiance.

A  different kind of radiation happens for rotating stars with ergoregions. In this case there is no horizon but a rotating
star suffers from a classical instability called the ergoregion instability \cite{Friedman:1978, Comins:1978}. Radiation comes out of the star
with amplitude increasing exponentially with time.  We will see in the later chapters that these seemingly unrelated phenomena, namely, Hawking radiation, superradiance, and ergoregion emission are in fact intimately interrelated.

\section{Hawking Information Paradox}
Hawking's discovery of the black hole radiance eluded to a deep connection between thermodynamics, gravitational physics, and quantum mechanics. Of particular importance is the formula for the (intrinsic) Bekenstein-Hawking entropy for a black hole
\be
S = \frac{A}{4}~,
\ee
where $A$ is the area of the event horizon in Plank units. The precise nature of the Bekenstein-Hawking entropy has been a topic of intense research since Hawking's remarkable discovery. Developments in string theory have provided significant insights into the nature of this entropy.

Black hole radiance, at the same time, raises some  serious puzzles.  These puzzles have  been much debated over the last four decades, and many of them have not yet been completely resolved.  One of the most debated puzzle of this nature is the Hawking's proposal that the usual rules of quantum mechanics do not apply to a process in which black holes form and evaporate completely. If this proposal is correct, then we face the formidable task of reformulating quantum physics as a new self-consistent framework that agrees with experiment. To begin with one might be inclined to dismiss Hawking's proposal---one might argue that the
proposal is founded on a semiclassical calculation in which gravitational back reaction effects are not properly taken into account. There is little prospect that a detailed calculation of back reaction effects can be done in the near future, as it would require the knowledge of quantum gravity (see \cite{Wald:1995yp} for a more detailed discussion of these issues).  On the other hand, it has been argued that Hawking's conclusion is completely robust and inescapable \cite{Mathur:2008wi, Mathur:2009hf}.

\subsection*{The Paradox}

As we discussed above, in the semiclassical calculation of black hole radiance, Hawking found that the emitted radiation is exactly thermal. In particular, the details of the radiation do not depend on the nature of the body that collapsed to form the black hole. The state of the radiation is completely determined by the geometry outside the horizon. The fact that the radiation outside the horizon is thermal (and hence a mixed state) is not puzzling. This is because the region outside the horizon is only a part of the full quantum system. The full quantum system consists of the infalling matter and degrees of freedom of quantum fields inside and outside the horizon. The degrees of freedom of quantum fields living outside and inside the horizon are correlated. However neither of them are correlated with the infalling matter. These correlations are responsible for the mixed nature of the radiation observed by observers outside the horizon.

A truly paradoxical situation arises when the black hole evaporates completely; since now the radiation \emph{is} the complete system, and there is no region behind the horizon.  So, if the radiation is  indeed in a thermal state as Hawking found, it appears that the initial pure quantum state, by collapsing into a black hole and evaporating completely, evolves in a thermal (mixed) state.  This is the Hawking information paradox.  In analyzing the evolution of black holes within the standard framework of general relativity and quantum mechanics we are led to conclude that a pure state can evolve into a mixed state. Such an evolution is forbidden by the usual rules of quantum mechanics.

\chapter{Fuzzballs}
\label{Fuzzballs}

As discussed in the previous chapter, black holes have entropy and they decay with a thermal spectrum. This leads to several puzzles, in particular, to the following, all of which have been intense topics of research in the last four decades:
\begin{itemize}
\item
\emph{What} are the microstates of a black hole?
\item
\emph{Where} are the microstates of a black hole?
\item
Does unitarity indeed break down in processes in which black holes form and evaporate completely?
\end{itemize}
String theory has provided significant insights in answering the first question. For a variety of extremal and
near-extremal black holes several authors have counted states at weak coupling using brane physics and have reproduced the Bekenstein-Hawking entropy of corresponding black holes.
However these state countings do not tell us anything about how the microstates look like in the strong coupling description, and in particular, in the
gravity description. These calculations also do not tell us much about the second and the third question. In the rest of the review  an approach to
resolve these questions in the framework of the so-called \emph{Fuzzball Proposal} is discussed. According to this proposal, a black hole geometry is
an effective description of an ensemble of states. In the framework of this proposal, quantum gravity effects are not confined to the Planck length.
Typical states of the ensemble have structure up to the scale of the horizon. In this proposal, radiation from non-extremal would-be black holes does
not happen by a pair production process from the vacuum, but rather happens from the surfaces of the black hole's microstates. Consequently, the emitted
radiation is capable of carrying information, and seemingly the information paradox does not arise.

The fuzzball proposal has met with most success in the case of the two-charge string-theory black holes. For the two charge system all states have been
identified in the gravity description and it is shown that  typical states have a size that scales as a function of charges as it would for a would-be black hole. For three and
four-charge string theory black holes large classes of  smooth supersymmetric geometries (see \cite{Bena:2007kg, Skenderis:2008qn, Balasubramanian:2008da, deBoer:2009un} for reviews)
and some non-supersymmetric geometries \cite{Jejjala:2005yu, Giusto:2007tt, AlAlawi:2009qe, Bena:2009qv, Bobev:2009kn} have been constructed.
Not all smooth supergravity states have been found. It has been argued that  all states may not admit a supergravity description \cite{deBoer:2009un}.
In this chapter we give a brief overview of the two- and three-charge solutions. In the three-charge case our discussion is particularly brief;
many topics of current interest are not addressed.

\section{Two Charge Solutions} \label{2Charge}

The simplest and the best understood example of quantum gravity effects scaling over distances much bigger than the
Planck length is the two-charge system. The black hole solution is obtained by compactifying type IIB string theory on $T^4 \times S^1$ and wrapping
$n_5$ D5 branes on the torus and the circle and $n_1$ D1 branes on the circle. The volume of the torus is taken to be $(2 \pi)^4  V$ and the the length
of $S^1$ is taken to be $2 \pi R$. At low energies the worldvolume dynamics of the branes is described
by an $\mathcal N=(4,4)$ super-conformal field theory (SCFT). This SCFT is discussed in section \ref{CFT}. Here we focus primarily on the supergravity solutions of the system.

\begin{figure}[ht] 
\begin{center}
\subfigure[~Fuzzballs]{\label{fig:fuzzball}
	\includegraphics[width=6.3cm]{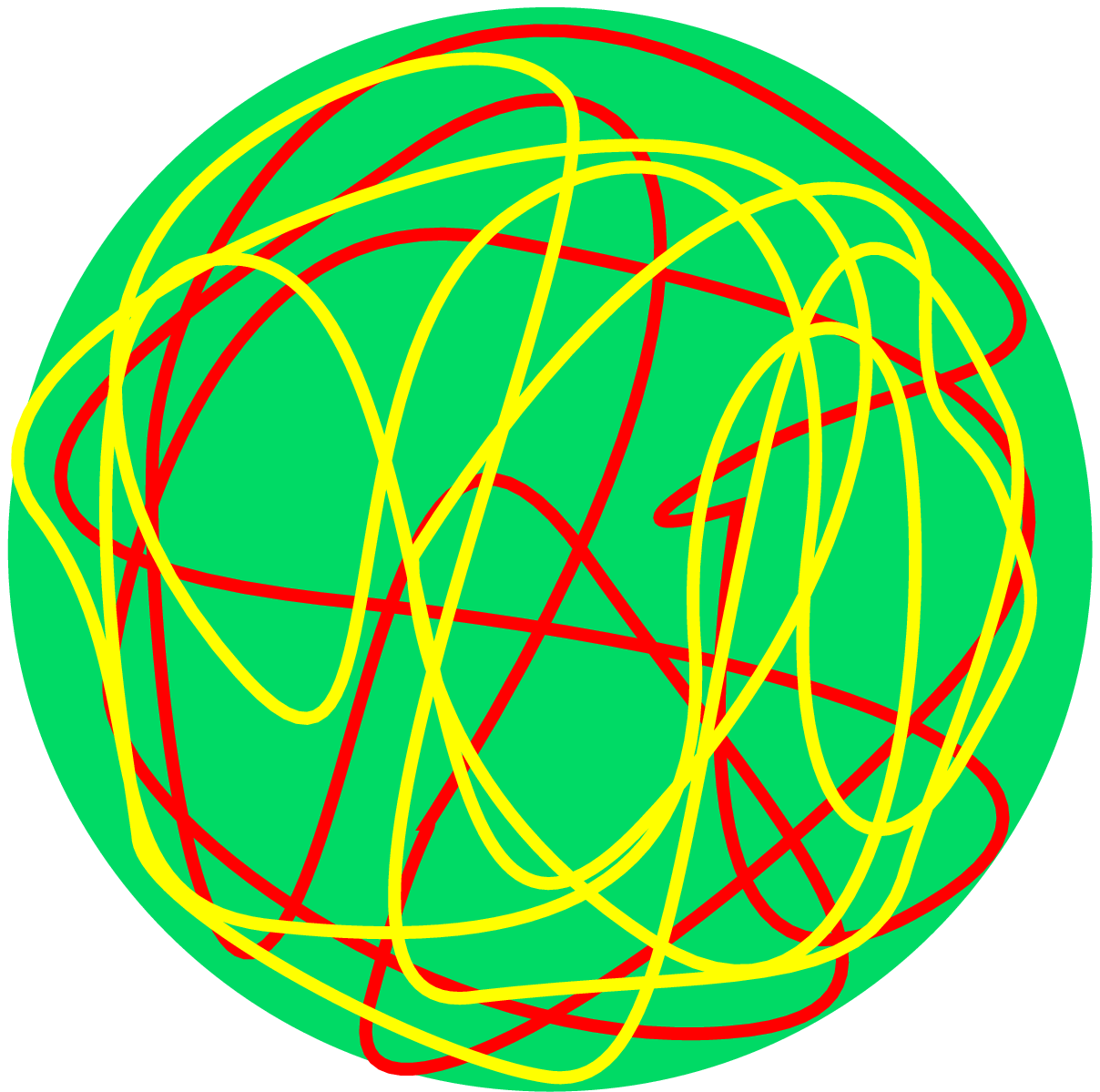}}
\hspace{15pt}
\subfigure[~Fuzzrings]{\label{fig:fuzzring}
	\includegraphics[width=6.3cm]{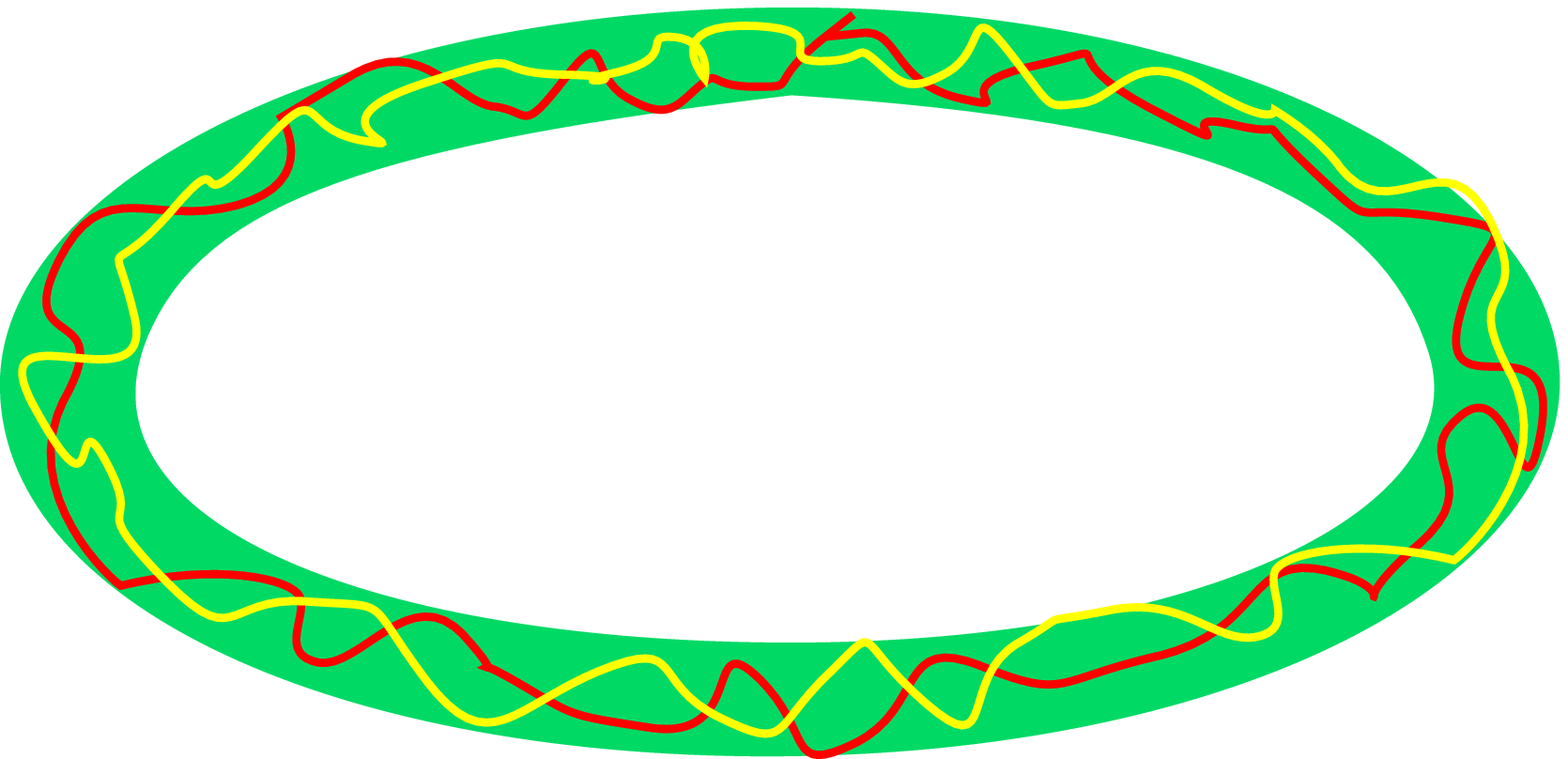}}
   \caption{Cartoons for (a)  fuzzball and (b)  fuzzring solutions for two profile functions each. The green (shaded) region is the region in which the typical solutions differ very much from each other. The metric for typical states is similar outside the green (shaded) region. If one puts a stretched horizon on the green (shaded) region one gets a coarse grained entropy that goes like $S_{\rom{stretched}} \sim \sqrt{n_1 n_5}$ for the fuzzballs and like $S_{\rom{stretched}} \sim \sqrt{n_1 n_5 - J}$ for the fuzzrings.}
   \label{fig:FuzzballAndFuzzring}
   \end{center}
\end{figure}

Naively the metric and the dilaton are obtained by the Harmonic superposition rule
\bea
ds^2_{\rom{naive}} &=& \f{1}{\sqrt{g_1 g_5}} ( -dt^2 + dy^2) + \sqrt{ g_1 g_5} \sum_{i=1}^4 dx^2_i + \sqrt{\f{g_1}{g_5}} \sum_{i=1}^4 dz_i^2 \nn
e^{2 \phi} &=& \f{g_1}{g_5}, \qquad g_1 = 1+ \f{Q_1}{r^2}, \qquad g_5 = 1+ \f{Q_5}{r^2} \label{D1D5Naive}~.
\eea
For simplicity we have not presented the RR gauge fields produced by the D-branes. The charge radii $Q_1$ and $Q_5$ are related to integer charges as
\be
Q_1 = \f{g l_s^6}{V} n_1 ,\qquad Q_5 = g l_s^2 n_5~.
\ee
This system has a zero size horizon  at $r=0$. Thus its Bekenstein-Hawking entropy is zero. It was argued in \cite{Lunin:2001fv} that the metric \bref{D1D5Naive} is not the correct description for the D1-D5 system. The correct description of the system depends on a closed curve in the non-compact space. The statistical mechanics of this curve gives the transverse size for a typical
curve and this sets the size for a typical solution as shown in figure \ref{fig:FuzzballAndFuzzring}. A subset of these solutions based on these curves were constructed in \cite{Lunin:2001fv, Lunin:2001jy} and the rest of the solutions were constructed in \cite{Lunin:2002iz, Taylor:2005db, Kanitscheider:2007wq}.

We can understand the statistical mechanics of this system by going to a dual frame where the D5 branes map to a fundamental string wrapping $S^1$ and the D1 branes
map to momentum running along $S^1$. The bound state of this system consists of a long string of winding $n_5$ on $S^1$ carrying $n_1$ units of momentum. The momentum  manifests itself as bosons and fermions on the string worldsheet.  There are 8 such bosons and 8 such fermions corresponding to the string bending in the $T^4$ and $R^4$ directions. Thus the system is a $1+1$ dimensional gas and from
its statistical mechanics (see \cite{Balasubramanian:2005qu} for a recent discussion) it follows that the entropy is
\be
S_\rom{{2-charge}} = 2 \pi \sqrt{2} \sqrt{n_1 n_5} \label{2chargeEntropy}~.
\ee
The generic quanta is in the harmonic
\be
k = \sqrt{n_1 n_5} \label{genericstatek}~,
\ee
with occupation
\be
n_k \approx O(1) \label{genericstatenk}~.
\ee

Now we have a puzzle. The weak coupling F1-P analysis shows that the system has an entropy that scales as $\sqrt{n_1 n_5}$. Since the system is supersymmetric, this quantity should be protected by supersymmetry as we increase the string coupling. However, the Bekenstein-Hawking entropy of the naive solution is zero.

The answer to this puzzle lies in the fact that the fundamental string carries \emph{only} transverse vibrations. This leads to a puffing up of the naive solution.  Supergravity solutions  for fundamental string carrying vibrations with an arbitrary transverse vibration profile  $\vec F(t-y)$ were explicitly constructed in \cite{Callan:1995hn, Dabholkar:1995nc}. These solutions correspond to bosonic excitations carrying momentum on a string in the $y$ direction. They were dualized to the D1-D5 frame in \cite{Lunin:2001fv} to get the solutions
\bea
ds^2_\rom{string} &=&  \f{1}{\sqrt{\tilde g_1 \tilde g_5}} ( -(dt- A_i dx^i)^2 + (dy+ B_i dx^i)^2) + \sqrt{ \tilde g_1 \tilde g_5} \sum_{i=1}^4 dx^2_i \nn & & \ \ + \  \sqrt{\f{\tilde g_1}{\tilde g_5}}  \sum_{i=1}^4 dz^2_i~,  \nn
e^{2 \phi} &=& \f{\tilde g_1}{\tilde g_5}, \quad \tilde g_5 (\vec x) = 1+ \f{Q_5}{L} \int_0^L \f{dv}{|\vec x - \vec F(v)|^2}, \nn   \tilde g_1 (\vec x) &=& 1+ \f{Q_5}{L} \int_0^L \f{ |\dot{ \vec F}(v)|^2 dv}{|\vec x - \vec F(v)|^2}~, \nn
A_i (\vec x) &=& - \f{Q_5}{L} \int_0^L \f{\dot F_i(v) dv}{|\vec x - \vec F(v)|^2}, \qquad d B = - \star_4 ~ d A \label{D1D5metric}~,
\eea
where $\star_4$ is taken with respect to the flat metric for the non-compact $x_i$ space. The length of integration is given by
\be
L = \f{2 \pi Q_5}{R}~,
\ee
and the charges are related by
\be
Q_1 = \f{Q_5}{L} \int_0^L dv (\dot F(v))^2~.
\ee
For these solutions we see that if $\vec F(v)$ is bounded from above, that is, if $|\vec F(v) | < b$, then at large distances, $r \gg b$, we recover the naive metric \ref{D1D5Naive}. Near $r \lessapprox b$, metrics for different profile functions $\vec F(v)$ differ from each other.
Furthermore, all these metrics seem to have a singularity at $\vec x = \vec  F(v)$, but it was shown in \cite{Lunin:2002iz} that this is just a coordinate
artifact; the solutions are completely smooth everywhere. These solutions were quantized using geometric quantization
in \cite{Rychkov:2005ji} and were found to give the entropy
\be
S_{\rom{Rychkov}} =2 \pi \sqrt{ \f{2}{3} n_1 n_5}~,
\ee
which is the correct contribution, when we take only four $R^4$ bosons into account in the microscopic counting. It is expected that geometric quantization of the full class of solutions constructed in \cite{Lunin:2002iz,Taylor:2005db,Kanitscheider:2007wq}  would give the entropy matching with the weak coupling result \bref{2chargeEntropy}.

There is another intuitive way of understanding the entropy of these solutions. One puts a `stretched horizon' at $r \approx b$ where the generic solutions starts to differ from  each other. The scale $b$ was found in \cite{Lunin:2002qf}.
It was obtained using the fact that in the F1-P frame the string oscillates transverse to
itself on the scale of the string length $l_s$. The dualities map this scale in the D1-D5 frame to
\be
b \sim \f{g l_s^4}{\sqrt{V} R}~.
\ee
The area of the stretched horizon is most easily calculated in six dimensions. This is because in six dimensions the Einstein and the string frame metrics are the same.
The area is then found to be
\be
A_{\rom{stretched}} = \int_{r=b} r^3 \sqrt{g_1 g_5} \ dy \ d \Omega_3 = 4 \pi^3 R b \sqrt{Q_1 Q_5}~,
\ee
where we have assumed $b \ll Q_1^\h , Q_5^\h$ which is true in the classical limit.
The six dimensional Newton's constant is
\be
G^{(6)}= \f{(2\pi)^3 g^2 l_s^8}{V}~.
\ee
The Bekenstein-Hawking entropy of the stretched horizon is therefore
\be
S_{\rom{stretched}} = \f{A_{\rom{stretched}}}{4 G^{(6)}}  \sim \sqrt{n_1 n_5}~.
\ee

The above analysis can be extended to the D1-D5 system with angular momentum. We let the angular momentum be in the $x_1 -x_2$ plane and of the order $J \sim O(n_1 n_5)$. The statistical mechanics of the system shows that excitations on the string splits into two parts \cite{Balasubramanian:2005qu}: (a) all the angular momentum is carried in the lowest harmonic, with this subsystem contributing no entropy, and (b) the rest of the energy is carried by a thermal distribution of excitation in higher harmonics. The entropy of the system is given by
\be
S = 2 \pi \sqrt{2} \sqrt{n_1 n_5-J}~,
\ee
with the generic quanta in the mode
\be
k \sim \sqrt{n_1 n_5 - J}~,
\ee
with occupation number
\be
n_k \sim O(1)~.
\ee
The profile function for this system is given by \cite{Lunin:2002qf}
\be
\vec F (v) = a \left( \vec e_1 \cos \f{2 \pi v}{L} + \vec e_2 \sin \f{2 \pi v}{L} \right) + \vec X (v), \quad \mbox{with} \quad |\vec{X}(v)| \lessapprox b~,  \label{blackring}
\ee
where
\be
a= \f{g l_s^4}{\sqrt{V} R} \sqrt{J}~.
\ee
Thus, all geometries described by these profile functions look essentially the same outside a donut shaped region.
The Bekenstein-Hawking entropy for the stretched horizon over this donut-shaped region is found to be
\be
S_{\rom{stretched}} \sim \sqrt{n_1 n_5 -J}~.
\ee
\begin{figure}[ht] 
\begin{center}
\subfigure[]{\label{fig:blackHoleEmbedding}
	\includegraphics[height=3.3cm]{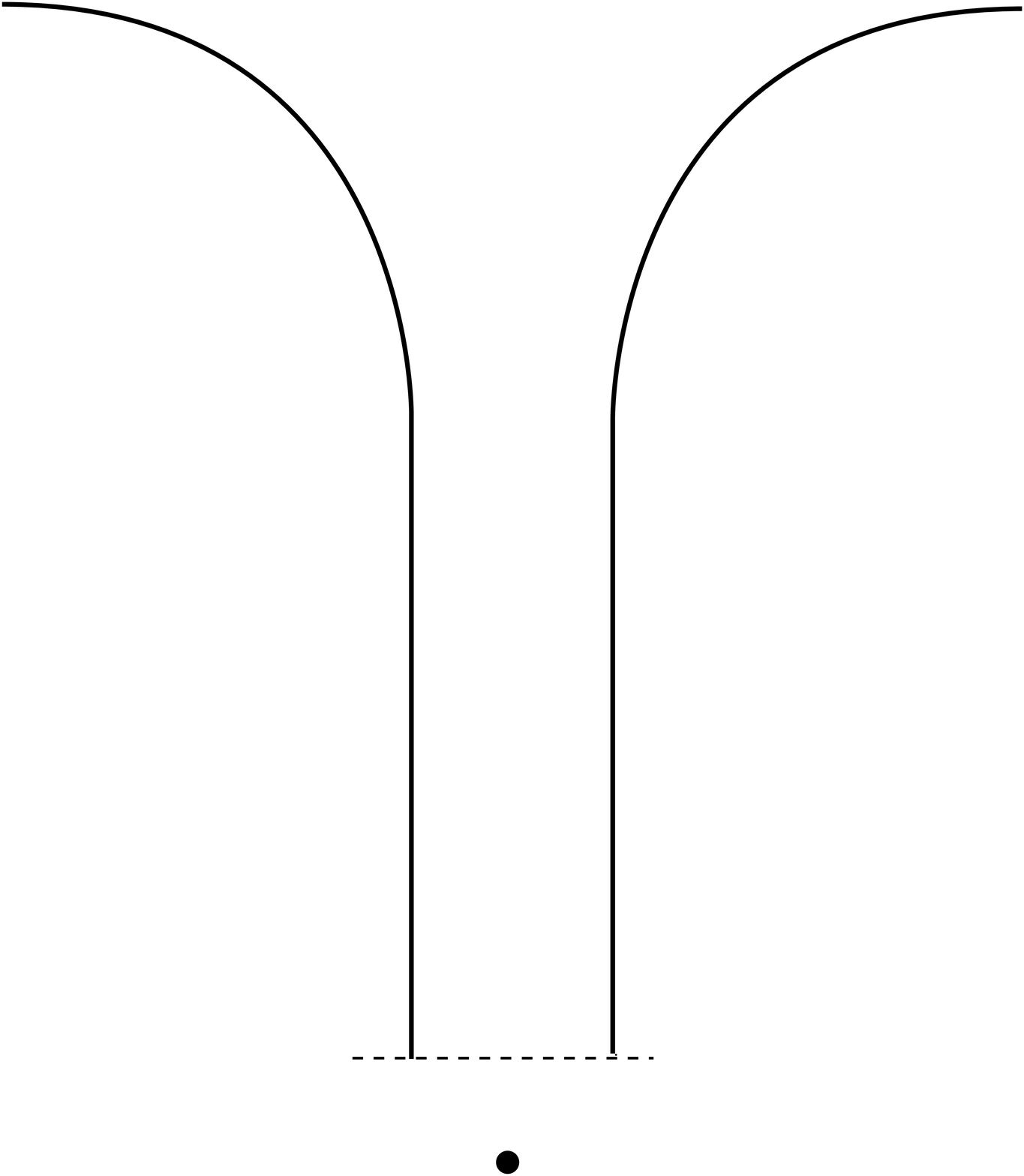}}
\hspace{35pt}
\subfigure[]{\label{fig:fuzzballEmbedding}
	\includegraphics[height=3.3cm]{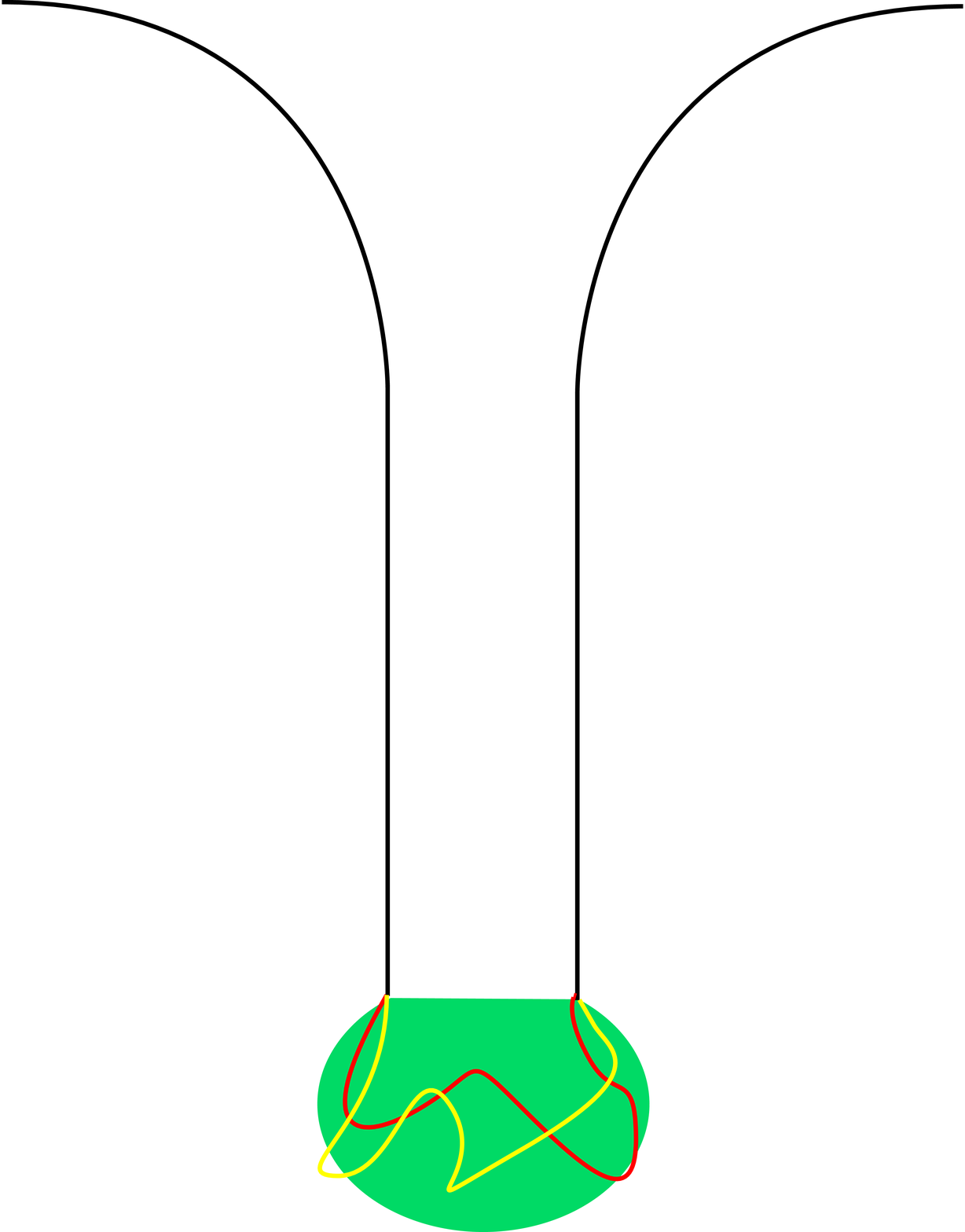}}
   \caption{(a): The geometry of a black hole has an outer flat space connected by a neck to a throat which ends in a horizon with a singularity hidden behind.
   (b) The geometry of a generic state has outer flat space connected by a neck to a throat which ends in a smooth cap without horizons and singularities.}
   \label{fig:BlackHoleNFuzzballEmbedding}
   \end{center}
\end{figure}

The fact that the real solutions of the D1-D5 system have a size which scales as $(n_1 n_5)^\f{1}{6} l_p$ suggests that probes with insufficient precision 
fail to capture the details of the system. 
The intuitive picture that the coarse graining of geometries gives a black hole (or a black ring for sufficiently large angular momentum) was put on a firmer technical footing in references \cite{Alday:2006nd,Balasubramanian:2005qu,Giusto:2005ag,Skenderis:2006ah,Kanitscheider:2006zf,Kanitscheider:2007wq,Mathur:2007sc,Das:2008ka}
for this system.

All these geometries have an inner region that is asymptotically $AdS_3 \times S^3 \times T^4$. As a result, we can have a description of the system in terms of a  dual conformal field theory. For the circular profile function \cite{Balasubramanian:2000rt,Maldacena:2000dr}
\be
F(v) = \f{\sqrt{Q_1 Q_5}}{R} \left( \vec e_1 \cos \f{2 \pi v}{L} + \vec e_2 \sin \f{2 \pi v}{L} \right) \label{SpecialProfile}
\ee
it was shown in \cite{Lunin:2001dt} that a minimal scalar falling into the $AdS_3 \times S^3$ throat takes the time $\Delta t = \pi R$ to emerge. This time matches the CFT travel time for the state dual to this geometry. 
If instead one uses the distribution \bref{genericstatek} for the CFT state one obtains a travel time to
be $ \Delta t_{\rom{CFT}} \sim e^{\sqrt{n_1 n_5}} R$ \cite{Balasubramanian:2005qu}. In the classical limit of large $n_1, n_5$ we see that a particle
falling-in never emerges back. This suggests that typical states of the D1-D5 system behave like a black hole in the classical limit.

\section{Three Charge Solutions}

Now we present a very brief introduction to  three charge solutions. This section is meant to be a starting tutorial on the subject. Many important topics of current interest are not addressed, and none of the three-charge solutions are discussed in detail. The aim here is to provide a flavor of the intense research program that is currently underway to see to what extent the correspondence between D1-D5  states and the smooth bulk solutions extends to the D1-D5-P system. In this section we primarily work in the M-theory frame. Results can be readily dualized to the IIB frame. Our presentation closely follows \cite{Bena:2007kg}. See also \cite{Skenderis:2008qn, Balasubramanian:2008da, deBoer:2009un}.

A detailed probe analysis of the three-charge system based on the DBI action strongly suggests the existence of a large number of BPS supergravity
configurations with the same supersymmetries as the three charge black hole. At the outset the set of configurations suggested by the DBI analysis
is so large that finding the corresponding supergravity solutions appears to be a formidable task. Fortunately, it turns out that the
entire problem of finding all these BPS configurations reduces to a linear problem in four dimensional Euclidean electrodynamics
\cite{Gauntlett:2002nw, Bena:2004de}. It is one of those instances where supersymmetry plays a crucial role in simplifying the problem.

\subsection{Supersymmetric Configurations}
\label{susyconfig}
Let us start with M-theory compactified on a six-torus $T^6$ and wrapping three sets of extremal M2 branes on three orthogonal two-tori in the $T^6$. This configuration is BPS. Rather surprisingly, one can further add three sets of M5 branes on top of this configuration and  find configurations that preserve the same set of supersymmetries as the original set of M2 branes \cite{Elvang:2004ds, Bena:2004de}. Each set of M5 branes can be thought of as the dual to a set of M2 branes. In each such pair the set of M5 branes wrap the four torus orthogonal to the two torus wrapped by the set of M2 branes. The fifth direction of the M5 branes wrap a contractible spacelike curve (denoted $\psi$) in the five-dimensional space orthogonal to $T^6$. This configuration is summarized in the following table.
\bea
\begin{array}{c c c c c c c c c c c c}
 & t &   x_5   &  x_6  & x_7& x_8 & x_9 & z_{10} & \psi & x_1 & x_2 & x_3 \\
  \rom{M2} &  \times & \times & \times & - & - & - & - & - &  &  &  \\
  \rom{M2} &  \times & - & -& \times & \times & - & - & - &  &  &  \\
  \rom{M2} &  \times & - & - & - & - & \times & \times  & - &  &  & \\
  \rom{M5} &  \times & - & - & \times & \times & \times & \times  & \times &  &  & \\
  \rom{M5} &  \times & \times & \times& - & - & \times & \times & \times &  &  &  \\
  \rom{M5} &  \times & \times & \times & \times & \times & - & - & \times &  &  &
\end{array}
\label{M2}
\eea

It is argued in \cite{Bena:2004de} that this is the most general supersymmetric configuration that has the same supersymmetries as the original three charge (M2-M2-M2) black hole. (Of course one can consider a configuration with multiple curves and black hole sources, which also has the same supersymmetries as the three charge black hole.) The eleven dimensional metric corresponding to these brane configurations takes the form
\bea
ds^2_{11} &=& ds^2_{5} + \left(\frac{Z_2 Z_3}{Z_1^2}\right)^{1/3}\left(dx_5^2 + dx_{6}^2\right)~ \nn & & \ \ \ \ \  + \  \left(\frac{Z_1 Z_3}{Z_2^2}\right)^{1/3}\left(dx_7^2 + dx_{8}^2\right)+ \left(\frac{Z_1 Z_2}{Z_3^2}\right)^{1/3}\left(dx_9^2 + dx_{10}^2\right)~,
\eea
where the five-dimensional space is
\be
ds^2_5 = \left(Z_1 Z_2 Z_3\right)^{-\frac{2}{3}}\left(dt + k\right)^2 + \left(Z_1 Z_2 Z_3\right)^{\frac{1}{3}} h_{\mu \nu}dx^{\mu}d x^{\nu}~,
\ee
and $x_5 \ldots x_{10}$ are the six torus directions.
The one form $k$ is defined on the spatial section of the five-dimensional metric. Supersymmetry requires that the metric $h_{\mu \nu}$ be hyper-K\"ahler. Two important examples of hyper-K\"ahler metrics are $(i)$ $\mathbf{R}^4$,
$(ii)$ multi-centered Taub-NUT metrics.
In this review we are only concerned with these two hyper-K\"ahler manifolds, and their ambipolar generalizations, which we discuss below.

The eleven-dimensional three-form $A_{(3)}$ is given by
\be
A_{(3)} = A^{(1)} \wedge dx_5 \wedge dx_6 +A^{(2)} \wedge dx_7 \wedge dx_8 + A^{(3)} \wedge dx_9 \wedge dx_{10}~.
\ee
The $A^{(I)}$'s for $I = 1, 2, 3$ are three abelian one-forms in the five-dimensional spacetime. We define
\be
\Theta^{(I)} \equiv dA^{(I)} + d (Z_I^{-1}(dt + k))~.
\ee
Then the following set
of ``sequentially linear'' {\emph{BPS equations}} determine the most general supersymmetric configuration \cite{Bena:2004de}:
\bea
\Theta^{(I)} &=& \star_4 \Theta^{(I)}~, \\
\n^2 Z_I &=& \frac{1}{2}C_{IJK} \star_4 \left(\Theta^{(J)} \wedge \Theta^{(K)} \right)~, \\
dk + \star_4 dk &=& Z_{I} \Theta^{(I)}~,
\eea
where $\star_4$ is the Hodge dual taken with respect to the four dimensional hyper-K\"ahler base and $C_{IJK} = |\epsilon_{IJK}|$. This system of equation is called ``sequentially linear'' because if these equations are solved in the order they are presented then at each step one is solving a linear problem.

\subsubsection*{Supersymmetric Black Ring}

As an illustration of the formalism presented above, now we very briefly discuss the supersymmetric black ring solution that can be constructed by  solving the BPS equations. We work with flat base space $\mathbf{R}^4$. One starts by choosing the circular profile for the M5 branes in the five-dimensional spacetime that sources $\Theta^{(I)}$'s. Then one has some freedom in choosing harmonic functions in the solution that add sources for the M2 branes. Finally, with these sources one solves for the one-form $k$.   The best coordinate system to solve these equations in this setting is the black ring coordinates. Here we do not review these coordinates. For details of the coordinate system we refer the reader to \cite{Emparan:2006mm, Bena:2007kg} or to the original references \cite{Emparan:2001wn,Emparan:2001wk}. In black ring coordinates the flat $\mathbf{R}^4$ metric has the form
\be
ds^2 = \frac{R^2}{(x-y)^2} \left(\frac{dy^2}{y^2 -1} + (y^2 -1) \, d\psi^2 + \frac{dx^2}{1-x^2} + (1-x^2) \, d\phi^2\right)~,
\ee
where $-1 \le x \le 1, \ -\infty \le y \le -1$, and the ring is located at $y = -\infty$.
The self-dual field strengths (with the orientation $\epsilon^{yx\psi\phi} = + 1$) sourced by the ring are
\be
\Theta^{(I)} = 2 q^I \left( dx \wedge d\phi - dy \wedge d\psi \right)~,
\ee
where $q^I$'s are directly related to the number of wrapped M5 branes. The warp factors take the form
\be
Z_I = 1 + \frac{Q_I}{R} (x-y) - \frac{2 C_{IJK}q^Jq^K}{R^2}(x^2 -y^2)~,
\ee
and the one-form $k$ is
\be
k = \left( (y^2-1)g(x,y) - 2\left(\sum_{I=1}^{3}q^I\right)(y+1)\right) d\psi + (x^2-1)g(x,y)d\phi~,
\ee
where
\be
g(x,y) = \left(- \frac{8C_{IJK}q^Iq^Jq^K}{3R^2} (x+y) + \frac{2}{R}\left( Q_I q^I\right) \right)~.
\ee
$Q_I$'s are directly related to the number of M2 branes. For a detailed study of physical properties of this solution we refer the reader to original references \cite{Elvang:2004rt, Elvang:2004ds, Bena:2004de, Gauntlett:2004wh, Gauntlett:2004qy} and to the reviews \cite{Emparan:2006mm, Bena:2007kg}.


\subsection{Gibbons-Hawking Base}
In section \ref{susyconfig} we mentioned that supersymmetry requires us to take the base space metric to be a four-dimensional hyper-K\"ahler manifold. Flat Euclidean space is the simplest example of a hyper-K\"ahler manifold. Then there are the multi-centered Gibbons-Hawking (GH) metrics. In this section we briefly review the GH metrics and there `ambipolar' generalizations. Our presentation closely follows \cite{Bena:2007kg}. GH metrics have a form of a $U(1)$ fibrations over a three dimensional Euclidean flat space $\mathbf{R}^3$:
\begin{equation}
ds^2 = h_{\mu \nu}dx^\mu dx^\nu = V^{-1}\left( d \psi + \vec{A}\cdot d\vec{y} \right)^2 + V \left( dx^2  + dy^2 + dz^2 \right)~,
\end{equation}
where $\vec{y} = {x,y,z}$. The function $V$ appearing in the metric is a harmonic function on $\mathbf{R}^3$. We take it to have a finite set of isolated sources. The connection $ A = \vec{A} \, . \, d\vec{y} \, $ is determined via $V$ through the equation
\begin{equation}
\vec{\nabla} \times \vec{A} = \vec{\nabla} V.
\end{equation}
Let $\vec{y}^{(j)}$ be the position of the point sources for the function $V$ in the three dimensional base $\mathbf{R}^3$ and let $r_j = |\vec{y} - \vec{y}^{(j)}|$, then
\begin{equation}
V = q_0 + \sum_{j=1}^{N} \frac{q_j}{r_j}~.
\end{equation}
Note that close to a GH center the fuction $V$ behaves as
$V \sim \frac{q_j}{r_j}.
$ Thus, to ensure that the metric is positive definite everywhere on the manifold one usually takes $q_j \ge 0.$  We relax this assumption. In considering negative GH charges $q_j$ we are allowing the base space to flip the signature from $(+,+,+,+)$ to $(-,-,-,-)$ in certain regions. Following \cite{Bena:2007kg} we call such metrics \emph{ambipolar}. At first sight these metrics may look completely unphysical, but this is not the case. Negative definite regions do not lead to any pathological features in the full five-dimensional spacetime \cite{Bena:2005va, Berglund:2005vb}. The ambipolar metrics, on the other hand, allow us to to construct explicit BPS solutions with extremely rich physics.

Next, we now introduce a set of frames for the GH base and the two-forms
\bea
\hat{e}^1 &=& V^{-\frac{1}{2}}(d\psi + A)~, \\
\hat{e}^{a+1} &=& V^{\frac{1}{2}}dy^a~,  \\
\Omega_{\pm}^{(a)} &=& \hat{e}^1 \wedge \hat{e}^{a+1} \pm \frac{1}{2} \epsilon_{abc} \hat{e}^{b+1}\wedge \hat{e}^{c+1}~, \qquad a = 1, 2, 3~.
\eea
The two-forms $\Omega_{+}^{(a)}$ are self-dual and can be used to construct harmonic fluxes $\Theta^{(I)}$ as follows
\begin{equation}
\Theta^{(I)} \equiv \sum_{a=1}^{3}\left(\partial_a (V^{-1} K^I)\right) \Omega_{+}^{(a)}~.
\label{theta}
\end{equation}
The two-forms $\Theta^{(I)}$  are closed if and only if $K^I$'s are harmonic functions on $\mathbf{R}^3$.

\subsection{Solutions on a Gibbons-Hawking Base}
Having discussed elementary properties of GH spaces, we are in the position to solve the BPS equations with a GH base. The first step is to construct the three self-dual two-forms $\Theta^{(I)}$. This is done by introducing three harmonic functions $K^I$ on $\mathbf{R}^3$
through \bref{theta}.
The next step is to substitute these two-forms in the equations for $Z_I$. Solving these equations one finds
\begin{equation}
Z_I = \frac{1}{2} C_{IJK}V^{-1}K^JK^K + L_I~,
\end{equation}
where $L_I$ are three more independent harmonic functions. We now write the one-form $k$ as:
\begin{equation}
k = \mu \, (d\psi + A) + \omega~,
\end{equation}
where $\mu$ is solved to be
\begin{equation}
\mu = \frac{1}{6}C_{IJK} \frac{K^I K^J K^K}{V^2} + \frac{1}{2 V} K^IL_I + M~,
\end{equation}
and where $M$ is another harmonic function on $\mathbf{R}^3$. $\omega$ is a one-form on $\mathbf{R}^3$. Using $M$ one can find $\omega$ as a solution of the equation
\begin{equation}
\vec{\nabla} \times \vec{\omega} = V \, \vec{\nabla}M - M \, \vec{\nabla}V + \frac{1}{2} \left( K^I\vec{\nabla}L_I - L_I \vec{\nabla}K^I\right)~.
\end{equation}
Since $K^I, L_I, M$ and $V$ are all harmonic, the above equation can be integrated to find $\vec{\omega}$ and thereby obtaining the complete solution. The full solution is characterized by eight harmonic functions on $\mathbf{R}^3$: $K^I, L_I,  M, V.$

At this point is is worth pointing out that there is a `gauge' invariance in the system of equations related to the fact that for arbitrary constants $c^I$ the shift
\be
K^I \rightarrow K^I + c^I V
\label{gauge}
\ee
does not change the self-dual two forms $\Theta^{(I)}$ \bref{theta}. Due to this gauge freedom the solutions are in effect characterized by seven independent harmonic functions. This functional freedom matches with the DBI analysis of \cite{Bena:2004wt}.

\subsection{Bubbled Geometries}
One of the main  purpose of this program is to be able to construct such smooth horizonless BPS configurations with no sources. We now say a few words about
smooth geometries. 
First recall that the black ring solution discussed above has brane sources. Geometries with no brane sources are topologically non-trivial. One of the simplest example of such a configuration is obtained by choosing the GH base with three co-linear centers with charges $(1, Q, -Q)$, where $Q$ is an integer. This three-center GH space  contains three topologically non-trivial two cycles. These cycles can be defined by considering a curve between any two GH centers and considering the $U(1)$ fibre along that curve. At the GH centers the $U(1)$ fibre shrinks to zero size. As a result, the curve together with the $U(1)$ fibre sweeps out a two-cycle. These non-trivial two-cycles are responsible for the bubbling nature of the five-dimensional solutions. In the limit when the $+Q$ GH center is moved on top of the GH center with $+1$ charge, the corresponding five dimensional  geometries reduce to the geometries constructed in \cite{Mathur:2003hj, Lunin:2004uu, Giusto:2004id, Giusto:2004ip, Giusto:2004kj}.

The bubbled geometries with three co-linear centers with GH charges $(1, Q, -Q)$ can be thought of as the geometric transition of a zero-entropy black ring. The picture is as follows \cite{Bena:2005va, Berglund:2005vb}: We can think of $\mathbf{R}^4$ as a trivial GH base with $V\sim \frac{1}{r}$. Now consider taking the black ring in the parameter regime where its entropy decreases and eventually approaches zero. The zero entropy black ring in such a limit is singular, but the singularity is not a curvature singularity. The singularity is  a ``null orbifold'' singularity. In the transitioned geometry the singularity is resolved by the nucleation of two equal, oppositely charged, GH centers $(Q, -Q)$. The transitioned geometry has no brane sources.  Since in the resolved geometries negative GH centers play a crucial role, ambipolar generalization of GH metrics becomes indispensable. Using the formalism outlined above harmonic functions for the bubbling solutions can be calculated \cite{Bena:2005va, Berglund:2005vb}. These solutions have extremely rich physics, and a detailed study of their properties will take us far beyond the scope of this review. We refer the interested reader for more details to the following literature \cite{Bena:2005va, Berglund:2005vb, Saxena:2005uk,Bena:2005zy,Giusto:2006zi, Bena:2006is, Balasubramanian:2006gi, Bena:2006kb, Cheng:2006yq,Ford:2006yb,Gimon:2007ps,  Bena:2007ju, Gimon:2007mha, Bena:2007qc,  Castro:2008ne, deBoer:2008fk, Bena:2008wt,  Bena:2008nh, Raeymaekers:2008gk, Bena:2008dw, Bena:2009en,Bena:2009fi, Bobev:2009zz}.

\chapter{Emission From the D1-D5 System}

\label{emission}

In previous chapter we saw how quantum gravity effects are not confined to the Planck length. In particular, for the two-charge system we explicitly saw how typical fuzzballs extend up to the horizon scale. A similar picture is expected to arise for the three-charge system. This builds up the picture of a black hole being an effective coarse grained geometry. In this chapter we will see an additional process that supports this idea. Here we understand the nature of Hawking radiation by studying emission from the D1-D5 system.  The main aim is to understand the interaction between excitations on the inner asymptotically AdS space and excitations on outer flat space from the CFT point of view.


In the CFT analysis, the D1-D5-P black hole is dual to a density matrix (or a thermal state). On the other hand, microstates of the black hole are dual to pure states, and, in particular the special non-extremal fuzzballs of reference \cite{Jejjala:2005yu} are dual to pure states. From the CFT analysis it will become clear that there is no qualitative difference between emission from a thermal state and a pure state. The  process dual to  emission from a thermal state is Hawking radiation (and superradiance for rotating black holes) that happens in the presence of horizons, while the  process dual to emission from the special non-extremal fuzzballs manifests itself as ergoregion emission that happens in the absence of horizons. The classical instability in ergoregion emissions found in \cite{Cardoso:2005gj} is seen to come from Bose-enhancement of quanta collecting in the cap of the highly symmetric fuzzball of reference \cite{Jejjala:2005yu}.


The study of emission from brane systems has been spread over several years. Here we give a brief account of the milestones and hope the references we miss can be found in the ones we provide. The emission of minimal scalars from the D1-D5 system without
angular momentum was studied in \cite{Callan:1996dv,Dhar:1996vu,Das:1996wn,Das:1996jy,Maldacena:1996ix,Klebanov:1997cx}. The emission of minimal scalars with angular momenta was studied in \cite{Mathur:1997et,Gubser:1997qr}. Universal dynamics of rotating black holes was studied in \cite{Maldacena:1997ih} while the specifics to the D1-D5-P black hole were studied in \cite{Dias:2007nj}. These results matched with the corresponding CFT results up to an overall constant. This constant was fixed for s-wave emission in \cite{David:1998ev} and for higher partial waves in \cite{Avery:2009tu} using the AdS/CFT correspondence. Emission of fixed scalars were studied in \cite{David:1998ev,Callan:1996tv}. Emission from three-branes was studied in \cite{Gubser:1997yh} while that from extremal Kerr solutions was studied in \cite{Bredberg:2009pv}. Emission from non-extremal fuzzball solutions found in \cite{Jejjala:2005yu} were studied in \cite{Cardoso:2005gj, Chowdhury:2007jx, Avery:2009tu, Chowdhury:2008bd, Chowdhury:2008uj, Avery:2009xr}.

This chapter is divided into two parts. The first part explains Hawking radiation, super-radiance, and ergoregion emission for the D1-D5 system from a gravitational or strong coupling point of view. Here the physics looks quite different for the black hole emissions and for the capped geometry emissions. In the second part we rederive these three emissions from the CFT point of view and show that all these emissions are in-fact only different manifestations of the same process.

\section{Emissions: Gravity Analysis}
\begin{figure}[ht]
\begin{center}
\subfigure[]
{\label{fig:throats-a}
	\includegraphics[width=6.3cm]{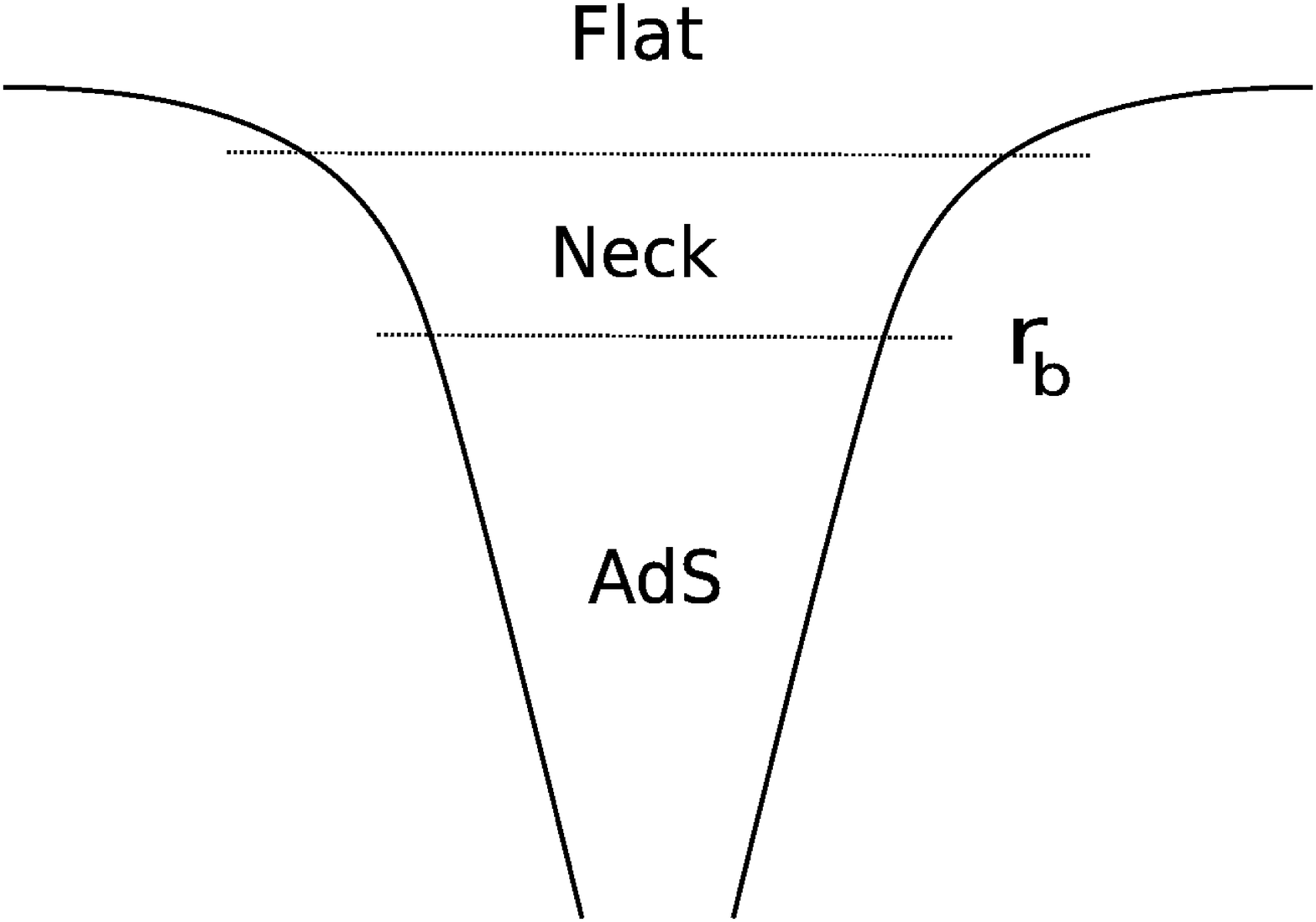}}
\hspace{15pt}
\subfigure[]{\label{fig:throats-b}
	\includegraphics[width=6.3cm]{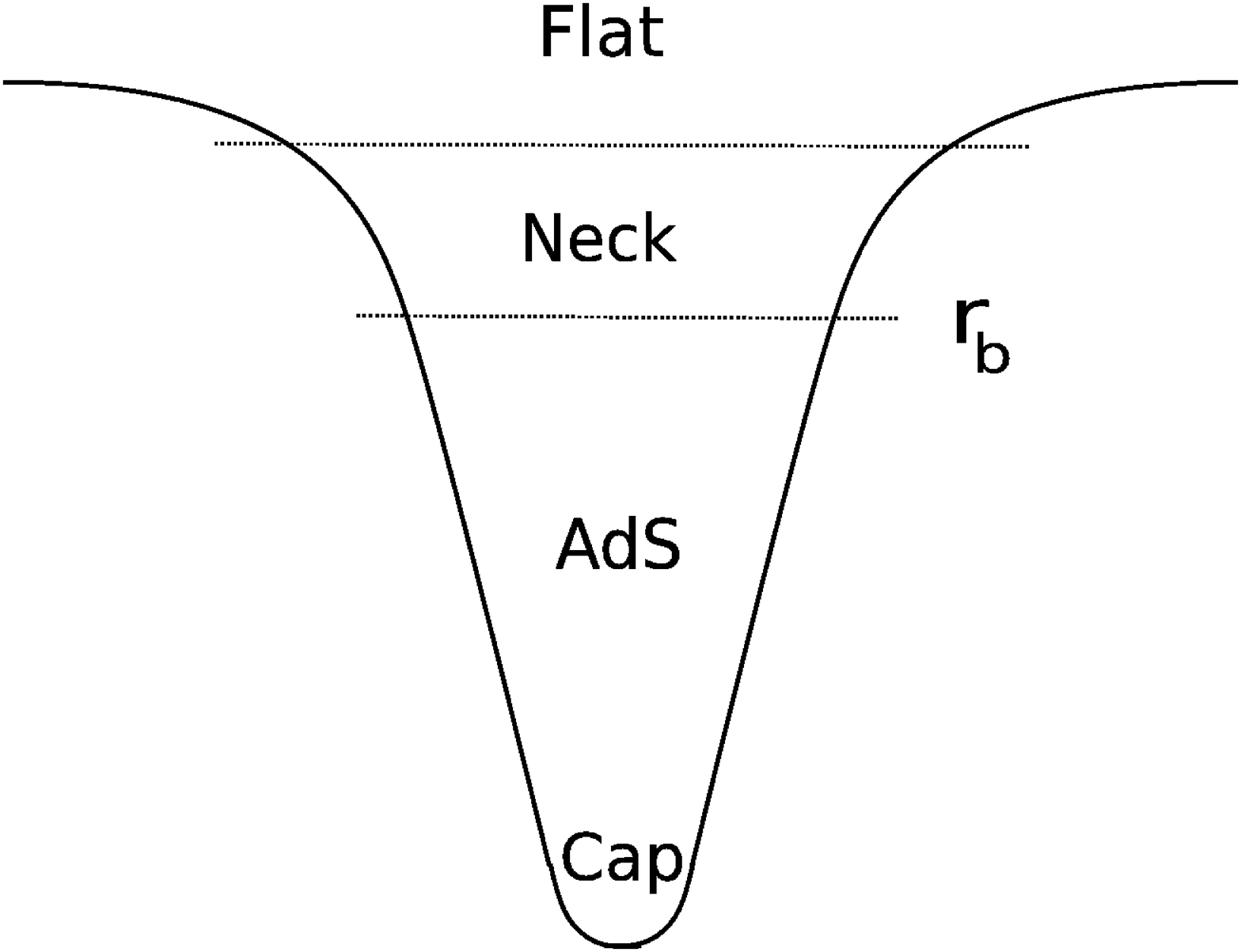}}
\caption{(a) The geometry of black hole is flat at infinity, then there is a `neck' region,  and further-in the geometry takes the form of $AdS_{3}\times S^3$. The $AdS_{3}\times S^3$ region is a part of the BTZ black hole. (b) The geometry of fuzzballs is also flat at infinity followed by a  `neck' region. Still further in, the geometry ends in a `fuzzball cap' whose structure is determined by the choice of microstate. For the states found in \cite{Jejjala:2005yu} the cap and the AdS region is
simply
the global AdS. \label{fig:throats}}
\end{center}
\end{figure}

We first review the supergravity D1-D5-P black hole \cite{Cvetic:1996xz,Cvetic:1997uw} and the non-extremal family of fuzzballs constructed in \cite{Jejjala:2005yu}. We show how the solutions simplify in the decoupling limit into an outer flat space region and and an inner region which is asymptotically AdS space. This is schematically shown in figure \ref{fig:throats}.

We look at the emission of minimally coupled scalars in these backgrounds\footnote{An example of a minimally coupled scalar in these backgrounds is a mixture of the dilaton and the graviton with both indices in the $T^4$ direction: $e^\f{-\phi}{2} h_{ab}$.}. For the black hole we first calculate the probability for the absorption of a minimal scalar. The emission rate is then related to the absorption probability by detailed balance.
This gives us the emission rate as a function of frequency, momentum along $S^1$, and angular momenta. We use spectral flow to relate this result to superradiance. Then we will look at the emission from the non-extremal fuzzballs. This results in a discrete spectrum and a classical instability.



\subsection{Gravity Solutions: Black Holes and Smooth Solutions}


The supergravity D1-D5-P black hole solution was found in \cite{Cvetic:1996xz,Cvetic:1997uw} and a non-extremal family of fuzzballs was constructed in \cite{Jejjala:2005yu}.
A cartoon of these geometries is shown in figure \ref{fig:throats}.
In both cases there is an inner (throat) region, an outer region, and a neck region that joins the inner and the outer regions.  The inner region ends up in a horizon in the black hole case, and has smooth cap in the case of fuzzballs. It turns out that the physics is much cleaner if we work in the near-decoupling limit. The solutions and their decoupling limits are discussed in detail in appendix \ref{App:D1D5P}.

The outer region metric is flat both in the case of the black hole and in the case of the capped geometries. It is simply given by
\be
ds_{\rom{outer}}^2= - dt^2 + dy^2 + dr^2 + r^2 (d\theta^2 + \sin^2 \theta d\psi^2 + \cos^2 \theta d\phi^2)~.
\ee
The inner region is analyzed separately for the two cases.
\subsubsection{D1-D5-P Black Hole}

For the black hole the inner region metric is BTZ times an $S^3$. The $S^3$ is fibered over the BTZ when angular momenta are present. This geometry has been extensively studied in \cite{Cvetic:1998xh}. The metric is given by
\bea
ds^2_\rom{inner,~BH} &=& \sqrt{Q_1 Q_5} \Big( - f^2 d \tau^2 + \rho^2 (d\varphi - \f{J_3}{2 \rho^2} d\tau)^2 + \f{d\rho^2}{f^2} \nn
&& + d\theta^2 + \sin^2 \theta (d \phi  -  \f{4 G^{(5)}}{\pi } \f{R}{Q_1 Q_5} (J_\psi d \varphi + J_\phi d \tau) )^2 \nn
&& +  \cos^2 \theta (d \psi  -  \f{4 G^{(5)}}{\pi }\f{R}{Q_1 Q_5}( J_\phi d \varphi + J_\psi d \tau))^2  \Big)~, \label{BHInner}
\eea
where
\bea
f^2 &=& \rho^2 - M_3 + \f{J_3^2}{4 \rho^2} =: \f{(\rho^2 - \rho_+^2)(\rho^2 - \rho_-^2)}{\rho^2}~, \nn
M_3 &=& \f{R^2}{Q_1 Q_5} [ (M-a_1^2 -a_2^2) \cosh 2 \delta_p + 2 a_1 a_2 \sinh 2 \delta_p ]~, \nn
J_3 &=& \f{R^2}{Q_1 Q_5} [ (M-a_1^2 -a_2^2) \sinh 2 \delta_p + 2 a_1 a_2 \cosh 2 \delta_p ]~.
\eea
Here the coordinates $\rho, \, \tau$ and $\varphi$ are related to the coordinates in the outer flat space as
\bea
\rho^2 &=& \f{R^2}{Q_1 Q_5} (r^2 + (M - a_1^2 -a_2^2) \sinh^2 \delta_p + a_1 a_2 \sinh 2 \delta_p)~, \nn
\tau &=& \f{t}{R}~, \nn
\varphi &=& \f{y}{R}~, \label{BHInnerOuterCoordinates}
\eea
and $M,a_1,a_2, \delta_p$ are parameters specifying the geometry. The mass above extremality, momentum charge radius, and angular momenta of the geometry are given as
\bea
\Delta M_\rom{ADM} &=& \f{\pi }{8 G^{(5)}} M \cosh 2 \delta_p~, \nn
Q_p &=&  \f{M}{2} \sinh 2 \delta_p~, \nn
J_\psi&=& - \f{\pi}{ 4 G^{(5)}} \sqrt{Q_1Q_5}( a_1 \cosh \delta_p - a_2 \sinh\delta_p)~, \nn
J_\phi&=& - \f{\pi}{4 G^{(5)}} \sqrt{Q_1Q_5}( a_2 \cosh \delta_p - a_1 \sinh\delta_p)~.
\eea
The entropy is given by
\bea
S &=&\f{\pi^2}{4 G^{(5)}} \sqrt{Q_1 Q_5} \Big[ \sqrt{ \left(\f{4 G^{(5)}}{\pi} \Delta M_\rom{ADM}+ Q_p\right) - \left( \f{4 G^{(5)}}{\pi} \f{1}{\sqrt{Q_1 Q_5}} (J_\phi - J_\psi) \right)^2} \nn
&& \ \  + \ \  \sqrt{ \left(\f{4 G^{(5)}}{\pi} \Delta M_\rom{ADM}- Q_p\right) - \left( \f{4 G^{(5)}}{\pi} \f{1}{\sqrt{Q_1 Q_5}} (J_\phi + J_\psi) \right)^2} \Big]~,
\eea
while the temperature and the velocity of the horizon along the $y$ direction  are given by
\bea
T_\rom{H} &=& \f{1}{2 \pi R} \f{\rho_+^2 - \rho_-^2}{\rho_+}~, \nn
\Omega_\rom{H} &=& \f{J_3}{2 \rho_+^2}  = \f{\rho_-}{\rho_+}~.
\eea
Our later expressions will be seen to cleanly split into a left and right sector. So in anticipation we also define a left and a right temperature
\be
T_L = \f{1}{2\pi R} (\rho_+ -\rho_-)~,    \qquad T_R= \f{1}{2 \pi R} (\rho_+ + \rho_-)~.
\ee

\subsubsection{Smooth Solutions}

These solutions have a capped inner region. In the  decoupling limit they are given by a global AdS with an $S^3$ fibered over it,
\bea
ds^2_\rom{inner, smooth} &=& \sqrt{Q_1 Q_5} \Big( - (\rho^2+1) d\tau^2 + \rho^2 d \varphi^2 + \f{d \rho^2}{\rho^2+1} \nn
&& + d\theta^2 + \sin^2 \theta (d \phi  -  \f{4 G^{(5)}}{\pi } \f{R}{Q_1 Q_5} (J_\psi d \varphi + J_\phi d \tau) )^2  \nn
&& +  \cos^2 \theta (d \psi  -  \f{4 G^{(5)}}{\pi }\f{R}{Q_1 Q_5}( J_\phi d \varphi + J_\psi d \tau))^2  \Big) \nn
\eea
Here the coordinate $\rho,\tau$ and $\varphi$ are related to the coordinates in flat space as
\bea
\rho^2 &=& \f{R^2}{Q_1 Q_5} \left(r^2 + \f{Q_1 Q_5}{R^2} \f{s^4}{1-s^4} \right)~, \nn
\tau &=& \f{t}{R}~, \nn
\varphi &=& \f{y}{R}~, \label{SmoothInnerOuterCoordinates}
\eea
where
\be
s^2= \Big | \f{\sqrt{n_L(n_L+1)} - \sqrt{n_R(n_R+1)} }{\sqrt{n_L(n_L+1)} + \sqrt{n_R(n_R+1)}}\Big |~.
\ee
The non-negative integers $n_L,n_R$ parameterize the system. The mass above extremality and charges are given by
\bea
\Delta M_\rom{ADM} &=&\f{n_L(n_L+1) + n_R(n_R+1) }{R} ~,  \nn
Q_p &=&   (n_L (n_L+1) - n_R (n_R+1)) \f{Q_1 Q_5}{R^2} ~,\nn
J_\psi&=& - (n_L+n_R +1) \f{\pi}{4 G^{(5)}} \f{Q_1Q_5}{R}~, \nn
J_\phi&=& -(n_L-n_R) \f{\pi}{4 G^{(5)}} \f{Q_1Q_5}{R}~. \label{SmoothSolutionCharges}
\eea
Since this is a smooth geometry without any horizon, it has no Bekenstein-Hawking entropy.
The extremal solution found in \cite{Balasubramanian:2000rt,Maldacena:2000dr} are obtained by taking $n_L=n_R=0$ while those found in \cite{Lunin:2004uu, Giusto:2004id} are obtained by taking $n_L \ne 0, n_R=0$.

\subsection{Emission from Black Holes: Hawking Radiation and Superradiance} \label{Section:BHGravityRadiation}

For the D1-D5-P black hole we first find the absorbtion cross-section for a minimally coupled scalar in this background and the calculate the decay rate using detailed balance. We start by solving the equation
\be
\Box \Phi =0~,
\label{boxphi}
\ee
with purely ingoing boundary conditions at the horizon. It is much simpler to work with a non-rotating black hole first. The results can be  very easily generalized to rotating black holes using spectral flow.

\subsubsection{Non-Rotating Black Holes: Hawking Radiation}
The non-rotating black hole is obtained by simply putting $J_\psi = J_\phi=0$ in the metric \bref{3charge}. We take as the ansatz for the scalar wave
\be
\Phi = e^{-i \omega t + i \lambda y} Y_{l,m_\psi,m_\phi}(\theta,\psi,\phi) h(r)~,
\ee
with the convention
\be
\int |Y|^2 d\Omega_3 = 1~.
\ee
Periodicity in the $y$ direction restricts $\lambda$ to be an integer multiple of $\f{1}{R}$.

We now solve for the wave equation in the inner BTZ $\times S^3$ region and in the outer flat space region separately. We then impose the boundary condition of being purely ingoing at the horizon and match the inner and outer solutions in the neck region. This  gives us the probability of absorption that can be related to the cross-section and hence to the decay rate. The angular part of the equation \bref{boxphi} separates\footnote{\label{footnote-decoupling}This is strictly true only in the decoupling limit. To account for the neck region joining the inner and outer region we have to replace $l \to l + \epsilon$ where $\epsilon$ is the parameter that controls the decoupling: $ \epsilon\approx O(\f{\sqrt{Q_1 Q_5}}{R} \sqrt{\omega^2- \lambda^2})$ . The decoupling limit corresponds to $\epsilon \to 0$. Further details on this issue can be found in \cite{Lunin:2001dt, Chowdhury:2008bd}.
To keep the notation simple we write $\epsilon$ explicitly only when we take the strict decoupling limit.} and the solutions are simply given by spherical harmonics on $S^3$. The radial part of the equation becomes
\be
\f{1}{\rho}\partial_\rho(\rho f^2 \partial_\rho h_{\rom{in}}) + \left[ \f{(\omega R)^2}{f^2} - \f{(\lambda R)^2}{\rho^2} \left( 1- \f{J_3^2}{4 f^2 \rho^2}\right) - \f{\omega \lambda R^2 J_3}{f^2 \rho^2} - l(l+2) \right]h_{\rom{in}}=0~.
\ee
With the substitution
\be
z=\f{\rho^2- \rho_+^2}{\rho^2-\rho_-^2}
\ee
the radial equation takes the form
\be
(1-z) \partial_z (z \partial_z h_{\rom{in}}) + \left( \f{A}{z}  - \f{l(l+2)}{4(1-z)} +B \right)h_{\rom{in}}=0~,
\ee
where
\bea
A&=& \left( \f{\omega-\lambda}{8\pi T_L} + \f{\omega+\lambda}{8 \pi T_R} \right)^2~, \nn
B&=& - \left( \f{\omega-\lambda}{8\pi T_L} - \f{\omega+\lambda}{8 \pi T_R} \right)^2~.
\eea
The solution of this equation with purely ingoing boundary conditions at the horizon $z=0$ is
\be
h_{\rom{in}}=z^{-i \sqrt{A}} (1-z)^{-\f{l}{2}} \phantom{a}_2 F_1 \left[-\f{l}{2} - \sqrt{B} -i \sqrt{A},-\f{l}{2} + \sqrt{B} -i \sqrt{A},1-2 i \sqrt{A}; z \right]~.
\ee

The outer region is flat spacetime and the solution of the scalar equation there is
\be
h_{\rom{out}} = \f{1}{\sqrt{\omega^2 - \lambda^2} r} \left( C_1 J_{1+l} ( \sqrt{\omega^2 - \lambda^2} r) + C_2 J_{-1-l} ( \sqrt{\omega^2 - \lambda^2} r) \right)~. \label{outer}
\ee
To find the complete solution we now need to match the above two solutions in the neck region  $r \approx (Q_1 Q_5)^\f{1}{4}$.
In the $z$ coordinate the neck region is at $z \approx 1$. Using the asymptotic form of hypergeometric functions, the inner solution in the neck region becomes
\bea
h_{\rom{in}} &\approx&  \f{ \Gamma(1- 2 i \sqrt{A}) \, \Gamma(l+1)}{\Gamma(1+\f{l}{2} + \sqrt{B} -i\sqrt{A}) \, \Gamma(1+\f{l}{2} - \sqrt{B} -i\sqrt{A})} \left( \f{\rho}{\sqrt{\rho_+^2 - \rho_-^2} }\right)^l \nn
&& \ + \ \  \f{ \Gamma(1- 2 i \sqrt{A}) \, \Gamma(-l-1)}{\Gamma(-\f{l}{2} - \sqrt{B} -i\sqrt{A}) \, \Gamma(-\f{l}{2}+ \sqrt{B} -i\sqrt{A})} \left( \f{\rho}{\sqrt{\rho_+^2 - \rho_-^2}} \right)^{-(l+2)}~.
\eea
In the neck region the radial coordinate $\rho$ using \bref{BHInnerOuterCoordinates} becomes
\be
\rho = \f{r R}{\sqrt{Q_1 Q_5}}~.
\ee
Using asymptotic form of the Bessel functions, the outer solution in the neck region takes the form
\bea
h_{\rom{out}} &\approx& C_1 \left( \rho \sqrt{\f{ (\omega^2 - \lambda^2) Q_1 Q_5}{R^2}} \right)^l \f{1}{2^{l+1} \Gamma(l+2)} \nn & & + \,  C_2 \left(\rho \sqrt{\f{ (\omega^2 - \lambda^2) Q_1 Q_5}{R^2}}  \right)^{-l-2} \f{2^{l+1}}{\Gamma(-l)}~. \label{outerNeck}
\eea
Matching the two solutions in the neck region gives
\bea
\f{C_2}{C_1} &=& \left[ \pi^2 (\omega^2 - \lambda^2) \, Q_1 \, Q_5 \,  T_L \, T_R \right]^{l+1} \nn & & \times \left[ \f{\Gamma(-l) \Gamma(-l-1)}{\Gamma(l+2)\Gamma (l+1) } \right] \left( \f{\Gamma(1+ \f{l}{2} - i \f{\omega + \lambda}{ 4 \pi T_R}) \Gamma(1+ \f{l}{2} - i \f{\omega - \lambda}{ 4 \pi T_L})}{\Gamma(- \f{l}{2}-  i \f{\omega + \lambda}{ 4 \pi T_R}) \Gamma(-\f{l}{2}  - i \f{\omega - \lambda}{ 4 \pi T_L})} \right)~.
\eea
At this point recalling the discussion in footnote \ref{footnote-decoupling} and using the identities
\be
\Gamma (-n-\epsilon) = -\f{1}{(-1)^n n! \epsilon}~, \qquad \Gamma(1-x) \Gamma(x) = \f{\pi}{\sin \pi x}~,
\ee
we get
\bea
\f{C_2}{C_1} &=& - \f{\kappa}{\epsilon^2} \sin \left(-\f{\pi l}{2} - \f{\pi \epsilon}{2} - i \f{\omega-\lambda}{4 T_L} \right) \sin \left(-\f{\pi l}{2} - \f{\pi \epsilon}{2} - i \f{\omega + \lambda}{4 T_R} \right)~,
\eea
where
\bea
\kappa &=& \left(\f{1}{\pi \, l!(l+1)!}\right)^2 \left[\pi^2 (\omega^2 - \lambda^2) Q_1 Q_5  T_L T_R \right]^{l+1}  \nn \ \ \ \ & &  \times \ \left| \Gamma\left(1+ \f{l}{2} - i \f{\omega + \lambda}{ 4 \pi T_R}\right) \right|^2 \left|\Gamma\left(1+ \f{l}{2} - i \f{\omega - \lambda}{ 4 \pi T_L}\right)\right|^2~.
\eea

The outer solution can also be written in terms of the ingoing and outgoing solutions using the appropriate asymptotic behavior of Bessel functions. The result is
\bea
h &\approx& \f{1}{\sqrt{2\pi}} \f{1}{(\sqrt{\omega^2 - \lambda^2} r)^\f{3}{2}} \Big( e^{ikr} e^{- i \f{\pi}{4}} \left(C_1 e^{-i (l+1) \f{\pi}{2}} + C_2 e^{i (l+1) \f{\pi}{2} }\right) \nn & & \ \ \ \ \ \ \ \  \ \ \ \ \ \ \ \   \   \ \ \ \ \ \ + \  e^{-ikr} e^{i \f{\pi}{4} }\left(C_1 e^{i (l+1) \f{\pi}{2}} + C_2 e^{-i (l+1) \f{\pi}{2} }\right) \Big)~. \label{outerAsymptotic}
\eea
Thus we get the ratio of the outgoing to ingoing flux to be
\be
 \f{\mathcal F_\rom{out}}{\mathcal F_\rom{in}} = \left |\f{1+ (-1)^{l+1} \f{C_2}{C_1} e^{i \epsilon \pi}}{(-1)^{l+1} e^{i \epsilon \pi}+ \f{C_2}{C_1} } \right |^2
\ee
The probability for absorption is therefore
\bea
\mathcal P &=& 1 -  \f{\mathcal F_\rom{out}}{\mathcal F_\rom{in}} \approx \left| 1+ (-1)^{l+1} \f{C_2}{C_1} e^{-i \epsilon \pi} \right|^2 -  \left| 1+ (-1)^{l+1} \f{C_2}{C_1} e^{i \epsilon \pi} \right|^2~, \nn
&=& (-1)^{l+1} (2i) \sin (\epsilon \pi) \left[ \left( \f{C_2}{C_1} \right)^* -  \left( \f{C_2}{C_1} \right) \right]~, \nn
&=& 2 \,\pi^2 \, \kappa \,\sinh \left( \f{\omega- \lambda}{4 T_L} + \f{\omega + \lambda}{4 T_R}\right)~. \label{probabsorption}
\eea
The probability of abosroption is related to the difference between the rates of absorption and emission as
\be
d\Gamma_\rom{absorption}-d\Gamma_\rom{emission} = \mathcal P \f{d \omega}{2 \pi}
\ee
The minimally coupled scalars being emitted are bosons. Using Bose-Einstein statistics we get the following condition from the detailed balance
\be
(1+\rho_L)(1+\rho_R) \Gamma_\rom{emission}=\rho_L \rho_R \Gamma_\rom{absorption}~,
\ee
where the $\rho_L$ and $\rho_R$ are, for the case of the D1-D5-P black hole, thermal population densities with left and right temperatures $T_L$ and $T_R$ respectively,
\be
\rho_L = \f{1}{e^{\f{\omega-\lambda}{2 T_L}}-1}, \qquad \rho_R = \f{1}{e^{\f{\omega+\lambda}{2 T_R}}-1}~.
\ee
Thus we get the rate of emission in terms of the probability of absorption as
\bea
d \Gamma_\rom{emission} &=& \mathcal P \left(\f{1}{e^{\f{\omega-\lambda}{2 T_L} + \f{\omega+ \lambda}{2 T_R} } -1} \right) \f{d\omega}{2 \pi} \nn
&=& \f{1}{2 \pi} \left(\f{1}{l!(l+1)!}\right)^2 \left[ \pi^2 (\omega^2 - \lambda^2) Q_1 Q_5  T_L T_R \right]^{l+1}  \nn
&& \  \times \  e^{-\f{\omega - \lambda}{4 T_L}  } e^{-\f{\omega + \lambda}{4 T_R} } \left| \Gamma(1+ \f{l}{2} - i \f{\omega + \lambda}{ 4 \pi T_R}) \right|^2 \nn
& & \ \times \  \left|\Gamma(1+ \f{l}{2} - i \f{\omega - \lambda}{ 4 \pi T_L})\right|^2 d \omega \label{EmmNonRotBH}~.
\eea
The probability of absorption is %
\be
d \Gamma_\rom{absorption} = \mathcal P \f{1}{1-e^{-\f{\omega-\lambda}{2 T_L} - \f{\omega+ \lambda}{2 T_R} }} \f{d\omega}{2 \pi} \label{AbsProb}~.
\ee
The cross section is related to the absorption probability through \cite{Gubser:1997qr}
\be
\label{sigma3}
\sigma = (l+1)^2 \f{4 \pi}{(\sqrt{\omega^2- \lambda^2})^3} \mathcal P~.
\ee

\subsubsection{Rotating Black Holes: Superradiance} \label{EmmisionSuperRadiance}
As promised, now we  use the above results to get the emission rate for rotating black holes. This is possible because in the decoupling limit the inner region is an $S^3$ fibered over the BTZ black hole. From the metric \bref{BHInner} we can see that when we use the transformed coordinates
\bea
\phi' &=& \phi -  \f{4 G^{(5)}}{\pi } \f{R}{Q_1 Q_5} (J_\psi  \varphi + J_\phi  \tau)~,  \nn
\psi' &=& \psi -   \f{4 G^{(5)}}{\pi }\f{R}{Q_1 Q_5}( J_\phi  \varphi + J_\psi  \tau)~, \label{FibrationBH}
\eea
with the rest of the coordinates going to themselves, the metric becomes that of BTZ $\times S^3$ without any fibration, hence that of the inner region of the non-rotating black hole. Under this coordinate transformation the coordinate derivatives transform as
\bea
\partial_{\tau'} &=& \partial_\tau + \f{4 G^{(5)}}{\pi } \f{R}{Q_1 Q_5} ( J_\phi \partial_\phi + J_\psi \partial _\psi)~, \nn
\partial_{\varphi'} &=& \partial_\varphi + \f{4 G^{(5)}}{\pi } \f{R}{Q_1 Q_5} ( J_\psi \partial_\phi + J_\phi \partial _\psi)~,
\eea
with the rest of the derivatives going to themselves. Therefore, we simply need to make the change
\bea
\omega &\to& \bar \omega = \omega-  \f{4 G^{(5)}}{\pi } \f{1}{Q_1 Q_5} ( J_\phi m_\phi + J_\psi m _\psi) \nn
\lambda &\to& \bar \lambda = \lambda +  \f{4 G^{(5)}}{\pi } \f{1}{Q_1 Q_5} ( J_\psi m_\phi + J_\phi m _\psi) \nn
\eea
{\em in the inner solution} to get the results for the rotating black hole. This results in
 \bea
\f{d \Gamma_\rom{emission}}{d\omega} &=& \f{1}{2\pi (l! (l+1)!)^2}\left(Q_1 Q_5 \f{\omega^2 -\lambda^2}{4} \right)^{l+1} (2 \pi T_L)^{l+1} (2 \pi T_R)^{l+1}  E_1 E_2 \nn
&& \times \  \left | \Gamma(\f{l}{2}+1+ i \left(\f{\omega-\lambda -\f{4 G^{(5)}}{\pi } \f{1}{Q_1 Q_5} (J_\psi + J_\phi) (m_\psi + m_\phi)}{4 \pi T_L}\right) \right|^2 \nn
&& \times \ \left | \Gamma(\f{l}{2}+1+ i \left( \f{\omega+\lambda - \f{4 G^{(5)}}{\pi } \f{1}{Q_1 Q_5} (J_\psi - J_\phi) (m_\psi - m_\phi)}{4 \pi T_R} \right) \right |^2 \, ,  \label{Eqn:GammaBHGravity}
\eea
where
\bea
E_1 &:=& e^{-\f{\omega-\lambda - \f{4 G^{(5)}}{\pi } \f{1}{Q_1 Q_5} (J_\psi + J_\phi) (m_\psi + m_\phi)}{4 T_L}}~, \\
E_2 &:=& e^{-\f{\omega+\lambda - \f{4 G^{(5)}}{\pi } \f{1}{Q_1 Q_5} (J_\psi - J_\phi) (m_\psi - m_\phi)}{4 T_R}}~. \label{E1E2}
\eea
From the expression for the probability \bref{probabsorption} and the cross-section \bref{sigma3} it follows  that the cross-section is proportional to
\be
\sigma \propto \sinh \left[ -\log E_1 - \log E_2  \right]~, \label{cross-section-prop}
\ee
with a positive proportionality constant.
Since $\omega^2 > \lambda^2$, for the non-rotating black hole it is not possible to have a negative cross-section. However, in the case of rotating black holes 
the cross section can be negative. To look at a much quoted simple case, take the left and right temperatures to be equal and the case of emission of quanta that have no motion along the $y$ and $\psi$ directions. Then we get
\be
\sigma \propto \sinh \left( \f{\omega - \f{4 G^{(5)}}{\pi} \f{R}{Q_1 Q_5} m_\phi J_\phi}{2T}\right)~,
\ee
which is negative if the scalar wave is co-rotating with the black hole $m_\phi J_\phi >0$ and the emission energy is below a threshold  $\omega < \f{4 G^{(5)}}{\pi} \f{R}{Q_1 Q_5} m_\phi J_\phi$. Thus if a scalar-wave falls on a rotating black hole with the above conditions met, it is reflected back with an increased amplitude. This is the same superradiance as discussed in section \ref{penrose}. When we study the microscopics of emission in the next subsection we will see that this is very similar to the lasing action. The superradiant emission happens even at zero temperature \cite{Dias:2007nj}. The emission at energies higher than the superradiant bound have a positive cross-section and behave just like Hawking radiation.

\subsection{Emission from Smooth Solutions with Ergoregions}
The emission calculation from smooth solutions with ergoregions is very similar to the  one presented above. The main difference is in the boundary conditions. This time we impose regularity in the inner part  as there is no horizon and purely outgoing boundary conditions at infinity. We can again first analyze the non-rotating case of $AdS_3 \times S^3$. Technically this is not a solution for the D1-D5 system as the fermions have to be periodic on the boundary. However by a coordinate change similar to \bref{FibrationBH} the solution can be mapped to D1-D5 solutions.

Following the discussion presented above we first solve the wave equation in the inner region. The angular part of the solution gives us the regular spherical harmonics\footnote{\label{footnote-decoupling-smooth} Just like in the black hole case explained in footnote \ref{footnote-decoupling} this is strictly true only in the decoupling limit. We take $l \to l+ \epsilon$, without explicitely writing so, and take $\epsilon \to 0$ in the end.} and the radial part becomes
\be
\f{1}{\rho} \partial_\rho(\rho (\rho^2+1)\partial_\rho h_{\rom{in}}) + \left[ \f{(\omega R)^2}{\rho^2+1} - \f{(\lambda R)^2}{\rho^2} - l(l+2) \right] h_{\rom{in}}=0~,
\ee
Defining the new radial coordinate
\be
x= \rho^2~,
\ee
the radial equation becomes
\be
4 \partial_x (x(x+1)\partial_x h_{\rom{in}}) + \left[ \f{\omega^2 R^2}{x+1} - \f{\lambda^2 R^2}{x} - l(l+2) \right] h_{\rom{in}}=0~.
\ee
The solution to this equation that is regular everywhere is
\be
h_{\rom{in}} = x^\f{|\lambda| R}{2} (1+x)^\f{\omega R}{2} \phantom{a}_2 F_1\left(\f{(\omega+|\lambda|)R +l+2}{2}, \f{(\omega+| \lambda| )R-l}{2},1+| \lambda | \, R,-x\right)~.
\ee
The outer region is flat so we get the same solution as \bref{outer}. Now we match the two solutions in the neck region. In the $x$ coordinate, the neck region corresponds to taking the limit $x \to \infty$. In this limit the inner solution takes the form 
\bea
h_{\rom{in}} &\approx& \Gamma(1+ |\lambda| R) \, \Big[  \f{\Gamma(-l-1)}{\Gamma\left( \frac{(\omega+ |\lambda|)R-l}{2}\right) \Gamma\left(\frac{-(\omega - |\lambda|)R -l}{2}\right)} \rho^{-(l+2)} \nn
&&  \  \  \ \ \ \ \ \ \ \ \ \ \ \ \  \  + \  \f{\Gamma(l+1)}{\Gamma\left(\f{(\omega+ |\lambda|)R+l+2}{2}\right) \Gamma\left(\f{-(\omega - |\lambda|)R +l+2 }{2}\right)} \rho^l  \Big]~.
\eea
In the neck region the radial coordinate $\rho$ using \bref{SmoothInnerOuterCoordinates} becomes
\be
\rho = \f{r R}{\sqrt{Q_1 Q_5}}~,
\ee
making the outer solution behave in the neck region in exactly the same way as it does for the black hole case \bref{outerNeck}.
Matching the two solutions  we get
\be
\f{C_1}{C_2} \left[ (\omega^2 - \lambda^2) \f{Q_1 Q_5}{4 R^2} \right]^{(l+1)} \f{\Gamma(-l)}{\Gamma(l+2)} =   \f{\Gamma(l+1)}{\Gamma(-l-1)} \Xi
\ee
where
\be
\Xi=
\f{\Gamma\left(\frac{(\omega + |\lambda|)R - l}{2}\right) \, \Gamma\left(\frac{-(\omega - |\lambda|)R - l}{2}\right)}{\Gamma\left(\f{(\omega + |\lambda|)R + l+2}{2}\right) \, \Gamma\left(\f{-(\omega - |\lambda|)R + l+2}{2}\right)}~.
\ee

By imposing purely outgoing boundary conditions on the outer  asymptotic solution \bref{outerAsymptotic} and recalling footnote \ref{footnote-decoupling-smooth} we get
\be
\f{C_1}{C_2} = (-1)^l e^{-i \epsilon \pi}~.
\ee
Thus the matching condition is
\be
(-1)^l e^{-i \epsilon \pi} \left[ (\omega^2 - \lambda^2) \f{Q_1 Q_5}{4 R^2} \right]^{l+1} \f{\Gamma(-l)}{\Gamma(l+2)} = \f{\Gamma(l+1)}{\Gamma(-l-1)} \Xi~. \label{Eqn:transcendental}
\ee

Now recall that, we solved the wave equation by imposing smoothness in the interior and purely outgoing boundary conditions at infinity. Since we have a second order equation and two boundary conditions, solving \bref{Eqn:transcendental} gives us a discrete spectrum. We will see that the resulting frequencies have a real and an imaginary part. We solve this using the method of \cite{Cardoso:2005gj}. Since we are looking at $\omega \f{(Q_1 Q_5)^\f{1}{4}}{R} \ll 1$  we can to zeroth order take the left hand side to be zero. We have two choices
\be
\omega R = l+ 2 + | \lambda | R + 2 \tilde N~,
\ee
and
\be
\omega R = -l -2 - | \lambda | R - 2 \tilde N~, \label{NegEnergySpectrum}
\ee
where $\tilde N$ is a non-negative integer. Note that the second solution does not make sense as it gives negative frequencies. However, we cannot simply throw it away, because our goal is to look at rotating solutions and right now we are only looking at the non-rotating ones. We will see in a moment that the second solution also gives positive frequencies when we look at rotating geometries by shifting the frequencies in a fashion similar to that described in section \ref{EmmisionSuperRadiance}.
We will also see that even though both choices lead to complex frequencies the first one gives negative imaginary contributions to the frequency. This represents decay of particles piled up in the core region tunneling out of the neck \cite{Lunin:2001dt, Chowdhury:2008bd}. Although this is a process by which excited D1-D5 systems can decay,  it is not a classical instability.  The second choice leads to positive imaginary contribution to the frequency which represents an exponential growth in the number of particles outside (as well as inside) \cite{Chowdhury:2008bd}. This instability is the ergoregion instability. To get the imaginary part of the frequency we now perturb the solution
\be
\tilde N \to \tilde N+ \delta \tilde N~.
\ee
Now we recall footnote \ref{footnote-decoupling-smooth} and reinstall  $\epsilon$. We also take $ \epsilon \gg \delta \tilde N$, which was shown to be the case in \cite{Chowdhury:2008bd}
\be
(-1)^l e^{-i \epsilon \pi} \left[ (\omega^2 - \lambda^2) \f{Q_1 Q_5}{4 R^2} \right]^{l+1} \f{\Gamma(-l - \epsilon) }{\Gamma(l+1)} = -  \f{1}{\phantom{a}^{\tilde N+l+1 + |\lambda|R} C_{l+1} \phantom{a}^{\tilde N+l+1 } C_{l+1} } \delta \tilde N~.
\ee
We thus get
\be
Im(\delta \tilde N) = -\f{\pi}{(l!)^2} \left [(\omega^2- \lambda^2) \f{Q_1 Q_5}{4 R^2} \right]^{l+1} \phantom{a}^{\tilde N+l+1+ |\lambda|R} C_{l+1} \phantom{a}^{\tilde N+l+1} C_{l+1}~,
\ee
and using \bref{NegEnergySpectrum} we get
\be
 \omega_I =\f{1}{R} \f{2\pi}{(l!)^2} \left [(\omega^2- \lambda^2) \f{Q_1 Q_5}{4 R^2} \right]^{l+1} \phantom{a}^{\tilde N+l+1+ |\lambda|R} C_{l+1} \phantom{a}^{\tilde N+l+1} C_{l+1}~.
\ee
In this equation $\omega$ on the right hand side is understood to be the real part of $\omega$, $\omega_R$. This is because to the leading order $\omega$ is real and is given by \bref{NegEnergySpectrum}.

This gives the rate of increase of the wavefunction. The emission rate is given by
\be
\Gamma = 2 \omega_I
\ee

We translate the results for rotating geometries in a fashion similar to what we did for black holes in section \ref{EmmisionSuperRadiance}.
In the {\em inner region}, it corresponds to the shift
\bea
\omega & \to &   \omega - \f{-(n_L+n_R+1) m_\psi - (n_L-n_R) m_\phi}{R}~, \nn
\lambda & \to &   \lambda + \f{-(n_L+n_R+1) m_\phi - (n_L-n_R) m_\psi}{R}~,
\eea
where we have used equations \bref{SmoothSolutionCharges} relating the angular momentum of the special states to integers.
The outer region is unaffected by this shift.
Thus the emission spectra now becomes
\bea
\omega_R R &=&  -l -2 -  (n_L+n_R+1) m_\psi - (n_L-n_R) m_\phi \nn & &  \ - | \lambda R - (n_L-n_R) m_\psi -(n_L+n_R+1) m_\phi |  - 2 \tilde N~, \nn
\omega_I &=& \f{1}{R} \f{2\pi}{(l!)^2} \left [(\omega_R^2- \lambda^2) \f{Q_1 Q_5}{4 R^2} \right]^{l+1} \nn
& & \times \  \phantom{a}^{\tilde N+l+1+ |\lambda R- (n_L-n_R) m_\psi -(n_L+n_R+1) m_\phi|} C_{l+1} \phantom{a}^{\tilde N+l+1} C_{l+1}~. \label{RossEmmission}
\eea

\section{Emissions: CFT Analysis}

In the previous subsection we  saw how the three different kind of emissions from gravitational systems --- Hawking radiation and superradiance for
black holes and ergoregion emission for rotating stars --- are present for the supergravity solutions for the D1-D5 system. Since these geometries
have an inner AdS region,  we can analyze these emissions from a CFT point of view.

Given the number of examples that confirm the AdS/CFT correspondence,  the importance of this calculation can hardly be to get another example of
AdS/CFT at work. Instead our aim in this section is to show that in the CFT description the three phenomena mentioned above are the same.
If we take our initial state to be a thermal state without R-charge we get the Hawking radiation; if we take to to be a thermal state with an R-charge
we get superradiance. If we take the initial state to be a very non-generic state, the one found in \cite{Jejjala:2005yu}, the same emission
process gives ergoregion instability.

In this subsection we briefly discuss the D1-D5 CFT, the vertex operator responsible for coupling the AdS region to  flat space,
amplitudes for emissions, spectral flow,  and  states in CFT involved in the emission processes. For simplicity, we do
calculations in the NS-sector instead of in the R-sector.  We then find the emission rates from black holes and non-extremal fuzzballs by relating the results to the R-sector by spectral flow.

\subsection{The D1-D5 CFT} \label{CFT}

Consider Type IIB string theory compactified on
\be
S^1 \times T^4
\ee
and wrap $n_1$ D1 branes on $S^1$ and $n_5$ D5 branes on $S^1 \times T^4$. The bound states of these branes is described by a field theory.
In the limit where the size of $S^1$ is large compared to the size of $T^4$, and at low energies, excitations live only in the $S^1$ directions.
This low energy limit of the field theory flows to a  $\mathcal N=(4,4) $ super-conformal field theory living on the circle $S^1$. A detailed introduction to the CFT and how it arises as a low energy limit of the D1-D5 field theory can be found in \cite{David:2002wn}.

The natural coordinates on the CFT for the time direction and the $S^1$ are the same as the one used for the inner region \bref{BHInnerOuterCoordinates} and \bref{SmoothInnerOuterCoordinates}
\be
\tau = \f{t}{R}, \qquad \sigma= \f{y}{R}~.
\ee
Note that $\sigma \in [0,2\pi)$. We also define light cone coordinates as
\be
u=\tau+ \sigma, \qquad v=\tau-\sigma~.
\ee
It is convenient to euclideanize the time coordinate $\tau_E=  i \tau$ and work in the coordinates on the complex plane
\be
z=e^{\tau_E + i \sigma} , \qquad \bar z= e^{\tau_E - i \sigma}~.
\ee
It has been conjectured that by varying moduli of string theory one can arrive at the `orbifold point' where the CFT is particularly
simple \cite{Seiberg:1999xz,Larsen:1999uk,deBoer:1998ip}. At the orbifold point the CFT is a 1+1 dimensional sigma model. The base
space is spanned by the coordinate $y$ and $t$ of string theory. The target space of the sigma model is the symmetrized product of $n_1 n_5$ copies of $T^4$,
\be
(T_4)^{n_1 n_5}/ S_{n_1 n_5}~,
\ee
with each copy of the torus giving four bosons and four left and right moving fermions.  For the  D1-D5 system the fermions are periodic on the $S^1$. 

The R-symmetry group for this CFT is $SU(2)_L \times SU(2)_R$ which in the dual picture is related to rotation under $SO(4)_E$. The subscript $E$ stands for external as the rotation is in the external space. In addition there is also the  (broken) rotation group in the $T^4$ which is $SO(4)_I = SU(2)_1 \times SU(2)_2$. The $I$ stands for internal.  We use the indices $\alpha,\dot \alpha$ for $SU(2)_L$ and $SU(2)_R$ and the indices $A,\dot A$ for $SU(2)_1$ and $SU(2)_2$ respectively.

The four real fermions of the left (right) sector are grouped under two complex fermions $\psi^{\alpha A}$ ($\bar \psi^{\dot \alpha A}$) .
The bosons are vectors under the $T^4$ and are labeled as $X^{A\dot A}$.

The superconformal generators are denoted by $L_n,G^{\alpha A}_r, J^{a}_n$ for the left sector and
$\bar L_n,\bar G^{\dot \alpha A}_r, \bar J^{a}_n$ for the right sector.

\begin{figure}[ht]
\begin{center}
	\includegraphics[width=6.3cm]{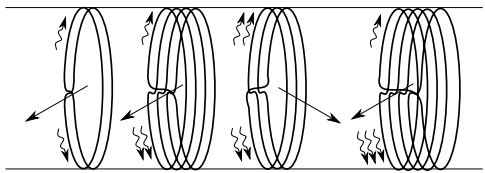}
      \caption{A pictorial representation of the CFT and its excitations.} \label{fig:CFT}
\end{center}
\end{figure}

Pictorially the CFT is described by a number of component strings of different windings wrapped on the circle $S^1$. This is shown in
figure \ref{fig:CFT}. The total winding is $n_1 n_5$.  Each component string also has an R-charge of the CFT and is called base spin
which is represented by the arrow in the figure. The bosonic and fermionic excitations are represented as arrows running along the strings.

\begin{figure}[ht]
\begin{center}
\subfigure[~Untwisted component strings]{\label{fig:twist1}
	\includegraphics[width=5.3cm]{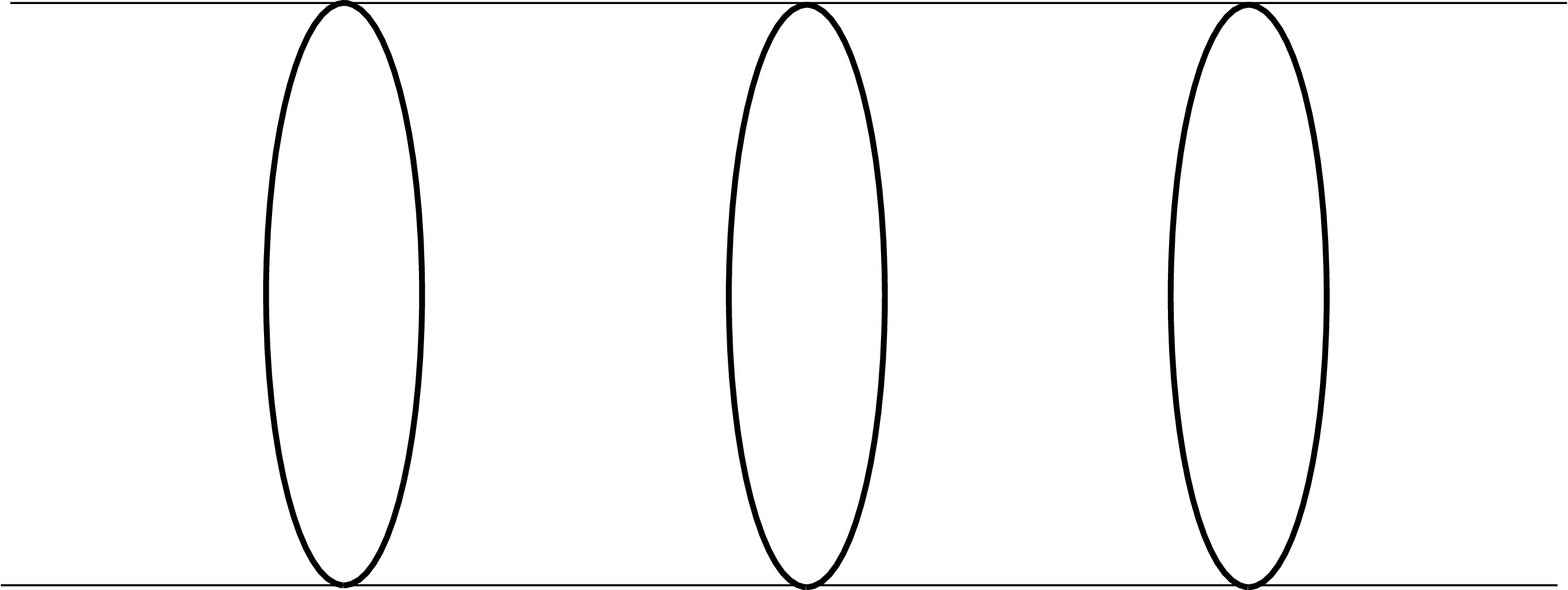}}
\raisebox{30pt}{$\xrightarrow{\hspace{5pt}{\displaystyle \sigma_3}\hspace{3pt}}$}
\subfigure[~The twisted component string]{\label{fig:twist2}
	\includegraphics[width=5.3cm]{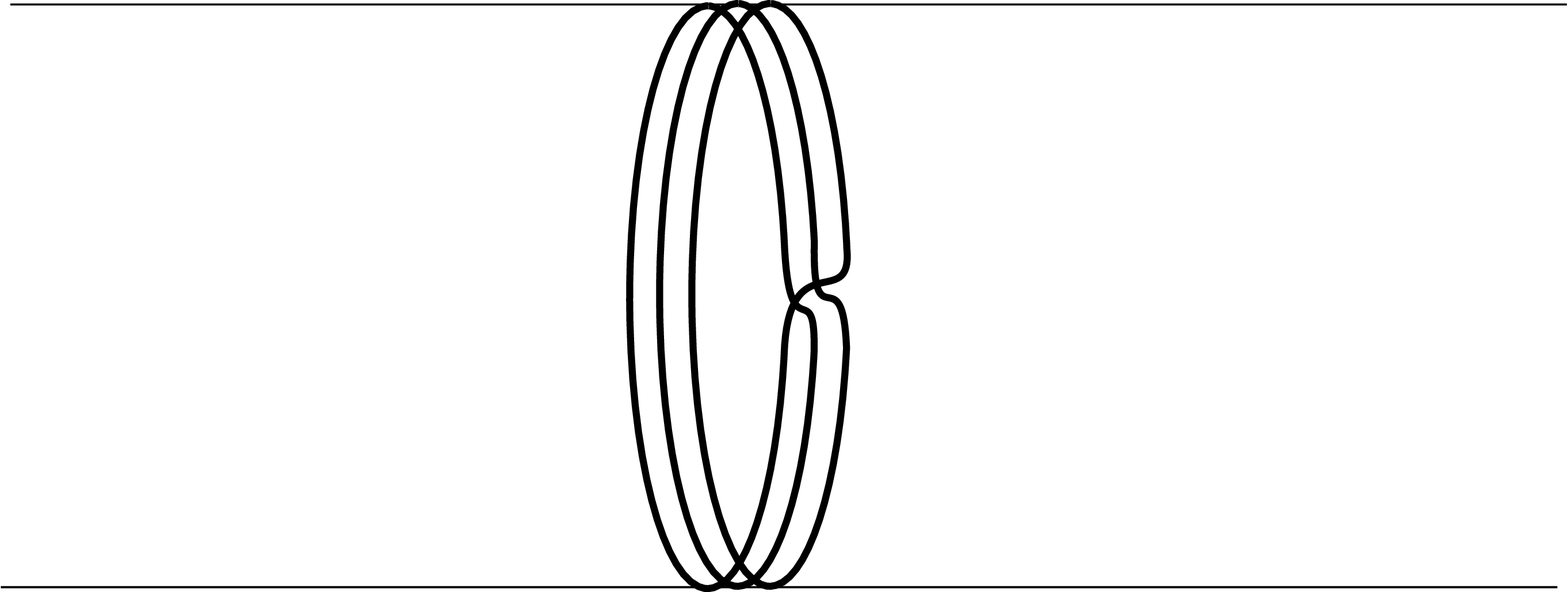}}
      \caption{The twist operator $\sigma_3$.   Each loop represents a
        `copy' of the CFT wrapping the $S^1$. The twist operator joins these copies into one single copy of the CFT living on a circle of three times the length of the original circle. \label{fig:twist}}
\end{center}
\end{figure}

In addition to the operators which can be made from the above field content we also have twist operators denoted by
$\sigma_n$ \cite{Lunin:2000yv,Lunin:2001pw,Pakman:2009zz,Pakman:2009ab}. The `n' gives the order of the twist operator.
The action of this operator is as follows: let the order 5 twist operator \be \sigma_{(24519)}\ee  be inserted in the complex plane, then every time one circles
the operator the fields of the theory get mapped as
\be
X^{A \dot A}_{(2)} \to X^{A \dot A}_{(4)} \to  X^{A \dot A}_{(5)} \to X^{A \dot A}_{(1)} \to X^{A \dot A}_{(9)}\to X^{A \dot A}_{(2)}
\ee
The other copies of $X^{A \dot A}_{m}$ remain unchanged. There is a similar action on fermionic fields. This twisting is schematically
depicted in the figure \ref{fig:twist}.

The connection between various quantities on the gravity side and on the CFT side is needed for the analysis below. The charge radii $Q_1$ and $Q_5$ are related with the number of D1 and D5 branes as
\be
Q_1 = \f{g {\alpha'}^3}{V} n_1 , \qquad Q_5 = g \alpha' n_5~, \label{ChargeRadius}
\ee
while the five dimensional Newton's constant is
\be
16 \pi G^{(5)} = \f{(2\pi)^7 g^2 {\alpha'}^4}{(2\pi)^4 V (2 \pi R)} = \f{(2\pi)^2 g^2 {\alpha'}^4}{ R V}~. \label{NewtonsConstant}
\ee

The mass above extremality $(M)$,  the momentum charge radius along the $S^1$ $(Q_P)$ and the angular momenta ($J_\psi, J_\phi$) are related to the left and right dimensions $(h,\bar h)$ and
R-charges $ (j ,\bar j)$ through the relations
\bea
M R &=& \left( h- \f{n_1 n_5}{4} \right)+\left(\bar h- \f{n_1 n_5}{4}\right)~, \nn
\f{\pi}{4 G^{(5)}} Q_p R &=&  \left( h- \f{n_1 n_5}{4} \right)-\left(\bar h- \f{n_1 n_5}{4}\right)~, \nn
J_\psi &=& - j -\bar j~, \qquad J_\phi = -j + \bar j~. \label{Eqn:dictionary}
\eea
We  subtracted $\f{n_1 n_5}{4}$ from the dimensions since this is the dimension of the Ramond sector ground states.

\subsection*{Vertex Operator} \label{VerOp}

We are considering the emission of a minimal scalar with angular quantum numbers $(l, m_\psi,m_\phi)$. The dimension and charge of the operator
dual to this are
\bea
h=\f{l}{2}+1 &,& \qquad k=\f{m_\psi + m_\phi}{2}~, \nn
\bar h = \f{l}{2}+1 &,& \qquad  \bar  k= \f{m_\psi - m_\phi}{2}~. \label{VertexOperatorCharges}
\eea
This gets fixed using the AdS/CFT correspondence \cite{Witten:1998qj}. Further details on the construction can be found in \cite{Avery:2009tu}. We denote the vertex operator for a minimal scalar as
\be
V_{l,k,\bar k}(\tau,\sigma)~.
\ee

\subsection*{Spectral Flow}

The CFT is naturally defined on the complex plane in the NS sector. However the D1-D5 system is described by the Ramond sector.
It turns out that often it is easier to do CFT calculations in the NS sector and relate them to the actual problem by 
spectral flow \cite{Schwimmer:1986mf}. This is like in the gravity analysis where we first calculated quantities for non-rotating geometries
and then related to rotating geometries via a coordinate transformation. In fact, the gravity dual to the spectral flow is precisely the coordinate transformation used in the gravity analysis.

If we have two states related by spectral flow
\be
\ket{\psi'} = \mathcal U_\alpha \ket{\psi}~, \label{SpectralFlowState}
\ee
then the dimensions and R-charges for the two are related by
\bea
h'&=&h + \alpha j + \alpha^2 \f{ c}{24}~, \nn
j' &=& j + \alpha \f{c}{12}~, \label{SpectralFlow}
\eea
where $\alpha~\in~ \f{\mathbb Z}{k}$, $k$ being the twist order of the sector we are working in \cite{Avery:2009xr,Martinec:2001cf}.
The CFT amplitudes are related under spectral flow by \cite{Avery:2009tu}
\be
\bra{f} V_{l,k,\bar k}(z,\bar z) \ket{i} = z^{-\alpha k} \bar z^{-\bar \alpha \bar k} \bra{f'} V_{l,k,\bar k}(z,\bar z) \ket{i'}~. \label{AmpUnderSpectFlow}
\ee

\subsection{States}

We want to describe the dual to the black hole geometry and duals to the non-extremal smooth solutions of \cite{Jejjala:2005yu}. In general finding the CFT dual
for a smooth geometry is not an easy task, but for geometries of \cite{Jejjala:2005yu} it turns out to be relatively straightforward.

\subsubsection{Non-Extremal Fuzzballs}

The mass, momentum and angular momenta for these geometries are given in \bref{SmoothSolutionCharges}. With the relation \bref{Eqn:dictionary} we get the CFT state with
\be
h = \left(n_L (n_L+1) + \f{1}{4} \right) n_1 n_ 5 , \qquad j =\left( n_L+\h \right) n_1 n_5~, \label{Eqn:JMaRT}
\ee
and similar expressions for the right sector. From the expression for the spectral flow \bref{SpectralFlow} we see that this state is related by a
spectral flow of $2n_L+1$ and $2 n_R+1$ units to the state with $h=\bar h=0$ and $j=\bar j=0$.  This state is just the vacuum in the NS sector. So we see that there is a unique state with the dimensions and charge given by \bref{Eqn:JMaRT} given by
\be
\ket{i'}=\ket{0}_{NS}=\mathcal U_{-\f{2j}{n_1 n_5},- \f{2 \bar j}{n_1 n_5}}|i \rangle=\mathcal U_{-(2 n_L+1),- (2 n_R+1)}|i \rangle
\ee
Even though its possible to write down the above state explicitly in the R-sector, the physics and presentation is cleaner if
we use the fact that its related to the NS vacuum by spectral flow.  The NS vacuum is  dual to the global AdS.

For a general amplitude of emission we need to calculate a 3-point function. However, since we spectral flowed the initial state to the NS vacuum
we now need to calculate a two point function. This restricts  the final states to be of the form
\be
|f \rangle =  \mathcal U_{(2 n_L+1, 2 n_R+1)} L_{-1}^{N_L} \bar L_{-1}^{N_R} V_{l,k,\bar k} |0\rangle~.
\ee
Equivalently,
\be
\ket{f'} = \mathcal U_{-(2n_L+1),-(2n_R+1)} \ket{f}= L_{-1}^{N_L} \bar L_{-1}^{N_R} V_{l,k,\bar k} |0\rangle~.
\ee

\subsubsection{Black Hole} \label{BlackHoleStates}

The black hole has an entropy and is thus dual to an ensemble in the CFT. We now show that the entropy and the temperature of the D1-D5-P black hole is
invariant under spectral flow. Thus the emission rates of non-rotating black holes with a certain left and right temperature are related by spectral flow
to emission rates for rotating black holes with the same left and right temperatures. Additionally we can work in the NS sector and relate the results to the R sector by spectral flow.

The left sector entropy is given by \cite{Cvetic:1998xh}
\be
s_L =2 \pi \sqrt{n_1 n_5} \sqrt{ \left( h - \frac{n_1 n_5}{4} \right) - \f{ j^2}{n_1 n_5}}
\ee
and the temperature by
\be
\f{1}{\tau_L} = \f{\partial s_L}{\partial h} \Big |_{j} = \f{1}{2 s_L}
\ee
Using \bref{SpectralFlow} we see that the left sector entropy and temperature (and similarly right sector entropy and temperature) are invariant under spectral flow.

Since the black hole is a thermal ensemble of pure states we can describe it by a thermal state. This state has the property
\be
\langle \psi | O | \psi \rangle_T = \f{\rom{Tr}(\rho_\rom{T} O)}{\rom{Tr}(\rho_\rom{T})}
\ee
where $\rho_\rom{T} = e^{-H/T}$ is the density matrix. In our case the thermal state also has angular momenta and is related by spectral flow to a thermal state wihout any angular momenta via
\be
\ket{\psi'}_{\tau_L,\tau_R} = \mathcal U_{-\f{2 j}{n_1n_5},-\f{2 \bar j}{n_1n_5}} \ket{\psi}_{\tau_L,\tau_R,j,\bar j} \label{BlackHoleSpectalFlowedStates}
\ee
where $j,\bar j$ are defined in \bref{Eqn:dictionary}.

\subsection{Amplitudes}

In \cite{Avery:2009tu} it was shown that the amplitude for emission is given as
\be
\mathcal A_\rom{emm} = -i  \kappa \int d\tau d\sigma ~ e^{iR(\omega \tau - \lambda \sigma)} ~ \bra{f} V_{l,k,\bar k} (\tau,  \sigma) \ket{i}~,
\ee
where
\be
\kappa = \left[\f{R}{2\pi^2 (l!)^2} \left( (\omega^2- \lambda^2) \f{Q_1 Q_5}{4 R^2} \right)^{l+1} \right]^\h~.
\ee
The amplitude for absorption is obtained by taking $\omega \to -\omega,$ and $\lambda \to -\lambda$. For later convenience we now define the light-cone momenta as
\be
p_L= R \f{\omega-\lambda}{2}, \qquad p_R =R \f{\omega+\lambda}{2}~. \label{lightconemomenta}
\ee
The amplitude in terms of the light-cone momenta and the coordinate $z$ takes the form
\bea
\mathcal A_\rom{emm} &=& -i \f{1}{2}  \kappa \int du dv~ e^{i(p_L u +p_R v)} ~ \bra{f} V_{l,k,\bar k} (u,v) \ket{i}~, \nn
&=& -i \f{1}{2} \kappa \int dz d \bar z ~ z^{p_L + \f{l}{2}} \bar z^{p_R + \f{l}{2}} ~ \bra{f} V_{l,k,\bar k} (z,\bar z) \ket{i}~,
\eea
where the extra factors of $z,\bar z$ come from the Jacobian of the conformal transformation. We can now use \bref{AmpUnderSpectFlow} to express the amplitude
in terms of the spectral flowed states \bref{SpectralFlowState} as
\bea
\mathcal A_\rom{emm} &=& -i \f{1}{2}  \kappa \int du dv~ e^{i(\tilde p_L u + \tilde p_R v)} ~ \bra{f'} V_{l,k,\bar k} (u,v) \ket{i'}~, \nn
&=& -i \f{1}{2} \kappa \int dz d \bar z ~ z^{\tilde p_L + \f{l}{2}} \bar z^{\tilde p_R + \f{l}{2}} ~ \bra{f'} V_{l,k,\bar k} (z,\bar z) \ket{i'}~, \label{Amp}
\eea
where the spectral flowed left and right light-cone momenta are
\be
\tilde p_L = p_L -\alpha k, \qquad \tilde p_R = p_R - \bar \alpha \bar k~. \label{ShiftedMomenta}
\ee
\subsection{Thermal States: Hawking Radiation and Superradiance}

We now want to show how the vertex operator of section \ref{VerOp} gives  Hawking radiation and superradiance for black holes. In the CFT language black holes
correspond to {\em thermal states} as described in section \ref{BlackHoleStates}. The probability after summing over all final states is given
by
\bea
|\mathcal A|^2_\rom{emm} &=& \f{1}{4} \kappa^2 \int du dv du' dv'~ e^{i (\tilde p_L (u-u') + \tilde p_R(v-v'))}  \langle V_{l,k,\bar k}^\dagger(u',v') V_{l,k,\bar k}(u,v) \rangle_{\tau_L,\tau_R}~, \nn
&=& \pi \tau  \kappa^2 \int d \tilde u d \tilde v ~ e^{-i (\tilde p_L \tilde u + \tilde p_R \tilde v)}  \langle V_{l,k,\bar k}^\dagger(\tilde u, \tilde v) V_{l,k,\bar k}(0,0) \rangle_{\tau_L,\tau_R}~.
\eea
Since the CFT lives on the $S^1$, we need the thermal two point function on a cylinder. For the states dual to a black hole the entropy is dominated by the long string sector (at least at the orbifold point). Therefore, we can take the spatial direction to be decompactified in the supergravity limit \cite{Maldacena:1997ih}. We then have
\be
\langle V^\dagger(w) V(0) \rangle_{\tau} =
\f{1}{i^{2 \Delta}} \left( \f{\pi \tau}{\sinh (\pi \tau (w-i\epsilon))} \right)^{2 \Delta}~,
\ee
Thus we can now write
\be
|\mathcal A|^2_\rom{emm} =   \pi \tau  \kappa^2 \int_{-\infty}^\infty d \tilde u \int_{-\infty}^\infty d \tilde v ~ e^{-i (\tilde p_L \tilde u + \tilde p_R \tilde v)}  \left( \f{\pi^2 \tau_L \tau_R}{\sinh (\pi \tau_L (\tilde u-i\epsilon)) \sinh (\pi \tau_R (\tilde v-i\epsilon))} \right)^{l+2}.
\ee
Using the identity
\be
\int_{-\infty}^\infty dx e^{-i\omega x} \left( \f{\pi \tau}{\sinh (\pi \tau (x \pm i \epsilon))} \right)^{2 \Delta} = i^{2\Delta} \f{(2\pi \tau)^{2 \Delta-1}}{\Gamma(2 \Delta)} e^{\pm \f{\omega}{2 \tau}} \left| \Gamma(\Delta+ i \f{\omega}{2 \pi \tau})\right|^2~,
\ee
we get
\be
|\mathcal A|^2_\rom{emm} = \pi \tau \kappa^2 \f{(4\pi^2 \tau_L \tau_R)^{l+1}}{\Gamma(l+2)^2} e^{- \f{\tilde p_L}{2 \tau_L}} \left| \Gamma(\f{l}{2}+1+ i \f{\tilde p_L}{2 \pi \tau_L})\right|^2 e^{- \f{\tilde p_R}{2 \tau_R}} \left| \Gamma(\f{l}{2}+1+ i \f{\tilde p_R}{2 \pi \tau_R})\right|^2.  \label{AmpSq}~
\ee
From \bref{BlackHoleSpectalFlowedStates} we see that $\alpha = - \f{2 j}{n_1 n_5}$ and $\bar \alpha = - \f{2 \bar j}{n_1 n_5}$. Further using \bref{ShiftedMomenta},  \bref{lightconemomenta}, \bref{VertexOperatorCharges} and \bref{Eqn:dictionary} we get
\bea
\tilde p_L &=& \h \left(R(\omega - \lambda)- \f{1}{n_1 n_5} (J_\psi + J_\phi) (m_\psi + m_\phi) \right)~, \nn
\tilde p_R &=& \h \left(R(\omega + \lambda)- \f{1}{n_1 n_5} (J_\psi - J_\phi) (m_\psi - m_\phi) \right)~.
\eea
Putting these together we have
\bea
\f{d \Gamma}{d\omega} &=& \f{1}{2\pi (l! (l+1)!)^2}\left(Q_1 Q_5 \f{\omega^2 -\lambda^2}{4} \right)^{l+1} (2 \pi T_L)^{l+1} (2 \pi T_R)^{l+1}  \nn
&& \times \  \tilde E_1 \, \tilde E_2 \ \left | \Gamma\left(\f{l}{2}+1 - i \log \tilde E_1 \right) \right|^2 \ \left | \Gamma\left(\f{l}{2}+1- i \log \tilde E_2 \right) \right |^2~,
\label{Eqn:GammaBHCFT}
\eea
where $\tilde E_1$ and $\tilde E_2$ are defined as 
\bea
\tilde E_1 &= & e^{-\f{\omega-\lambda - \f{1}{n_1 n_5 R} (J_\psi + J_\phi) (m_\psi + m_\phi)}{4 T_L}}~,\nn
\tilde E_2 & = & e^{-\f{\omega+\lambda - \f{1}{n_1 n_5 R} (J_\psi - J_\phi) (m_\psi - m_\phi)}{4 T_R}}~.
\eea
Using \bref{ChargeRadius} and \bref{NewtonsConstant}, we immediately see that this matches the gravity answer \bref{Eqn:GammaBHGravity}.

It is instructive to see  in more detail why we get an exponential growth of emission for superradiance while not for Hawking radiation.
The physics is exactly the same as the lasing action. As the emitted particles are bosons, the net rate of production of bulk quanta is
\be
\f{dn}{d\omega dt} = (n+1) \f{d \Gamma_\rom{emission}}{d\omega} - n \f{d \Gamma_\rom{absorption}}{d\omega}
\ee
For large number of quanta we can approximate $n+1 \approx n$ . We can get the expression for $\Gamma_\rom{absorption}$ by taking
$\tilde p_L \to - \tilde p_L$ and $\tilde p_R \to -\tilde p_R$ in \bref{Amp} and hence in \bref{AmpSq}. We see that
\be
\Gamma_\rom{emission} - \Gamma_\rom{absorption} \propto  \sinh \left[ -\log E_1 - \log E_2 \right]~,
\ee
where the proportinality constant is positive. For cases when this quantity is positive (superradiance) we get an exponentially growing number of quanta (black hole bomb \cite{Cardoso:2004nk}). When
the above is negative we have a steady Hawking radiation.

\subsection{Special States: Ergoregion Emission}

We now evaluate the rate of emission for special states dual to \cite{Jejjala:2005yu}. In particular, we want to calculate
\be
\mathcal A = - i \f{\kappa}{2} \int dz d \bar z z^{\tilde p_L+ \f{l}{2}} \bar z^{\tilde p_R + \f{l}{2}} \bra{f'} V_{l,k,\bar k}(z,\bar z) \ket{i'}~,
\ee
where normalized states are
\be
|i' \rangle = |0 \rangle, \qquad |f'\rangle = \sqrt{ \f{(l+1)!}{N! (N+l+1)!}}  \sqrt{ \f{(l+1)!}{\bar N! (\bar N+l+1)!}} (L_{-1}^N \bar L_{-1}^{\bar N}) V_{l,k,\bar k} |0\rangle
\ee
It can be shown that
\be
\langle f' | V_{l,k,\bar k} (z,\bar z) |i' \rangle = \sqrt{ \phantom{a}^{N+l+1} C_N}   \sqrt{ \phantom{a}^{\bar N+l+1} C_{\bar N}} ~ z^N~ \bar z^{\bar N}~.
\ee
Thus we get
\be
\mathcal A = - i \f{\kappa}{2} \sqrt{ \phantom{a}^{N+l+1} C_N}   \sqrt{ \phantom{a}^{\bar N+l+1} C_{\bar N}} \int dz d \bar z ~z^{\tilde p_L+ \f{l}{2}+N} \bar z^{\tilde p_R + \f{l}{2}+\bar N}~.
\ee
Going back to $\tau$ and $\sigma$ coordinates we get
\bea
\mathcal A &=& - i \kappa \sqrt{ \phantom{a}^{N+l+1} C_N}   \sqrt{ \phantom{a}^{\bar N+l+1} C_{\bar N}} \int d\tau d \sigma ~e^{i(\tilde p_L+ \tilde p_R + l+2 + N + \bar N) \tau} e^{i (\tilde p_L - \tilde p_R + N -\bar N) \sigma} \nn
&=& - i \kappa  \sqrt{ \phantom{a}^{N+l+1} C_N}   \sqrt{ \phantom{a}^{\bar N+l+1} C_{\bar N}} (2\pi) \delta(\tilde p_L+ \tilde p_R + l+2 + N + \bar N) (2\pi) \delta_{\tilde p_L + N, \tilde p_R + \bar N} \nn
&=& - i \kappa  \sqrt{ \phantom{a}^{N+l+1} C_N}   \sqrt{ \phantom{a}^{\bar N+l+1} C_{\bar N}} \nn
&& \times \ (2\pi) \ \delta(R \omega +(n_L+n_R+1) m_\psi + (n_L-n_R) m_\phi  + l+2 + N + \bar N) \nn
&& \times \ (2\pi) \ \delta_{\lambda R - (n_L-n_R) m_\psi -(n_L+n_R+1) m_\phi +\bar N - N,0}~.
\eea
Squaring to amplitude we get the probability 
\be
|\mathcal A|^2 = \f{T}{R^2} (2\pi)^3 \kappa^2 \p^{N+l+1} C_N    \p^{\bar N+l+1} C_{\bar N} \delta (\omega - \omega_0) \delta_{\lambda,\lambda_0}~,
\ee
where
\bea
\omega_0 &=& -l-2  - (n_L+n_R+1) m_\psi -(n_L-n_R) m_\phi - N - \bar N~, \nn
\lambda_0 &=& (n_L-n_R) m_\psi + (n_L+n_R+1) m_\phi + N- \bar N~,
\eea
Thus we get
\be
\f{d \Gamma}{d\omega} =  \f{4 \pi}{ (l!)^2 R}  \left( Q_1 Q_5 \f{\omega^2 - \lambda^2}{4 R^2} \right)^{l+1} \p^{N+l+1} C_N \p^{\bar N+l+1} C_{\bar N} ~ \delta_{\lambda, \lambda_0} \delta (\omega-\omega_0).
\ee
Comparing this result with the gravity answer \bref{RossEmmission} and taking into account the fact that $\lambda R$ is an integer and $\omega_0$ is the real part of $\omega$, we see that the two match exactly.

So far we have only gotten the emission rate. This by itself cannot lead to a classical instability. However recall from section \ref{CFT} that the target space of the CFT is
\be
(T^4)^{n_1 n_5}/ S^{n_1 n_5}
\ee
Thus all the states we make have to be symmetrized over all $n_1n_5$ strands. Also the vertex operator has to symmetrized in all its indices and its action should be symmetrized on all strands. We refer the reader to \cite{Avery:2009tu,Avery:2009xr} for the computation. The result is that if $\nu$ quanta have already been emitted then we have
\be
\f{d \Gamma}{d\omega} =  \f{4 \pi (1+\nu) }{  (l!)^2 R}  \left(Q_1 Q_5 \f{\omega^2 - \lambda^2}{4 R^2} \right)^{l+1} \p^{N+l+1} C_N \p^{\bar N+l+1} C_{\bar N} ~ \delta_{\lambda, \lambda_0} \delta (\omega-\omega_0).
\ee
For large $\nu$ this leads to an exponentially growing number of emitted particles. On the gravity side it is interpreted as the classical ergoregion instability \cite{Avery:2009tu,Avery:2009xr}.

\section{Understanding the Physics of Emissions}

In this section we studied emission from the D1-D5 system from both the gravity and the CFT perspective.
We saw that on the gravity side black holes emit Hawking radiation and if they are rotating they also  exhibit superradiance.
The special non-extremal fuzzballs of reference \cite{Jejjala:2005yu} on the other hand exhibit ergoregion instability. Non-extremal fuzzballs do not have a horizon or singularity
and thus have more in common with stars than black holes. On the CFT side we saw that all these emission processes---Hawking radiation, superradiance, and ergoregion instability---are infact different manifestations of the
same phenomenon. If the CFT is taken to be in a thermal state with no rotation we get  Hawking radiation; if it is taken in a thermal state with rotation
(which amounts to having a chemical potential) we get superradiance. For pure states dual to geometries constructed in \cite{Jejjala:2005yu} we get ergoregion instability. Thus from the CFT perspective a unified picture of
various emission phenomena emerges.  This unification strongly suggests that the D1-D5-P black hole is nothing but an
effective description of an ensemble of smooth geometries.

One might wonder if a pure state exhibits a classical instability \cite{Cardoso:2005gj} how can an ensemble not exhibit it. In
 \cite{Cardoso:2007ws} the authors studied the  classical stability of  the rotating D1-D5-P black hole. No instabilities were found.
We can now understand how this comes about. We first explain this on the CFT side and then propose a picture  as to how it might look on the gravity side.
\begin{figure}[ht]
\begin{center}
\subfigure[]{\label{fig:RossInitial}
	\includegraphics[width=5.3cm]{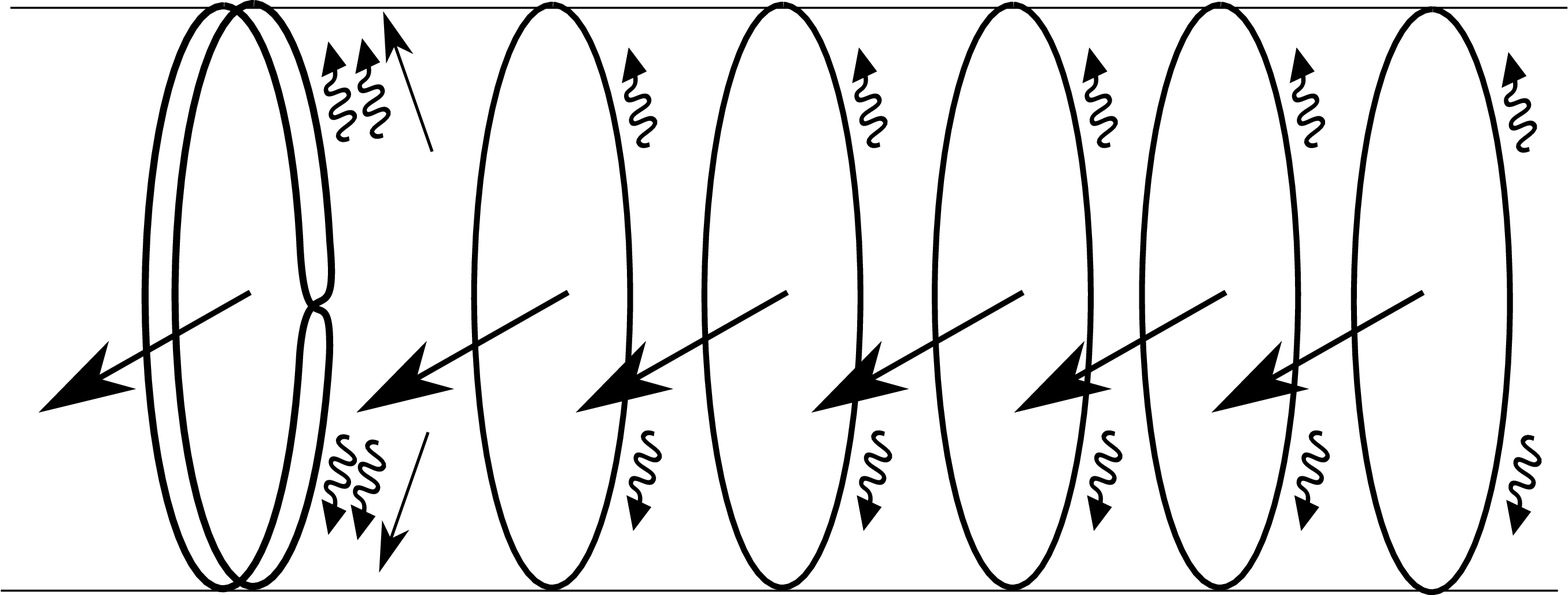}}
\hspace{15pt}
\subfigure[]{\label{fig:RossFinal}
	\includegraphics[width=5.3cm]{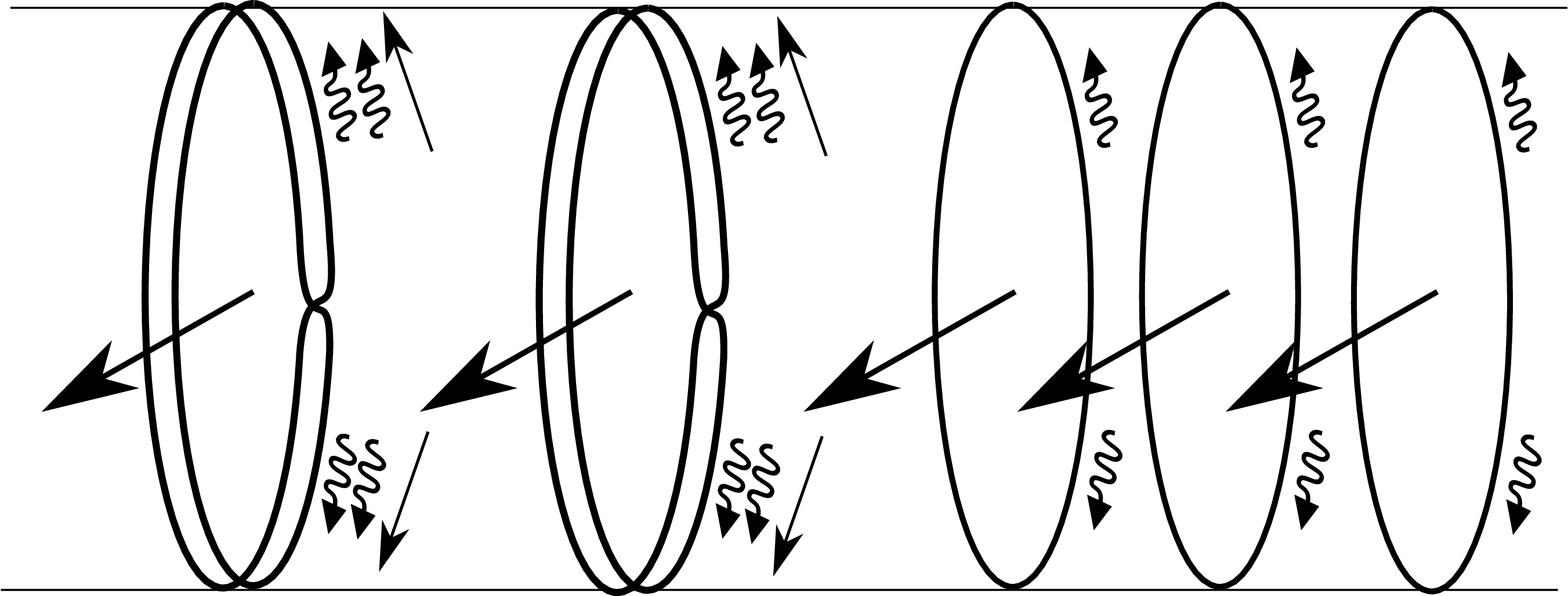}} \caption{The states after emission of one and two particles. The pictures correspond to
	 $l=1$ emission.\label{fig:AlphaDecay} }
\end{center}
\end{figure}

The states dual to the special non-extremal fuzzballs we studied have all the
component strings in the same state. We refer to this as the background state. The twist operator then twists $l+1$ of the component strings together
during the emission process. In figure \ref{fig:RossInitial} we show the CFT state after one particle has been emitted. In the large $n_1 n_5$ limit this is like a quantum sitting in the background state. The next emission twists a
new set of $l+1$ strands together (the process involving the original strings is highly suppressed). This is shown in figure \ref{fig:RossFinal}. Thus after the
second emission we have two quanta {\em in the same state}. This coherence of these quanta leads to a Bose-enhancement of the emission
rate and we get a classical instability.

\begin{figure}[ht] 
\begin{center}
\subfigure[]{\label{fig:RossInitialGravity}
	\includegraphics[height=3cm]{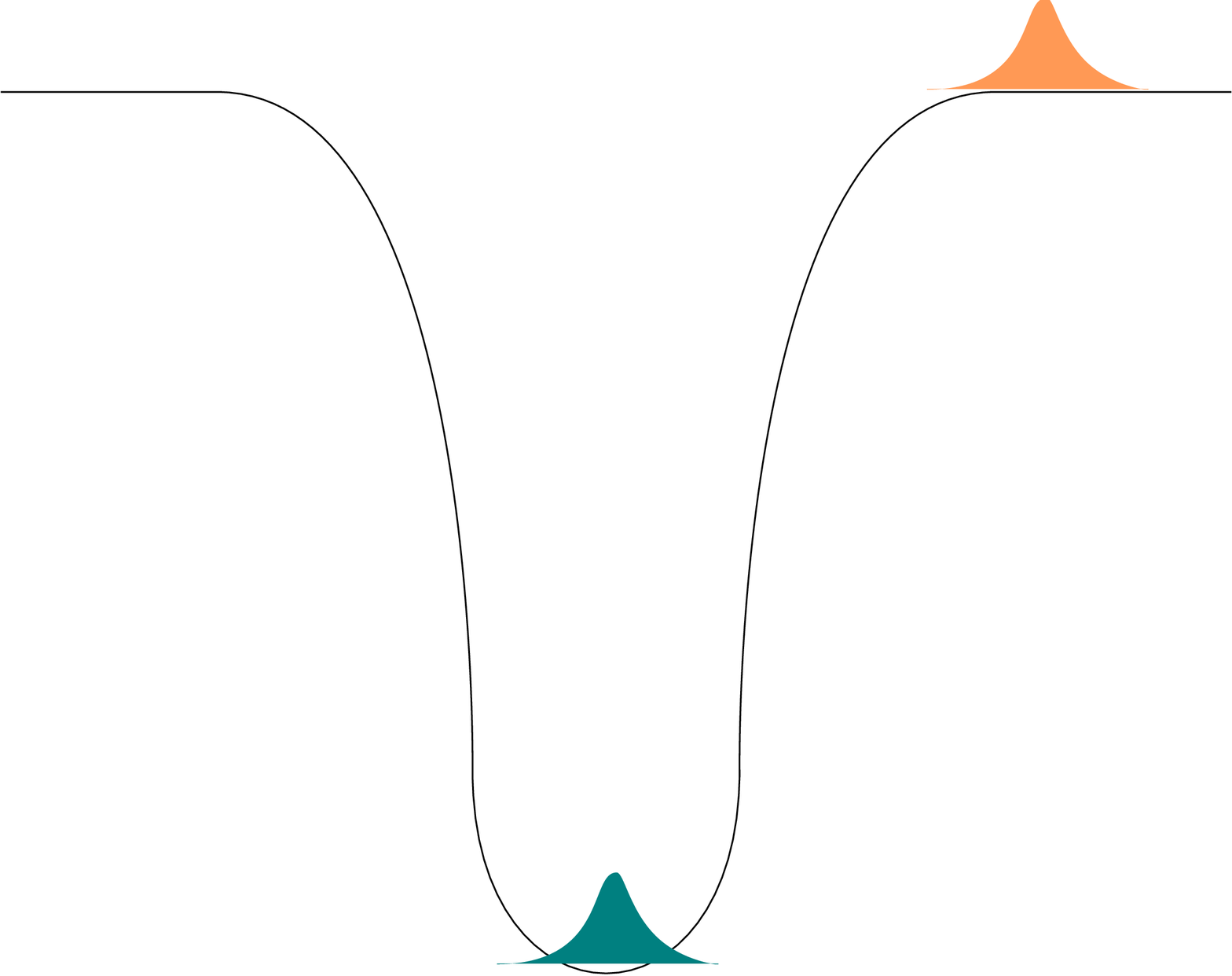}}
	\hspace{15pt}
\subfigure[]{\label{fig:RossFinalGravity}
	\includegraphics[height=3cm]{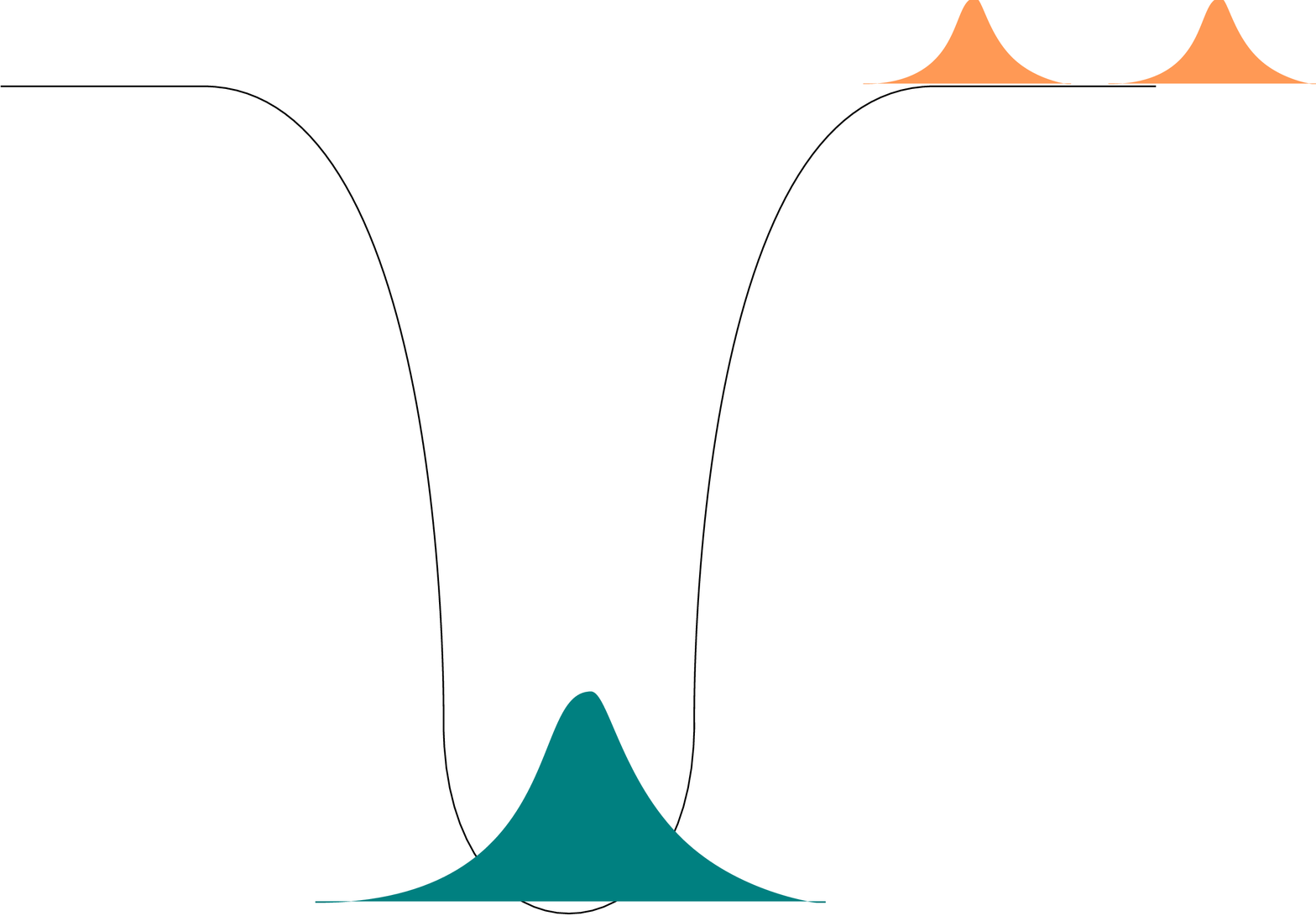}}
   \caption{The emission from special non-extremal fuzzballs. The smooth global AdS$_3$ of the core region has a discrete spectrum. We see the state after (a) one emission and after (b) two emissions. Subsequent emissions happen in the same state and the emission rates gets Bose-enhanced leading to a classical instability.}
   \label{fig:EmissionRoss}
   \end{center}
\end{figure}

On the gravity side the picture is as follows. The non-extremal fuzzballs all have a core AdS$_3$ region. Global AdS$_3$ has a discrete spectrum with energy levels separated by $\f{1}{R}$ and the process of emission produces an excitation in this region in a certain harmonic. This is shown in figure \ref{fig:RossInitialGravity}. The second emission causes another excitation in the same harmonic as shown in figure \ref{fig:RossFinalGravity}. The coherence of these quanta in the core AdS region leads to a Bose enhancement and hence to a classical instability.

\begin{figure}[ht] 
\begin{center}
\subfigure[]{\label{fig:GenericGravityOne}
	\includegraphics[height=3cm]{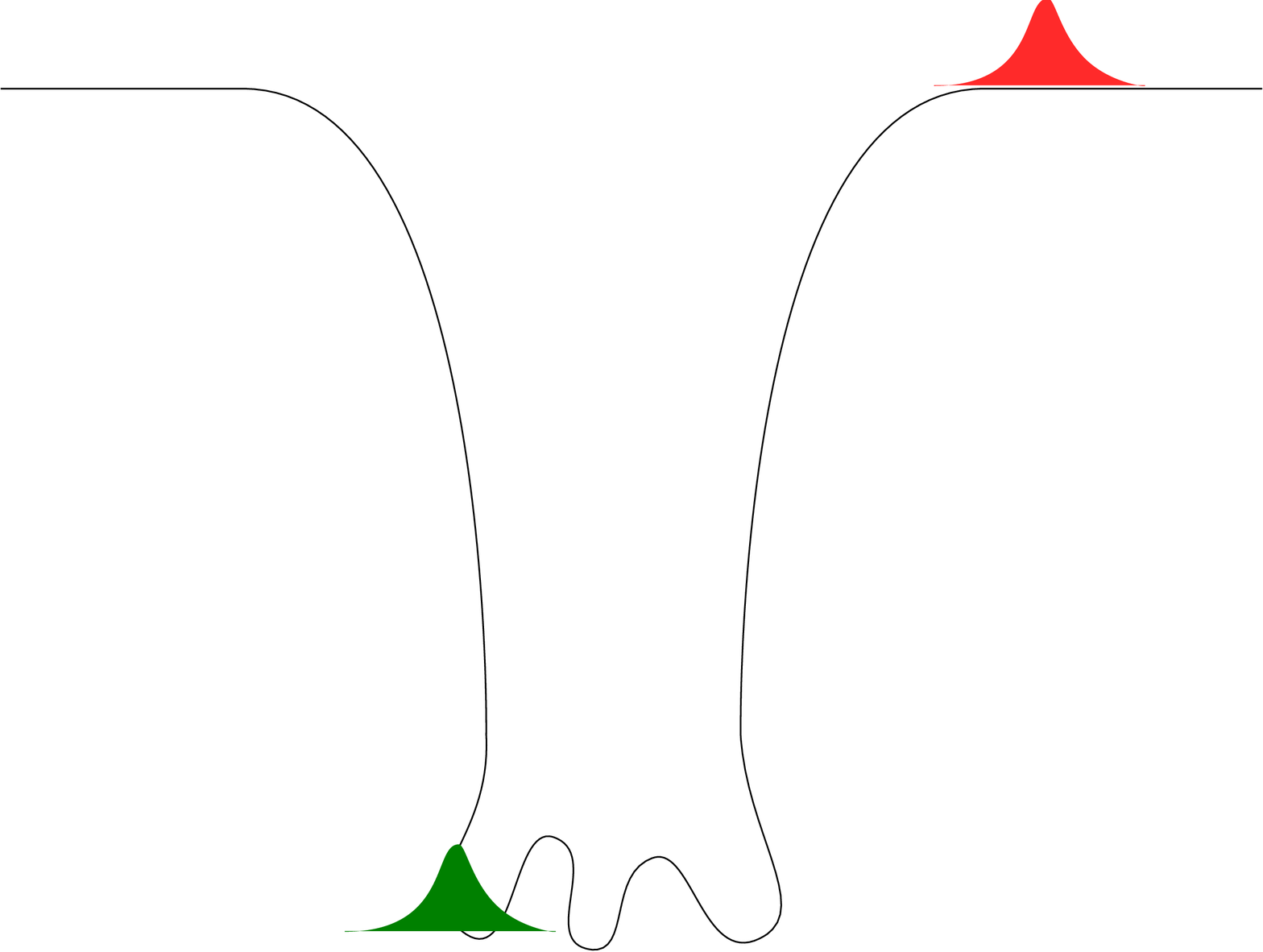}}
	\hspace{15pt}
\subfigure[]{\label{fig:GenericGravityTwo}
	\includegraphics[height=3cm]{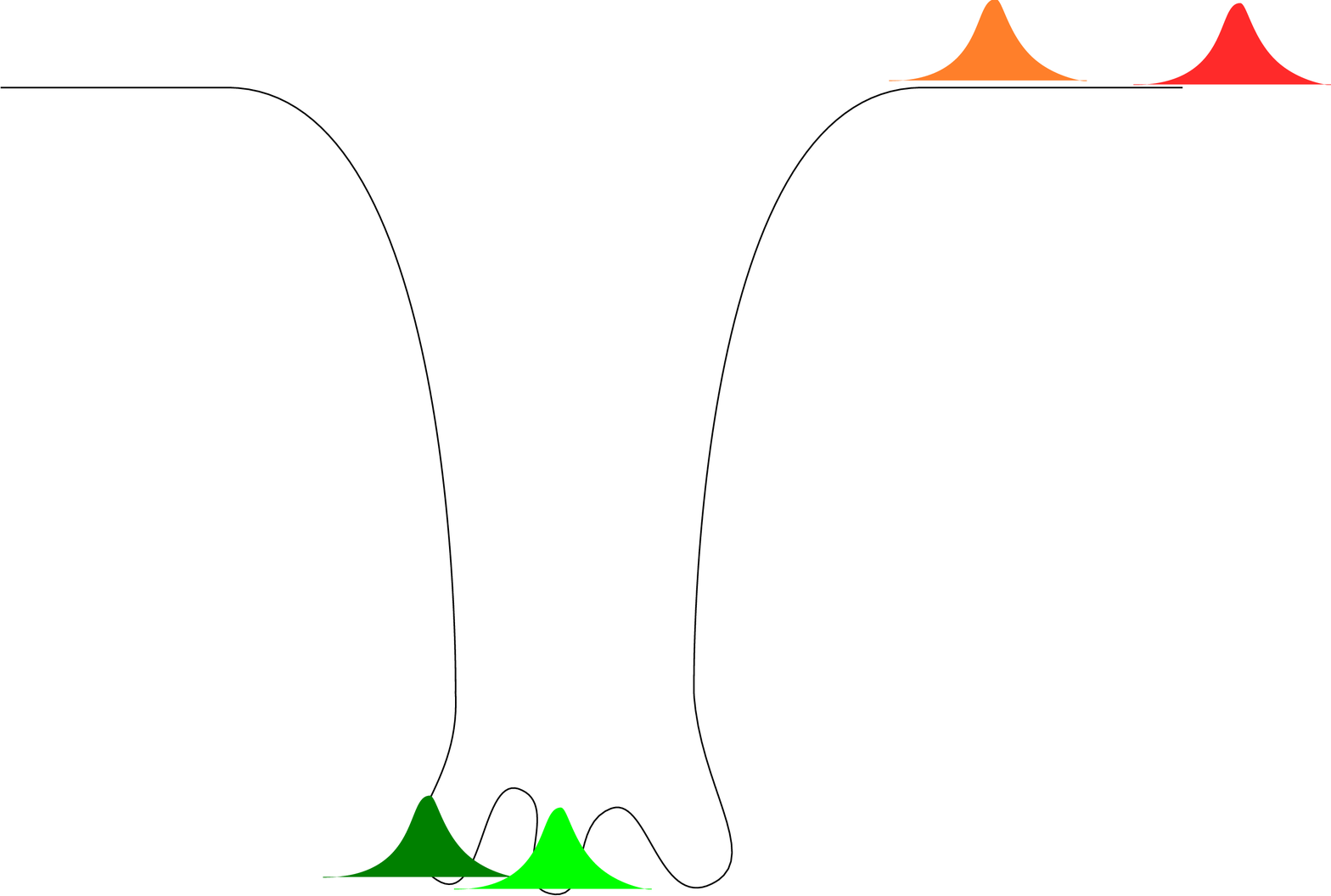}}
   \caption{The emission from typical non-extremal fuzzballs. The complicated cap in the core region has a band spectrum. We see the state after (a) one emission and after (b) two emissions. Subsequent emissions happening in the same state are highly suppressed and the emission rate  does {\em not} get Bose-enhanced. Thus there is no classical instability for typical non-extremal fuzzballs.}
   \label{fig:EmissionGeneric}
   \end{center}
\end{figure}

The generic state of the CFT has component strings of different lengths with different excitations.  The probability of the second particle being produced in the same state is suppressed by $O(\f{1}{n_1 n_5})$. Thus there is no Bose enhancement and no classical instability. On the gravity side a typical state is expected to have complicated throat structure and hence the energy levels in the throat region will be very close to each other. In figure \ref{fig:EmissionGeneric} we show a heuristic picture of this.  The first emission causes a particle to be produced in one corner of the cap (figure \ref{fig:GenericGravityOne}) and the second emission produces a particle  in another corner of the cap (figure \ref{fig:GenericGravityTwo}). Thus for a generic fuzzball there will not be any Bose enhancement and hence no classical instability. This is consistent with the results of  \cite{Cardoso:2007ws} that the rotating D1-D5-P black holes do not suffer from classical instability.

\begin{figure}[ht] 
\begin{center}
\subfigure[]{\label{fig:HawkingRadiation}
	\includegraphics[height=3.3cm]{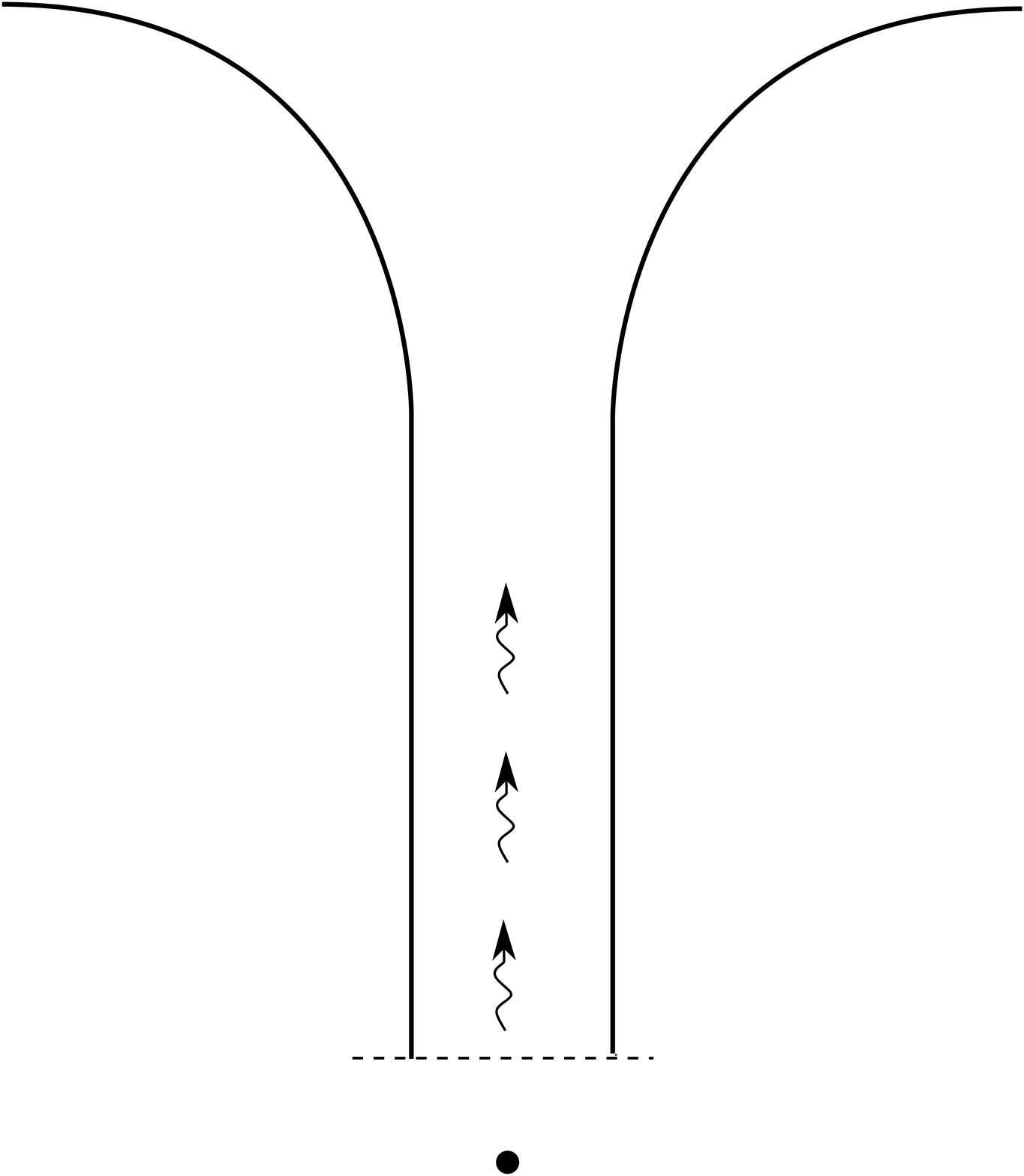}}
\hspace{35pt}
\subfigure[]{\label{fig:FuzzballRadiation}
	\includegraphics[height=3.3cm]{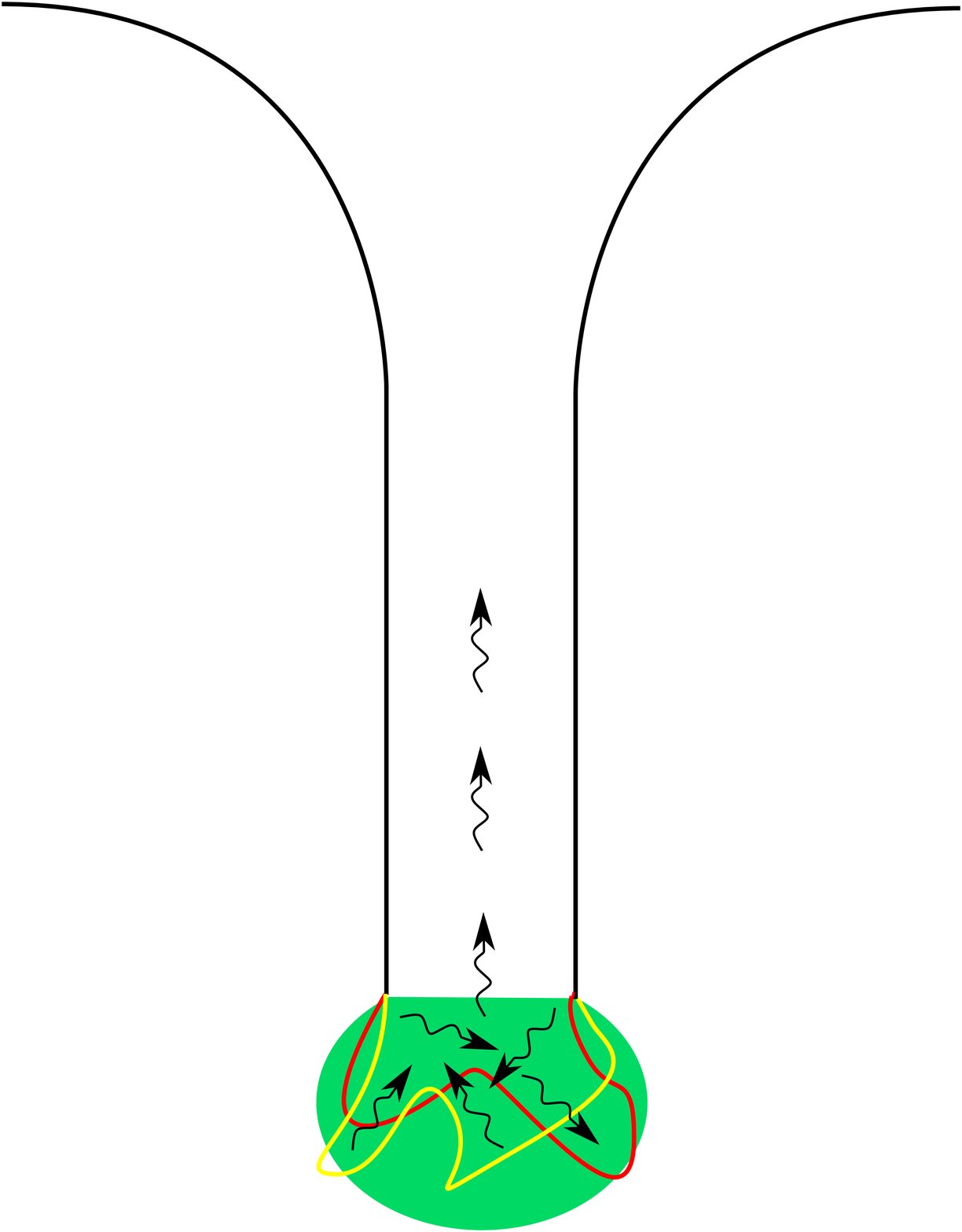}}
   \caption{The traditional picture of black hole evaporation (a) has particles produced as a result of pair creation near the horizon. Information of the state is localized near the singularity so the radiation does not carry any information. The analysis of this section supports the alternate picture (b) of radiation from fuzzballs. The microstates of black holes have a structure all the way to the horizon scale. In general, quanta are exchanged amongst different regions and some leak out to infinity carrying information.}
   \end{center}
\end{figure}

This suggests a modification of the traditional picture of black hole evaporation. In the traditional picture, shown in figure \ref{fig:HawkingRadiation},
Hawking quanta are produced as pair creation in the vicinity of the black hole. One particle of the pair has negative energy and it falls inside the horizon,
while the other particle  has positive energy and it stays outside and escapes to infinity.
Since such pairs are created out of vacuum fluctuations near the horizon,  they do not carry any information about the state that collapsed to form the black hole.
The fuzzball proposal modifies this picture.  In the fuzzball proposal black holes are  effective descriptions of ensemble of fuzzballs.
Typical fuzzball states have structure up to the horizon scale as shown in figure \ref{fig:FuzzballRadiation}.  The analysis of this section suggests
that inside a fuzzball different parts continuously exchange quanta; only some of these quanta escape to infinity from the
surface of the fuzzball. The average behavior of a typical fuzzball is captured by the corresponding black hole.

\newpage

\subsection*{Acknowledgements}
AV thanks the organizers of the Fifth
International Modave Summer School on Mathematical Physics held in
Modave, Belgium,  August 2009,
where the lectures on which these notes are based were given. The work of AV was supported by IISN Belgium conventions 4.4511.06 and
4.4514.08 and by the Belgian Federal Science Policy Office through the Interuniversity Attraction Pole P6/11.
The work of BDC was supported by the Foundation for Fundamental Research on Matter (FOM).
AV would like to thank Andrea Puhm and Cl\'ement Ruef for useful discussions at the summer school. BDC would like to thank Steven G. Avery, Samir D. Mathur and Yogesh K. Srivastava for many useful discussions.
%
%
\appendix

\chapter{D1-D5-P Black Hole} 

\renewcommand{\theequation}{\thechapter.\arabic{equation}}
\pagestyle{plain}

\label{App:D1D5P}
In this appendix we give the explicit metric for the rotating non-extremal D1-D5-P black hole. Explicit metric is needed for certain calculations presented in the main text.
Non-extremal fuzzball states of \cite{Balasubramanian:2000rt, Maldacena:2000dr, Lunin:2004uu, Giusto:2004id,Giusto:2004ip,Jejjala:2005yu}  
 are also closely related to the D1-D5-P black hole.  In the following, we discuss relevant properties of the D1-D5-P black hole and some salient features of the non-extremal
fuzzballs of \cite{Jejjala:2005yu} . 
Our presentation follows \cite{Cvetic:1996xz,Cvetic:1997uw,Jejjala:2005yu}.

Consider type IIB string theory compactified on
$
T^4\times S^1
$.
Let the volume of $T^4$ be $(2\pi)^4 V$ and the length of $S^1$ is $(2\pi) R$. The $T^4$ is described by coordinates $z_i$ and the $S^1$ by a coordinate $y$.
The noncompact $M_{4,1}$ is described by a time coordinate $t$, a radial coordinate $r$, and  angular $S^3$  coordinates $\theta, \psi, \phi$. The rotating non-extremal
D1-D5-P
black
hole
has angular momenta along $\psi, \phi$, parameterized by $a_1, a_2$.
The solution also carries three kinds of charges: $n_1$ units of D1 charge along $S^1$, $n_5$ units of D5 charge wrapped on $T^4\times S^1$, and $n_p$ units of momentum charge (P) along $S^1$.
These charges are parameterized by $\delta_1, \delta_5, \delta_p$, respectively. To avoid notational clutter we use the abbreviations
\be
s_i=\sinh\delta_i, ~~~c_i=\cosh\delta_i, ~~~~(i=1, 5, p)~.
\ee
The non-extremality of the solution is captured by the parameter $M$. With these preliminaries, we are in position to write the black hole solution of interest:
\begin{eqnarray} \label{3charge}
\mathrm{d}s^2&=&-\frac{f}{\sqrt{\tilde{H}_{1} \tilde{H}_{5}}}(
\mathrm{d}t^2 - \mathrm{d}y^2) +\frac{M}{\sqrt{\tilde{H}_{1}
\tilde{H}_{5}}} (s_p \mathrm{d}y - c_p
\mathrm{d}t)^2 \nonumber \\ &&  + \ \sqrt{\tilde{H}_{1} \tilde{H}_{5}}
\left(\frac{ r^2 \mathrm{d}r^2}{ (r^2+a_{1}^2)(r^2+a_2^2) - Mr^2}
+\mathrm{d}\theta^2 \right)\nonumber \\ && + \ \left( \sqrt{\tilde{H}_{1}
\tilde{H}_{5}} - (a_2^2-a_1^2) \frac{( \tilde{H}_{1} + \tilde{H}_{5}
-f) \cos^2\theta}{\sqrt{\tilde{H}_{1} \tilde{H}_{5}}} \right) \cos^2
\theta \mathrm{d} \psi^2 \nonumber \\ && + \ \left( \sqrt{\tilde{H}_{1}
\tilde{H}_{5}} + (a_2^2-a_1^2) \frac{(\tilde{H}_{1} + \tilde{H}_{5}
-f) \sin^2\theta}{\sqrt{\tilde{H}_{1} \tilde{H}_{5}}}\right) \sin^2
\theta \mathrm{d} \phi^2 \nonumber \\ && + \
\frac{M}{\sqrt{\tilde{H}_{1} \tilde{H}_{5}}}(a_1 \cos^2 \theta
\mathrm{d} \psi + a_2 \sin^2 \theta \mathrm{d} \phi)^2 \nonumber \\ &&
+ \ \frac{2M \cos^2 \theta}{\sqrt{\tilde{H}_{1} \tilde{H}_{5}}}[(a_1
c_1 c_5 c_p -a_2 s_1 s_5 s_p) \mathrm{d}t + (a_2 s_1
s_5 c_p - a_1 c_1 c_5 s_p) \mathrm{d}y ] \mathrm{d}\psi \nonumber \\
&& + \ \frac{2M \sin^2 \theta}{\sqrt{\tilde{H}_{1} \tilde{H}_{5}}}[(a_2
c_1 c_5 c_p - a_1 s_1
s_5 s_p) \mathrm{d}t + (a_1
s_1 s_5 c_p - a_2 c_1 c_5 s_p) \mathrm{d}y] \mathrm{d}\phi \nonumber
\\ && + \  \sqrt{\frac{\tilde{H}_1}{\tilde{H}_5}}\sum_{i=1}^4
\mathrm{d}z_i^2~,
\end{eqnarray}
where
\begin{eqnarray}
\tilde{H}_{i}=f+M\sinh^2\delta_i, \quad
f=r^2+a_1^2\sin^2\theta+a_2^2\cos^2\theta~.
\end{eqnarray}
The dilaton has the profile
\be
e^{2 \phi} = \f{\tilde H_1}{\tilde H_5}~.
\ee
The D1 and D5 charges of the solution produce a RR two-form gauge field. It is given by \cite{Giusto:2004id}
\begin{eqnarray}
C_2 &=& \frac{M \cos^2 \theta}{\tilde H_1} \left[ (a_2 c_1
  s_5 c_p - a_1 s_1 c_5
  s_p) dt + (a_1 s_1 c_5 c_p - a_2 c_1 s_5 s_p) dy \right] \wedge d\psi
  \nonumber \\
&& + \frac{M \sin^2 \theta}{\tilde H_1} \left[  (a_1 c_1
  s_5 c_p - a_2 s_1 c_5
  s_p) dt  + (a_2 s_1 c_5 c_p - a_1 c_1 s_5 s_p) dy \right] \wedge d \phi
  \nonumber \\
&& - \frac{M s_1 c_1}{\tilde H_1} dt \wedge dy -
  \frac{M s_5 c_5}{\tilde H_1} (r^2 + a_2^2 + M
  s_1^2) \cos^2 \theta d\psi \wedge d\phi~.
\end{eqnarray}
The angular momenta are given as
\bea
J_\psi &=& -  \f{\pi M}{4 G^{(5)}} (a_1 c_1 c_5  c_p - a_2 s_1 s_5 s_p)~,  \\
J_\phi &=& - \f{\pi M}{4 G^{(5)}} (a_2 c_1 c_5 c_p - a_1 s_1 s_5 s_p)~,
\eea
and the ADM mass is
\be
M_\rom{ADM} = \f{\pi M}{4 G^{(5)}} (s_1^2 + s_5^2 + s_p^2 + \f{3}{2})~. \label{Eqn:ADM_Mass}
\ee
It is convenient to define charge radii
\be
Q_1=M\sinh\delta_1\cosh\delta_1, ~~Q_5=M\sinh\delta_5\cosh\delta_5, ~~Q_p=M\sinh\delta_p\cosh\delta_p~.
\label{qdef}
\ee
Extremal SUSY solutions are reached in the limit
\be
M\rightarrow 0, ~~\delta_i\rightarrow\infty, ~~Q_i ~~{\rm fixed}~,
\ee
whereupon we get the BPS relation
\be
M_\rom{extremal,~ADM}=\frac{\pi}{4 G^{(5)}} [Q_1+Q_5+Q_p]~.
\ee
The integer charges of the solution are related to the $Q_i$'s through
\bea
Q_1&=& \frac{g \Regge^3}{V} n_1 \label{Eqn:Q1}~, \\
Q_5 &=& g \Regge n_5 \label{Eqn:Q5}~, \\
Q_p &=& \f{g^2 \Regge^4}{V R^2} n_p~.
\label{q1q5qp}
\eea

We can now see that metric splits into an outer flat space region and an inner `core' region. The core region asymptote to $AdS_3\times S^3 \times T^4$.
To see the outer flat region it suffices to take $r^2 \gg Q_i$. In this limit  we get flat space
\be
ds^2 = -\mathrm{d}t^2 + \mathrm{d}r^2 + r^2 \mathrm{d}\Omega_3^2 +\sum_{i=1}^4 \mathrm{d}z_i^2~.
\ee

To see the core part of the geometry we can use the Killing symmetries $\partial_t$ and $\partial_y$  to rewrite the metric
\eqref{3charge} as a fibration of these two directions over a
four-dimensional base space \cite{Jejjala:2005yu}. This gives
\begin{eqnarray} \label{fibred}
\mathrm{d}s^2 &=& \frac{1}{\sqrt{ \tilde{H_1}\tilde{H_5} } } \times \nn
& & \left\{
-(f-M) \left[\mathrm{d}\tilde{t}- (f-M)^{-1} M
\cosh\delta_1\cosh\delta_5 (a_1 \cos^2\theta \mathrm{d}\psi + a_2 \sin^2\theta
\mathrm{d}\phi) \right]^2 \right. \nonumber \\
&& + \ f \left.\left[
\mathrm{d}\tilde{y}+f^{-1} M \sinh\delta_1\sinh\delta_5 (a_2
\cos^2\theta \mathrm{d}\psi + a_1 \sin^2\theta \mathrm{d}\phi) \right]^2
\right\}\nonumber\\
&& + \ \sqrt{\tilde{H_1}\tilde{H_5}}\Bigg{\{}
\frac{r^2\mathrm{d}r^2}{ (r^2+a_1^2)(r^2+a_2^2)-Mr^2 }
+\mathrm{d}\theta^2 \nonumber \\
&& + \ (f(f-M))^{-1}\left[
  \left(f(f-M)+f a_2^2\sin^2\theta - (f-M)a_1^2\sin^2\theta
  \right)\sin^2\theta \mathrm{d}\phi^2 \right. \nonumber \\
  && + \  2  M
  a_1 a_2 \sin^2 \theta \cos^2 \theta \mathrm{d}\psi \mathrm{d}\phi
  \nonumber \\
  && + \ \left.\left(f(f-M)+fa_1^2\cos^2\theta -
  (f-M)a_2^2\cos^2\theta\right)\cos^2\theta \mathrm{d}\psi^2
  \right] \Bigg{\}} \nn & & + \ \sqrt{\frac{\tilde{H}_1}{\tilde{H}_5}}\sum_{i=1}^4
\mathrm{d}z_i^2,
\end{eqnarray}
where $\tilde t = t \cosh \delta_p - y \sinh \delta_p$, $\tilde y = y
\cosh \delta_p -t \sinh \delta_p $.

The near-horizon limit is obtained by assuming that $Q_1, Q_5
\gg M, a_1^2, a_2^2$, and focusing on the region $r^2 \ll Q_1, Q_5$. This amounts to taking $\tilde H_1 \approx Q_1$,
$\tilde H_5 \approx Q_5$, and approximating $M \sinh \delta_1
\sinh \delta_5 \approx M \cosh \delta_1 \cosh \delta_5 \approx
\sqrt{Q_1 Q_5}$. This gives us the
asymptotically AdS$_3 \times S^3$ geometry:
\begin{eqnarray}
\mathrm{d}s^2 &=& \frac{1}{\sqrt{ Q_1 Q_5 } }\left\{ -(f-M)
[\mathrm{d}\tilde{t}- (f-M)^{-1} \sqrt{Q_1 Q_5} (a_1 \cos^2\theta
\mathrm{d}\psi + a_2 \sin^2\theta \mathrm{d}\phi) ]^2
\right. \nonumber \\   &&   + \ f \left.[ \mathrm{d}\tilde{y}+f^{-1}
\sqrt{Q_1 Q_5} (a_2 \cos^2\theta \mathrm{d}\psi + a_1 \sin^2\theta
\mathrm{d}\phi) ]^2 \right\}\nonumber\\
&& + \  \sqrt{Q_1
  Q_5} \left. \Bigg{\{} \frac{r^2\mathrm{d}r^2}{ (r^2+a_1^2)(r^2+a_2^2)-Mr^2 }
+\mathrm{d}\theta^2 \right.\nonumber \\
&& + \ (f(f-M))^{-1}\left[
  \left(f(f-M)+f a_2^2\sin^2\theta - (f-M)a_1^2\sin^2\theta
  \right)\sin^2\theta \mathrm{d}\phi^2 \right. \nonumber \\
  && + \ 2 M
  a_1 a_2 \sin^2 \theta \cos^2 \theta \mathrm{d}\psi \mathrm{d}\phi
  \nonumber \\
  && + \  \left.\left(f(f-M)+fa_1^2\cos^2\theta -
  (f-M)a_2^2\cos^2\theta\right)\cos^2\theta \mathrm{d}\psi^2
  \right]  \Bigg{\}}\nn & &  + \ \sqrt{\f{Q_1}{Q_5}} \sum_{i=1}^4
\mathrm{d}z_i^2~.
\end{eqnarray}
This can be rewritten as
\bea
ds^2_\rom{inner,~BH} &=& \sqrt{Q_1 Q_5} \Bigg( - \left(\rho^2 - M_3 + \f{J_3^2}{4 \rho^2}\right) \mathrm{d} \tau^2 + \rho^2 \left(\mathrm{d}\varphi - \f{J_3}{2 \rho^2} \mathrm{d}\tau\right)^2 \nn
&& + \ \left(\rho^2 - M_3 + \f{J_3^2}{4 \rho^2}\right)^{-1}\mathrm{d}\rho^2 \nn
&& + \ \mathrm{d}\theta^2 + \sin^2 \theta (\mathrm{d} \phi  + \f{R}{\sqrt{Q_1 Q_5}}  \left((a_1 c_p-a_2 s_p) \mathrm{d} \varphi + (a_2 c_p -a_1 s_p) \mathrm{d} \tau)\right) )^2 \nn
&& + \ \cos^2 \theta (\mathrm{d} \psi  + \f{R}{\sqrt{Q_1 Q_5}}  \left((a_2 c_p-a_1 s_p) \mathrm{d} \varphi + (a_1 c_p -a_2 s_p) \mathrm{d} \tau)\right) )^2 \Bigg) \nn
&& + \ \sqrt{\f{Q_1}{Q_5}} \sum_{i=1}^4
\mathrm{d}z_i^2~,
\eea
where
\bea
M_3 &=& \f{R^2}{Q_1 Q_5} [ (M-a_1^2 -a_2^2) \cosh 2 \delta_p + 2 a_1 a_2 \sinh 2 \delta_p ]~, \nn
J_3 &=& \f{R^2}{Q_1 Q_5} [ (M-a_1^2 -a_2^2) \sinh 2 \delta_p + 2 a_1 a_2 \cosh 2 \delta_p ]~.
\eea
The coordinates $\rho,\tau$ and $\varphi$ are related to the coordinates in flat space as
\bea
\rho^2 &=& \f{R^2}{Q_1 Q_5} (r^2 + (M^2 - a_1^2 -a_2^2) \sinh^2 \delta_p + a_1 a_2 \sinh 2 \delta_p)~, \nn
\tau &=& \f{t}{R}~, \nn
\varphi &=& \f{y}{R}~.
\eea
In this limit the mass above extremality, $S^1$ momentum, and angular momenta become
\bea
\Delta M_\rom{ADM} &=& \f{\pi }{8 G^{(5)}} M \cosh 2 \delta_p~, \nn
Q_p &=&  \f{M}{2} \sinh 2 \delta_p~, \nn
J_\psi&=& - \f{\pi}{ 4 G^{(5)}} \sqrt{Q_1Q_5}( a_1 \cosh \delta_p - a_2 \sinh\delta_p)~, \nn
J_\phi&=& - \f{\pi}{4 G^{(5)}} \sqrt{Q_1Q_5}( a_2 \cosh \delta_p - a_1 \sinh\delta_p)~.
\eea

In  \cite{Jejjala:2005yu} non-extremal smooth solutions were constructed from the above black hole solution by taking a limit where the $y$ circle shrinks to zero at the
larger root of $g^{rr}$ without a conical defect. Each such state has an inner AdS region. The AdS region ends in a smooth cap. In the decoupling limit the metric for these states
is given by
a global AdS with an $S^3$ fibered over it:
\bea
ds^2 &=& \sqrt{Q_1 Q_5} \Bigg( - (\rho^2+1) d\tau^2 + \rho^2 d \varphi^2 + \f{d \rho^2}{\rho^2+1} \nn
&& + \ d\theta^2 + \sin^2 \theta (d \phi  -  \f{4 G^{(5)}}{\pi } \f{R}{Q_1 Q_5} (J_\psi d \varphi + J_\phi d \tau) )^2  \nn
&& + \  \cos^2 \theta (d \psi  -  \f{4 G^{(5)}}{\pi }\f{R}{Q_1 Q_5}( J_\phi d \varphi + J_\psi d \tau))^2  \Bigg) \nn
&& + \ \sqrt{\f{Q_1}{Q_5}} \sum_{i=1}^4
\mathrm{d}z_i^2~,
\eea
The coordinates $\rho,\tau$ and $\varphi$ are related to the coordinates in flat space as
\bea
\rho^2 &=& \f{R^2}{Q_1 Q_5} \left(r^2 + \f{Q_1 Q_5}{R^2} \f{s^4}{1-s^4} \right)~, \nn
\tau &=& \f{t}{R}~, \nn
\varphi &=& \f{y}{R}~,
\eea
where
\be
s^2= \left | \f{\sqrt{n_L(n_L+1)} - \sqrt{n_R(n_R+1)} }{\sqrt{n_L(n_L+1)} + \sqrt{n_R(n_R+1)}}\right |~.
\ee
The non-negative integers $n_L,n_R$ parameterize these states. The mass above extremality and various charges are given by
\bea
\Delta M_\rom{ADM} &=&\f{n_L(n_L+1) + n_R(n_R+1) }{R}~,   \nn
Q_p &=&   (n_L (n_L+1) - n_R (n_R+1)) \f{Q_1Q_5}{R^2}~, \nn
J_\psi&=& - (n_L+n_R +1) \f{\pi}{4 G^{(5)}} \f{Q_1Q_5}{R}~, \nn
J_\phi&=& (n_L-n_R) \f{\pi}{4 G^{(5)}} \f{Q_1Q_5}{R}~.
\eea
Since all these geometries are smooth geometries without horizons they do not have entropy and temperature. Further details can be found in \cite{Jejjala:2005yu}.

\cleardoublepage
\pagestyle{plain}
\def\href#1#2{#2}
\bibliographystyle{BibliographyStyle}
\addcontentsline{toc}{chapter}{Bibliography}

\bibliography{FuzzballBibliography}


\end{document}